\pdfoutput=1
\documentclass[a4paper,11pt]{article}

\newif\ifjhepstyle
\IfFileExists{jheppub.sty}{\jhepstyletrue}{\jhepstylefalse}
\ifjhepstyle
  \usepackage{jheppub}
\else
  \usepackage[margin=28mm]{geometry}
  \usepackage{amsmath,amssymb}
  \usepackage[colorlinks=true,linkcolor=blue,citecolor=blue,urlcolor=blue]{hyperref}
  \renewcommand{\abstract}[1]{\gdef\paperabstracttext{#1}}
  \newcommand{\printabstract}{\begin{center}\bfseries Abstract\end{center}%
    \begin{quotation}\small\paperabstracttext\end{quotation}}
\fi
\usepackage[T1]{fontenc}
\usepackage{mathtools}
\usepackage{mathrsfs}
\usepackage{array}
\usepackage{bm}
\usepackage{enumitem}
\usepackage{todonotes}
\usepackage{tikz-cd}
\usetikzlibrary{arrows.meta}

\newcommand{\slTwo}{\mathfrak{sl}_{2}}

\newcommand{\cF}{\mathcal{F}}
\newcommand{\cT}{\mathcal{T}}
\newcommand{\dd}{\mathrm{d}}

\numberwithin{equation}{section}
\hypersetup{pdfauthor={Sibasish Banerjee, Babak Haghighat, Jie Sun},
  pdftitle={Quantum Langlands correspondence for Painlev\'e II}}

\title{\boldmath Quantum Langlands correspondence for Painlev\'e II}
\ifjhepstyle
  \author[a]{Sibasish Banerjee,}
  \author[b,c]{Babak Haghighat,}
  \author[d]{Jie Sun}
  \affiliation[a]{Hetao Institute of Mathematics and Interdisciplinary Sciences (HIMIS),
    Shenzhen, China}
  \affiliation[b]{Yau Mathematical Sciences Center, Tsinghua University,
    Beijing 100084, China}
  \affiliation[c]{Beijing Institute of Mathematical Sciences and Applications (BIMSA),
    Huairou District, Beijing 101408, China}
  \affiliation[d]{Qiuzhen College, Tsinghua University, Beijing 100084, China}
  \emailAdd{sibasishbanerjee@himis-sz.cn}
  \emailAdd{babakhaghighat@tsinghua.edu.cn}
  \emailAdd{sun-j25@mails.tsinghua.edu.cn}
\else
  \author{Sibasish Banerjee\thanks{Hetao Institute of Mathematics and
    Interdisciplinary Sciences (HIMIS), Shenzhen, China.
    Email: \href{mailto:sibasishbanerjee@himis-sz.cn}{sibasishbanerjee@himis-sz.cn}}
    \and Babak Haghighat\thanks{Yau Mathematical Sciences Center,
    Tsinghua University, Beijing 100084, China; Beijing Institute of
    Mathematical Sciences and Applications (BIMSA), Huairou District,
    Beijing 101408, China.
    Email: \href{mailto:babakhaghighat@tsinghua.edu.cn}{babakhaghighat@tsinghua.edu.cn}}
    \and Jie Sun\thanks{Qiuzhen College,
    Tsinghua University, Beijing 100084, China.
    Email: \href{mailto:sun-j25@mails.tsinghua.edu.cn}{sun-j25@mails.tsinghua.edu.cn}}}
\fi
\date{}

\abstract{
We construct an irregular KZ--BPZ correspondence for quantum Painlev\'e II
from a rank-three Wakimoto--Whittaker module. The irregular KZ equation
and the quantum isomonodromy equation form a joint affine system whose
Drinfeld--Sokolov reduction yields scalar BPZ equations. We obtain
matrix-integral solutions of this system for one fundamental insertion. An exact
quantum Langlands duality exchanges the fundamental insertion with the zero
of the dual Whittaker differential and allows affine solutions to be
reconstructed from the BPZ equations. This duality interchanges the
classical and Nekrasov--Shatashvili limits, which yield the
Painlev\'e II Lax connection and an irregular Gaudin oper, respectively.

}

\begin{document}

\maketitle
\ifjhepstyle\else\printabstract\fi
\flushbottom

\section{Introduction}
\label{section:Introduction}

Conformal blocks in the presence of irregular singularities play a
central role at the interface of four-dimensional $\mathcal{N}=2$
supersymmetric gauge theories, two-dimensional conformal field theory,
and quantum integrable systems.  Under the AGT correspondence
\cite{Alday:2009aq}, colliding regular punctures produces higher-order
poles of the stress tensor or affine currents, mapping irregular
Virasoro and Kac--Moody states to four-dimensional Argyres--Douglas
superconformal field theories \cite{Argyres:1995jj,Gaiotto:2009ma,Bonelli:2011aa,Gaiotto:2012sf}
whose partition functions in the presence of surface defects govern
both the Painlev\'e/gauge-theory correspondence \cite{Bonelli:2016qwg,GrassiGu2019}
and non-semisimple three-dimensional topological quantum field
theories \cite{Dedushenko:2018bpp}.  On the affine side, flat
Knizhnik--Zamolodchikov (KZ) connections provide a complementary
route to braiding and knot invariants.  Earlier gauge-theoretic and
Yang--Yang constructions relate knot polynomials to monodromy and
braiding of integration cycles
\cite{Gaiotto:2011nm,Hu2016}, while irregular Liouville conformal
blocks make the associated Stokes phenomena explicit
\cite{Gu:2023bjg,HaghighatLiuReshetikhin2023}.
KZ connections with irregular singularities have recently been
shown to yield generalized braid-group
representations and topological invariants of links and tangles in
the configuration space of points, controlled by the interplay of
braiding and Stokes phenomena \cite{Gu:2026mlf}.  In the prototypical
example of the $H_{1}$ Argyres--Douglas theory, the relevant quantum
Painlev\'e II system can be formulated either through such irregular
affine current-algebra equations---retaining finite spin degrees of
freedom and higher current modes at infinity---or through scalar
Virasoro BPZ equations with degenerate surface-defect insertions.
Relating these two descriptions requires matching both the position
equations and the quantum isomonodromic deformation of the irregular
time.

Two complementary mechanisms bridge affine $\mathfrak{sl}_{2}$ current
algebra and Virasoro conformal field theory, and both extend naturally
when regular punctures collide to form irregular singularities.  The
first is the WZW--Liouville correspondence of Ribault and Teschner
\cite{RibaultTeschner2005,HikidaSchomerus2007} and its extension to
irregular punctures \cite{GaiottoLamyPoirier2013}, which introduces
additional degenerate operators on the Liouville side to capture the
spin degrees of freedom on the WZW side; from the four-dimensional
gauge-theory perspective, this correspondence implements the duality
between codimension-two monodromy defects and canonical surface defects
\cite{FrenkelGukovTeschner2015}.  The second mechanism is quantum
Drinfeld--Sokolov reduction
\cite{DrinfeldSokolov1985,FeiginFrenkel2007}, which passes directly
from affine Kac--Moody representations to Virasoro (or $W$-algebra)
modules.  In our irregular setting, implementing this reduction locally
requires twisting the affine currents by spectral flow and the
coordinate-dependent gauge transformations identified in
\cite{BanerjeeHaghighatLatifi2026}, while the quantum Langlands duality
of the resulting $W$-algebra representations exchanges weights and
coweights under $\kappa^{\vee}=\kappa^{-1}$ \cite{ArakawaFrenkel2019}.
Combining these two perspectives for a rank-three irregular module at
infinity allows us to reduce the joint affine KZ and quantum
isomonodromy system directly to a pair of coupled scalar BPZ equations.

Several prior works on irregular KZ and BPZ equations and quantized
Painlev\'e systems provide the immediate background for our
construction.  The interpretation of KZ equations as a deformation
of the isomonodromy problem goes back to \cite{Resh-KZ};
explicit confluent KZ equations and their relation to quantum
Painlev\'e systems were developed in \cite{JimboNagoyaSun2008}.
Confluent primary fields and higher-rank confluent KZ equations
were constructed in \cite{Nagoya_2010,NagoyaSunConfluent2011}.
Irregular Virasoro modules and their vertex operators were studied
in \cite{Gaiotto:2009ma,Gaiotto:2012sf,Nagoya:2015cja},
while induced singular modules for affine Lie algebras and the
associated irregular WZNW conformal blocks were developed in
\cite{Felder:2020ugk}.  At the critical level, irregular Gaudin
models provide the corresponding description in terms of opers
with higher-order poles \cite{FeiginFrenkelToledanoLaredo2010}.
In \cite{Nagoya2011}, hypergeometric solutions for
quantum Painlev\'e Schr\"odinger equations and the actions of
confluent KZ and quantized Painlev\'e Hamiltonians on finite polynomial
sectors were obtained using confluent Verma modules for truncated
current algebras.  The screening-integral realization of irregular
Virasoro blocks as $\beta$-deformed matrix integrals was developed in
\cite{Dijkgraaf:2009pc,Nishinaka:2012kn,Rim:2012tf,Choi:2014qha},
while the related $\beta$-ensemble formulation of
quantum Painlev\'e II was studied in \cite{Rumanov2014}.  Modified BPZ
equations and Lefschetz-integral flat connections for rank-one
Virasoro irregular vectors (with nonzero irregular data through
$L_{2}$) were derived in
\cite[Section~2.1]{HaghighatLiuReshetikhin2023}, and higher-degree
irregular Kac--Moody modules together with explicit degree-one
screening integrals and their constrained-WZW reduction were developed
in \cite{GukovHaghighatLiuReshetikhin2024,Haghighat2025}.  More
recently, \cite[Sections~2.2, 3.3 and~4.6]{BanerjeeHaghighatLatifi2026}
quantized the classical Painlev\'e Lax phase space into a Weyl-algebra
valued flat connection and connected it to irregular KZ and BPZ
equations.  Here we start directly from a completed Wakimoto module
for a rank-three irregular state characterized by
$a_{3}|W\rangle=-2|W\rangle$ (a fourth-order Cartan current pole),
whose nonconstant Whittaker character simultaneously generates the
irregular KZ connection and the quantum isomonodromy equation and
yields explicit matrix-integral solutions of the joint system.

From a broader perspective, our construction provides an explicit
irregular, chiral realization of ideas central to the Quantum Analytic
Langlands Correspondence recently formulated in
\cite{GaiottoTeschner2024}.  Away from the critical level
($\kappa=k+2\neq0$), \cite{GaiottoTeschner2024} interprets the Ribault--Teschner relation
between $H_{3}^{+}$ WZNW correlators (in the isotopic spin
$x$-representation) and Liouville correlators with additional
degenerate insertions as a quantum deformation of Sklyanin's
separation of variables (SoV) intertwining KZ/KZB and quantized
isomonodromic (BPZ) equations.  In that regular-puncture setting, the
raising current acts on the $x$-variables as a first-order differential
operator, so passing to Sklyanin's separated variables
$\bm{y}=(y_{1},\ldots,y_{n-3})$ (the zeros of $e(z)$ where the
degenerate Virasoro fields sit) requires an integral transform
\begin{equation}
	\Psi_{\mathrm{WZW}}(\bm{x};\bm{z})
	=\int\dd^{2}y_{1}\cdots\dd^{2}y_{n-3}\,
	\mathcal{K}_{\mathrm{SOV}}(\bm{x},\bm{y};\bm{z})\,
	\Phi_{\mathrm{Vir}}(\bm{y};\bm{z}).
\end{equation}
In the critical-level limit $\kappa\to0$, their correlators factorize
into a heavy-puncture exponential generating the Gaudin eigenvalues
times light-degenerate wave functions solving the scalar oper equation,
while discrete shifts generated by Verlinde line operators reduce to
grafting of opers (with irregular real opers and the Airy oper
discussed in \cite[Appendix~A.2]{GaiottoTeschner2024}).  In our
rank-three irregular setting on $\mathbb{P}^{1}$, the
Wakimoto--Whittaker coherent state at infinity is already in the
separated-variable polarization, since it diagonalizes the positive
modes of the raising current as the Whittaker differential
\eqref{equation:Whittaker meromorphic differential},
\begin{equation}
	\left.e_{z}(z)\,\dd z\right|_{\mathrm{Whittaker}}
	=\psi(z)\,\dd z,\qquad
	\psi(z)=u(z-q),
\end{equation}
whose simple zero $q$ is Sklyanin's separated variable from the
outset.  Consequently, passing from the affine KZ block $\cF$ to the
reduced Virasoro BPZ block $\mathcal{G}$ requires no SoV integral
transform over $q$, but only the local algebraic oper-gauge dressing
$\Phi=\bigl(u(z-q)\bigr)^{-1/2}\cF_{1}=Z_{\mathrm{irr}}(q)\,\mathcal{G}$
of \eqref{equation:PII integral oper Virasoro dictionary}:
Drinfeld--Sokolov reduction maps the fundamental insertion at $z$ and
the Whittaker zero at $q$ to the dual Virasoro degenerate defects
$\Phi_{1,2}(z)$ and $\Phi_{2,1}(q)$, whose $\kappa\to0$ (NS) limit
produces the exact same factorization into a heavy-defect Yang--Yang
action $\mathcal{W}_{\mathrm{NS}}(z,t)$ and a light-defect oper wave
function $\chi_{N}(q;z,t)$, while discrete shifts of the screening
number $N$ act as polynomial Darboux/B\"acklund transformations along
the Airy $\tau$-hierarchy of dual Painlev\'e II.

\subsection{Summary of results}
\label{subsection:summary_of_results}

We conclude this introduction with an overview of the main results of the paper and how the affine, Virasoro, and matrix-integral descriptions fit together.

\paragraph{Joint irregular KZ--isomonodromy system and twisted BPZ reduction.}
In Section~\ref{section:PII affine BPZ}, we construct a rank-three Wakimoto--Whittaker module at infinity characterized by a fourth-order Cartan pole ($a_{3}|W\rangle=-2|W\rangle$, $a_{1}|W\rangle=-t|W\rangle$) and the nonconstant raising-current character $\psi(z)=u(z-q)$.  Evaluating this state against a spin-$\frac{1}{2}$ primary insertion at $z$ produces a two-component affine conformal block $\cF(z,t;q,u)$ governed by the irregular KZ equation \eqref{equation:physical evaluation KZ} in $z$, which directly quantizes the classical Painlev\'e II Lax matrix $\mathbf{A}(z,t)$ of \eqref{eq:Laxpaircl}:
\begin{equation}
	\kappa\partial_{z}\cF
	=
	\begin{pmatrix}
		z^{2}+\dfrac{t}{2}-\partial_{q} & u(z-q)\\[2mm]
		\dfrac{2(z+q)}{u}\partial_{q}-2\partial_{u} & -z^{2}-\dfrac{t}{2}+\partial_{q}
	\end{pmatrix}\cF
	\;\xrightarrow[\;\partial_{q}\to -p,\; u\partial_{u}\to\theta\;]{\text{classical limit}}\;
	\mathbf{A}(z,t)\,\Psi,
\end{equation}
together with a compatible quantum isomonodromy equation in $t$ given in \eqref{equation:PII physical matrix isomonodromy equation} that quantizes the $t$-connection $\mathbf{B}(z,t)$.
Because $\psi(z)=u(z-q)$ vanishes at $z=q$, passing to the scalar oper gauge eliminates the lower component $\cF_{2}$ algebraically and introduces an apparent singularity at $q$.  Implementing twisted quantum Drinfeld--Sokolov reduction (with explicit spectral flow at $q$ and $\infty$) maps the spin-$\frac{1}{2}$ insertion at $z$ and the Whittaker zero at $q$ to the dual degenerate Virasoro fields $\Phi_{1,2}(z)$ and $\Phi_{2,1}(q)$ of weights $\Delta_{z}=(3\kappa^{-1}-2)/4$ and $\Delta_{q}=(3\kappa-2)/4$.  Dressing the top affine component into the reduced Virasoro conformal block $\mathcal{G}(q;z,t,u)$, the joint affine system reduces to the coupled BPZ equations \eqref{equation:q BPZ from twisted module}--\eqref{equation:z BPZ from twisted module} in $q$ and $z$.  Conversely, any solution $\mathcal{G}$ of the coupled BPZ system uniquely reconstructs a joint solution $\cF$ of the affine KZ and quantum isomonodromy equations.

\paragraph{Matrix-integral representation.}
In Section~\ref{section:PII integral solutions}, we obtain explicit solutions of the joint system in the integer screening sectors $N=\theta+\frac{3}{2}\in\mathbb{Z}_{\geq0}$ via the Wakimoto screening prescription.  With the cubic potential $V(x)=\frac{2}{3}x^{3}+tx$, the master $\beta$-ensemble eigenvalue integral ($\beta=2/\kappa$) takes the form
\begin{equation}
	\mathcal{I}_{N}(q,t;z)
	=\frac{1}{N!}\int_{\Gamma}\prod_{a=1}^{N}\dd x_{a}
	\prod_{a<b}(x_{a}-x_{b})^{\frac{2}{\kappa}}
	\prod_{a=1}^{N}(x_{a}-q)(z-x_{a})^{-\frac{1}{\kappa}}
	\exp\left[-\frac{1}{\kappa}\sum_{a=1}^{N}V(x_{a})\right],
\end{equation}
where $\Gamma$ is a twisted rapid-decay cycle.  Notice how the asymmetric roles of the two defects are built directly into the integrand: the Whittaker zero $q$ enters with a polynomial factor $(x_{a}-q)$ of exponent $1$, whereas the spin-$\frac{1}{2}$ puncture $z$ enters with exponent $-1/\kappa$.  The top component $\cF_{1}$ and the reduced Virasoro block $\mathcal{G}$ are proportional to $\mathcal{I}_{N}$ up to elementary exponential and algebraic prefactors, while the lower component $\cF_{2}$ is given by a $1/(z-x_{a})$ resolvent insertion inside the same integral.  Integration by parts along $\Gamma$ yields coupled Schwinger--Dyson equations in $z$ and $q$ that reproduce both the affine KZ--isomonodromy system and the two BPZ equations.

\paragraph{Classical and Nekrasov--Shatashvili limits.}
Section~\ref{section:PII limits} analyzes the two degeneration limits of the matrix integral \eqref{equation:PII matrix integral}:
\begin{itemize}[leftmargin=1.5em]
	\item \emph{Classical limit ($\kappa\to\infty$):} Rescaling $(z,q,t)$ and keeping the filling fraction $\sim N/\kappa$ finite, the WKB asymptotics of the affine block $\cF$ recover the classical Painlev\'e II Lax pair $(\mathbf{A},\mathbf{B})$ of \eqref{eq:Laxpaircl}.  Compatibility along the WKB characteristics yields the Painlev\'e II equation \eqref{eq:PIIscl} for the separated variable $q$, while the scalar oper reduction produces the classical Painlev\'e II spectral curve.
	\item \emph{Nekrasov--Shatashvili limit ($\kappa\to0$ at fixed $N$):} The integral $\mathcal{I}_{N}$ is dominated by a stationary-phase configuration of $N$ Bethe roots solving the irregular Gaudin Bethe equations.  The conformal block factorizes into an exponential of the heavy-defect Yang--Yang action, with exponent proportional to $\mathcal{W}_{\mathrm{NS}}(z,t)/\kappa$, times an oper wave function $\chi_{N}(q;z,t)$, which solves an irregular Gaudin oper equation and whose polynomial factor has degree $N$ in the light variable $q$.  For generic parameters, the Bethe eigenstates furnish a complete joint eigenbasis of the $(2N+1)$-dimensional polynomial sector.
\end{itemize}

\paragraph{Exact quantum Langlands duality.}
In Section~\ref{section:PII Langlands interpretation}, we show that the level-inversion symmetry $\kappa^{\vee}=\kappa^{-1}$ of the central charge $c(\kappa)=13-6(\kappa+\kappa^{-1})$ lifts to an exact, invertible equivalence between the original system at level $\kappa$ and a Langlands-dual system at level $\kappa^{\vee}$.  Fixing $a^{3}=\kappa^{-1}$, the duality map $\mathcal{T}_{\mathrm{BPZ}}$ acts on the conformal blocks by rescaling and exchanging the spin-$\frac{1}{2}$ insertion $z$ with the Whittaker zero $q$:
\begin{equation}
	\begin{gathered}
		\kappa^{\vee}=\kappa^{-1},\qquad
		q^{\vee}=az,\qquad
		z^{\vee}=aq,\qquad
		t^{\vee}=a^{2}t,\qquad
		\theta^{\vee}+\frac{5}{2}=\frac{\theta+\frac{5}{2}}{\kappa},\\[1mm]
		\mathcal{G}^{\vee}(q^{\vee};z^{\vee},t^{\vee},u^{\vee})
		=(u^{\vee})^{\frac{3}{2}(\kappa^{-1}-1)}\,
		\mathcal{G}\bigl(a^{-1}z^{\vee};a^{-1}q^{\vee},a^{-2}t^{\vee},(u^{\vee})^{1/\kappa}\bigr).
	\end{gathered}
\end{equation}
This transformation intertwines the two BPZ operators ($\mathcal{E}_{q^{\vee}}^{\vee}\mathcal{T}_{\mathrm{BPZ}}=a^{-2}\mathcal{T}_{\mathrm{BPZ}}\mathcal{E}_{z}$ and $\mathcal{E}_{z^{\vee}}^{\vee}\mathcal{T}_{\mathrm{BPZ}}=a^{-2}\mathcal{T}_{\mathrm{BPZ}}\mathcal{E}_{q}$) and lifts via affine reconstruction to an isomorphism $\mathcal{T}_{\mathrm{aff}}$ of the full joint KZ--isomonodromy solution spaces.  At the free-field level, the reduced Virasoro stress tensor is common to the magnetic and electric bosonic screenings $V_{-}=:e^{-\sqrt{2/\kappa}\varphi}:$ and $V_{+}=:e^{\sqrt{2\kappa}\varphi}:$, respectively, exchanged by level inversion as explained in Section~\ref{sec:hecke-separation-langlands}.  These reduced operators are distinct from the full Wakimoto screening current $\beta V_{-}$.  Because $\kappa\to0$ maps to $\kappa^{\vee}\to\infty$, quantum Langlands duality interchanges the Nekrasov--Shatashvili and classical limits: the finite-$N$ Gaudin Bethe sectors at $\kappa\to0$ map directly to the special integer-screening sectors of the dual classical Painlev\'e II system, where characteristic determinants define spectral curves preserved by the Painlev\'e II flow and shifts $N\mapsto N\pm1$ act as polynomial Darboux/B\"acklund transformations.

\paragraph{Seiberg--Witten geometry in the large-$N$ limit.}
Finally, in Section~\ref{section:PII planar SW geometry}, we take the correlated large-$N$ planar limit ($\epsilon_{1}\to 0$, $N\to\infty$ with $\Theta=\epsilon_{1}(N-1)$ fixed and $\kappa=-\epsilon_{2}/\epsilon_{1}$) of the Nekrasov--Shatashvili Bethe saddle.  On a two-cut branch $C_{+}\cup C_{-}$, the discrete Bethe roots condense onto the genus-one Seiberg--Witten curve of the $H_{1}$ Argyres--Douglas theory,
\begin{equation}
	\Sigma:\quad y^{2}=\zeta^{4}+T\zeta^{2}-2\Theta\zeta+u_{\mathrm{CB}},
	\qquad
	\lambda_{\mathrm{SW}}=y(\zeta)\,\dd\zeta,
\end{equation}
where the pullback of $\lambda_{\mathrm{SW}}$ at the light defect $\zeta=Q$ is recovered directly from the leading WKB momentum $-\dd S_{0}(Q)$ of the oper wave function $\chi_{N}$, while the filling fractions and the on-shell planar Yang--Yang action $\mathcal{W}_{0}$ reproduce the special coordinates $(a,a_{D})$ and the Seiberg--Witten prepotential $\mathcal{F}_{\mathrm{SW}}(a,T,\Theta)$.  Complementing this formal large-$N$ expansion, we also show in Section~\ref{subsection:PII exact finite reconstruction} that the exact finite-$N$ NS action $\mathcal{W}_{\mathrm{NS}}$ and wave action $S$ can be reconstructed in closed form from ratios of adjacent determinant $\tau$-functions $\widehat{\tau}_{N}$ and $\widehat{\tau}_{N+1}$.

\section{Irregular KZ and quantum isomonodromy: BPZ reduction}
\label{section:PII affine BPZ}

We use the Wakimoto free-field realization and the Sugawara
construction in the conventions of
\cite[Lectures~2 and~5]{EtingofFrenkelKirillov1998}.
The screening construction follows the hypergeometric approach to
KZ equations \cite{SchechtmanVarchenko1990,Kohno2014}.
Our classical-limit and NS parameter conventions follow
\cite{FrenkelGukovTeschner2015,NekrasovShatashvili2009}.
\subsection{The starting point: The Painlev\'e II connection}

The Painlev\'e connections constitute a family of integrable linear ODEs given by
\begin{equation}
    \frac{d}{dz} \Psi(z) = \mathbf{A}(z) \Psi(z),
\end{equation}
where $z \in \mathcal{C}_{0,n}$, with $\mathcal{C}_{0,n}$ being n-punctured sphere, and $\mathbf{A}(z) \in \mathfrak{sl}_2(\mathbb{C})$ complex and traceless. In general, there is a whole family of such connections with prescribed monodromy properties around punctures of $\mathcal{C}$ which one usually parametrized by one (or several) time variable $t$. The time derivative is then the generator of isomonodromic (monodromy preserving) deformations of the system which is captured by the following extended system of differential equations
\begin{align}
\label{eq:compatibilitycl}
    \partial_z \Psi(z,t) &= \mathbf{A}(z,t) \Psi(z,t), \\
    \partial_t \Psi(z,t) &= \mathbf{B}(z,t) \Psi(z,t),
\end{align}
satisfying the compatibility condition
\begin{equation} \label{eq:Pflatness}
    \partial_t \mathbf{A}(z,t) = \partial_z \mathbf{B}(z,t) + [\mathbf{B}(z,t), \mathbf{A}(z,t)].
\end{equation}
For the case of the Painlev\'e II with which we will be interested in this paper, the corresponding connections are given by
\begin{equation}
\label{eq:Laxpaircl}
    \mathbf{A} = \left(\begin{array}{cc}
        z^2 + p + t/2 & u(z-q) \\
        -\frac{2}{u}(pz + \theta + pq) & - (z^2 + p + t/2) 
    \end{array}\right), \quad \mathbf{B} = \left(\begin{array}{cc}
        z/2 & u/2 \\
        -p/u & -z/2
    \end{array}\right)~,
\end{equation}
where $p$, $q$ and $u$ separately depend on $t$ in such a way that the compatibility condition \eqref{eq:Pflatness} is satisfied, while $\theta$ is a constant parameter specifying the monodromy at infinity. The Painlev\'e II system is of fundamental importance for the superconformal $H_1$ Argyres-Douglas theory \cite{Bonelli:2016qwg}. This theory is reached within the moduli space of the $\mathcal{N}=2$ supersymmetric $SU(2)$ Yang-Mills theory by tuning parameters such that the Seiberg-Witten curve approaches a singular limit where two mutually non-local cycles collapse.

We also note that the compatibility equation \eqref{eq:compatibilitycl} implies the following equations of motion : 
\begin{equation}
\begin{split}
&\dot{u} = -q u , \quad \dot{q} = p + q^2+ \frac{t}{2}, \quad \dot{p} = -2pq - \theta
\end{split}
\end{equation}
From this one recovers the Painleve\'e II equation as 
\begin{equation}
\label{eq:PIIscl}
\ddot{q} = 2q^3 + q t + \alpha,
\end{equation}
where $\alpha = \frac12 - \theta$.

Taking the trace 
\begin{equation}
\label{eq:spectralcurve}
\frac12{\mathrm{Tr}}\, {\bf{A}}^2
= z^4 + t z^2 - 2\theta z + \frac{t^2}{4} + 2H_{II},
\end{equation}
where 
\begin{equation}
\label{eq:Hamiltoniancl}
H_{II} (t) = \frac{p^2}{2} + pq^2 + \frac{pt}{2} + q\theta
\end{equation}
which satisfies 
\begin{equation}
\ddot{H}^2_{II} = 2\dot{H}_{II}(H_{II}- t\dot{H}_{II}) - 4\dot{H}^3_{II} + \frac{\theta^2}{4}
\end{equation}
and the Hamiltonian $H_{II}(t)$ is related to the $\tau$-function through $H_{II}(t) = \frac{d}{dt}\log\tau_{II}$ \cite{Bonelli:2016qwg}. 

In \cite{BanerjeeHaghighatLatifi2026}, it was shown that when quantizing the phase space of the isomonodromic flow problem, one arrives at Knizhnik-Zamolodchikov (KZ) equations for $\mathfrak{sl}_2(\mathbb{C})$ conformal blocks with an irregular Whittaker module at infinity. For our Painlev\'e II problem, we use the differential operators
$\hat p=-\partial_q$ and $\hat\theta=u\partial_u$, which satisfy
\footnote{The sign convention for the irregular current contribution
differs from that of \cite{BanerjeeHaghighatLatifi2026}. Here the
positive modes are defined in the local coordinate $w=1/z$ at infinity,
with $J_z(z)=-z^{-2}J_w(1/z)$; the mode signs are chosen accordingly
to reproduce the same Painlev\'e II connection.}
\begin{equation}
    [\hat p,q]=-1,\qquad [\hat\theta,\phi]=1,
\end{equation}
where $\phi=\log u$. In the local coordinate $w=1/z$ at infinity,
the $\mathfrak{sl}_2$ currents are expanded as
\begin{equation}
    e(w) = \sum_{n \in \mathbb{Z}} e_n w^{-n-1}, \quad f(w) = \sum_{n \in \mathbb{Z}} f_n w^{-n-1}, \quad
    h(w) = \sum_{n \in \mathbb{Z}} h_n w^{-n-1}.
\end{equation}
We use $[h,e]=2e$, $[h,f]=-2f$, and $[e,f]=h$.
The positive current modes act on the rank-three irregular family as
\begin{align}
    h_1|W\rangle&=(-t-2\hat p)|W\rangle,
    &h_2|W\rangle&=0,
    &h_3|W\rangle&=-2|W\rangle,\\
    e_1|W\rangle&=uq|W\rangle,
    &e_2|W\rangle&=-u|W\rangle,
    &e_3|W\rangle&=0,\\
    f_1|W\rangle&=\frac{2}{u}(q\hat p+\hat\theta)|W\rangle,
    &f_2|W\rangle&=\frac{2}{u}\hat p|W\rangle,
    &f_3|W\rangle&=0.
\end{align}
The differential operators on the right describe derivatives of the
irregular-state family, as established by the Wakimoto construction
below. The coordinate transformation $J_z(z)=-z^{-2}J_w(1/z)$,
together with the transpose on the finite-spin evaluation column, gives
\begin{equation*}
    \widehat A(z)
    =-\sum_{n=1}^{3}z^{n-1}
    \begin{pmatrix}
        h_n/2&e_n\\
        f_n&-h_n/2
    \end{pmatrix}.
\end{equation*}
Substituting the mode actions and using the irregular-family insertion
identities reproduces the quantized Painlev\'e II connection
\eqref{equation:physical evaluation KZ}.

In the following, our task will be to rewrite these representations in terms of Wakimoto generators. As we will see, this will illuminate the connections between the KZ equation and the isomonodromic flow equations.  Based on this, in section \ref{subsection:PII original time connection}, we shall give the definition of
the {\it{quantum isomonodromy equations}} for Painlev\'e II system. 

\subsection{Affine Wakimoto module and the irregular KZ connection}
\label{subsection:PII original affine construction}

Set $\kappa=k+2\neq0$, with $[h,e]=2e$, $[h,f]=-2f$ and
$[e,f]=h$.  The invariant form is normalized by $(h,h)=2$ and
$(e,f)=1$.  We take $u\neq0$ and $z\neq q$, fixing
branches of the fractional powers used in the gauge transformations.
We use the standard spin-$\frac{1}{2}$ representation, with
$h=\operatorname{diag}(1,-1)$ and $e^{T}=f$, $f^{T}=e$.
On a conformal-block evaluation column the finite-point generators act
by transpose; for higher spin we retain the transpose explicitly.

In the coordinate $w=1/z$ at infinity, the Wakimoto fields are expanded as
\begin{equation}
	\beta(w)=\sum_{n\in\mathbb{Z}}\beta_{n} w^{-n-1},
	\qquad
	\gamma(w)=\sum_{n\in\mathbb{Z}}\gamma_{n} w^{-n},
	\qquad
	a(w)=\sum_{n\in\mathbb{Z}}a_{n} w^{-n-1}.
\end{equation}
Their nonzero commutators are
\begin{equation}
 [\beta_{n},\gamma_{m}]=\delta_{n+m,0},\qquad
 [a_{n},a_{m}]=2\kappa n\delta_{n+m,0}.
\end{equation}
We use the Wakimoto convention
\begin{equation}
	e(w)=\beta(w),
	\qquad
	h(w)=a(w)-2:\gamma\beta:(w),
\end{equation}
and
\begin{equation}
	f(w)=-:\gamma^{2}\beta:(w)+\gamma(w)a(w)+k\,\partial_{w}\gamma(w).
\end{equation}

The rank-three irregular state is defined by the following eigenvalue conditions:
\begin{equation}
	\beta_{3}|W\rangle=0,
	\qquad
	\beta_{2}|W\rangle=-u|W\rangle,
	\qquad
	\beta_{1}|W\rangle=uq|W\rangle,
	\qquad
	\gamma_{0}|W\rangle=0,
\end{equation}
and
\begin{equation}
	a_{3}|W\rangle=-2|W\rangle,
	\qquad
	a_{2}|W\rangle=0,
	\qquad
	a_{1}|W\rangle=-t|W\rangle,
	\qquad
	a_{0}|W\rangle=2\theta|W\rangle.
\end{equation}
The remaining annihilation conditions are
\begin{equation}
 \beta_{n}|W\rangle=0\quad(n\geq3),\qquad
 \gamma_{n}|W\rangle=0\quad(n\geq1),\qquad
 a_{n}|W\rangle=0\quad(n\geq4).
\end{equation}
These conditions are realized in the oscillator-degree completion
of a Fock module.  If $|2\theta\rangle$ is annihilated by
$\beta_{n\geq1}$, $\gamma_{n\geq0}$ and $a_{n\geq1}$, with
$a_{0}|2\theta\rangle=2\theta|2\theta\rangle$, then
\begin{equation}
 |W\rangle
 =\exp\left[
 uq\gamma_{-1}-u\gamma_{-2}
 -\frac{t}{2\kappa}a_{-1}-\frac{1}{3\kappa}a_{-3}
 \right]|2\theta\rangle.
\end{equation}
The oscillator commutators verify the eigenvalue conditions and
the horizontal-family identities used below.  Normal ordering is
defined in this polarization.  In particular,
$\gamma_{0}\beta_{0}|W\rangle=-|W\rangle$.
The required positive affine modes act as
\begin{align}
	e_{3}|W\rangle&=0,
	&
	e_{2}|W\rangle&=-u|W\rangle,
	&
	e_{1}|W\rangle&=uq|W\rangle,
	\\
	h_{3}|W\rangle&=-2|W\rangle,
	&
	h_{2}|W\rangle&=0,
	&
	h_{1}|W\rangle&=\left(-t+2u\gamma_{-1}\right)|W\rangle,
	\\
	f_{3}|W\rangle&=0,
	&
	f_{2}|W\rangle&=-2\gamma_{-1}|W\rangle,
	&
	f_{1}|W\rangle&=-2\gamma_{-2}|W\rangle.
\end{align}

\paragraph*{Irregular KZ equation.}

With $J_{z}(z)=-z^{-2}J_{w}(1/z)$, the finite-point KZ equation is generated by the Sugawara translation
operator $L_{-1}$ at the spin-$\frac{1}{2}$ insertion.  Before passing
to the evaluation column, the reduced operator is
\begin{equation}
	A(z)
	=
	-\sum_{n\geq 1}z^{n-1}
	\begin{pmatrix}
		h_{n}/2&f_{n}\\
		e_{n}&-h_{n}/2
	\end{pmatrix}.
\end{equation}
In the present convention this is the single-insertion form of the reduced irregular KZ Hamiltonian: if positive modes through $J_{p}$ may act nontrivially at infinity, their contributions occur with coefficients $1,z,\ldots,z^{p-1}$.

Substituting the rank-three positive-mode data gives the module-side matrix
\begin{equation}
	A(z)
	=
	\begin{pmatrix}
		z^{2}+\dfrac{t}{2}-u\gamma_{-1}
		&
		2z\gamma_{-1}+2\gamma_{-2}
		\\[2mm]
		u(z-q)
		&
		-z^{2}-\dfrac{t}{2}+u\gamma_{-1}
	\end{pmatrix}.
\end{equation}
The affine zero modes $e_{0},f_{0},h_{0}$ remain part of the representation and of the
global $\slTwo$ Ward constraints.

\paragraph*{The irregular-state connection.}

Write the conformal block as
\begin{equation}
	\cF(q,u;z)
	=
	\langle W(q,u)|
	V(v,z)|0\rangle,
\end{equation}
where $V(v,z)$ denotes the vertex operator corresponding to a spin $\frac{1}{2}$ field. The evaluations on the two finite-spin basis vectors $v = v_+$ and $v= v_-$ form the column
$\cF$.  A notation such as $\cF(\gamma_{-1}v)$ denotes a mode
insertion in the irregular factor via the corresponding contragredient action on $\langle W |$.  The parameters $q,u$ act through
the family of irregular states.

The two varying $\beta$-eigenvalues are $\beta_1 = uq$ and $\beta_2=-u$.  Using
\begin{equation}
	[\beta_{1},\gamma_{-1}]=1,
	\qquad
	[\beta_{2},\gamma_{-2}]=1,
	\qquad
	[\beta_{1},\gamma_{-2}]=[\beta_{2},\gamma_{-1}]=0,
\end{equation}
the natural flat connection on the family of irregular states is
\begin{equation}
	\nabla^{M}
	=
	\dd-\gamma_{-1}\,\dd(uq)+\gamma_{-2}\,\dd u.
\end{equation}
In $(u,q)$ coordinates,
\begin{equation}
	\nabla_{q}^{M}=\partial_{q}-u\gamma_{-1},
	\qquad
	\nabla_{u}^{M}=\partial_{u}-q\gamma_{-1}+\gamma_{-2}.
\end{equation}
For a horizontal irregular state,
\begin{equation}
	\nabla_{q}^{M}|W\rangle=0,
	\qquad
	\nabla_{u}^{M}|W\rangle=0,
\end{equation}
so that
\begin{equation}
	\partial_{q}|W\rangle
	=
	u\gamma_{-1}|W\rangle,
	\qquad
	\partial_{u}|W\rangle
	=
	q\gamma_{-1}|W\rangle-\gamma_{-2}|W\rangle.
\end{equation}
For a fixed conformal-block pairing, differentiation along the
horizontal family gives the insertion identities
\begin{equation}
	\label{equation:physical family insertion identities}
	\cF(\gamma_{-1}v)
	=
	\frac{1}{u}\partial_{q}\cF(v),
	\qquad
	\cF(\gamma_{-2}v)
	=
	\left(
	\frac{q}{u}\partial_{q}-\partial_{u}
	\right)\cF(v).
\end{equation}
Passing to the conformal-block evaluation column transposes the finite-spin matrix. Using the insertion identities above, the KZ equation becomes
\begin{equation}
	\label{equation:physical evaluation KZ}
	\kappa\partial_{z}\cF
	=
	\begin{pmatrix}
		z^{2}+\dfrac{t}{2}-\partial_{q}
		&
		u(z-q)
		\\[3mm]
		\dfrac{2(z+q)}{u}\partial_{q}-2\partial_{u}
		&
		-z^{2}-\dfrac{t}{2}+\partial_{q}
	\end{pmatrix}
	\cF.
\end{equation}

\subsection{Whittaker differential and local twists}

The geometric interpretation of the affine Whittaker conditions provides a natural starting point for understanding the irregular KZ–BPZ correspondence. In an affine Whittaker module, the prescribed eigenvalues of the positive modes of the raising current determine the singular structure of an associated meromorphic one-form. This differential encodes the irregular data of the affine representation and plays a central role in relating the current-algebra description to its Virasoro counterpart. In the $H_3^+$–Liouville correspondence, the zeros of the meromorphic differential associated with the raising current are identified with the positions of additional degenerate insertions in the Liouville theory \cite{RibaultTeschner2005}. This geometric correspondence extends naturally to irregular representations, where the higher-order poles of the differential reflect the Whittaker eigenvalues and encode the irregular singularities, as illustrated by the collision-limit construction of \cite{GaiottoLamyPoirier2013}. Consequently, the pole structure and zeros of the Whittaker differential provide the essential geometric data for the subsequent Drinfeld–Sokolov reduction and the emergence of the corresponding BPZ equations.


In the conventions used above, the conditions
$\beta_{1}=u q$, $\beta_{2}=-u$, and $\beta_{n\geq3}=0$ on the irregular
cyclic state prescribe, in the local coordinate $w=1/z$ at infinity,
\begin{equation}
 \left.e_{w}(w)\,\dd w\right|_{\mathrm{Whittaker}}
 =\left(\frac{u q}{w^{2}}-\frac{u}{w^{3}}\right)\dd w.
\end{equation}
Since an affine current has conformal weight one,
$e_{z}(z)=-z^{-2}e_{w}(1/z)$.  The same prescribed one-form therefore takes
the form
\begin{equation}
 \label{equation:Whittaker meromorphic differential}
 \left.e_{z}(z)\,\dd z\right|_{\mathrm{Whittaker}}
 =\psi(z)\,\dd z,
 \qquad \psi(z)=u(z-q).
\end{equation}
Thus $q$ is the zero of the Whittaker differential, while its pole at
infinity has order three.  Its simple zero creates the additional degenerate Virasoro module
at $q$ under reduction.  The rank-three designation refers to the
Cartan datum $a_{3}=-2$, whose current has a fourth-order pole;
the raising-current one-form displayed here has a third-order pole.

To relate the affine KZ system to a scalar BPZ equation, we first transform the matrix-valued connection into oper form through a suitable diagonal gauge transformation. More precisely, we introduce
$g_C=\psi(z)^{-h^{(z)}/2}, $
where $h^{(z)}$ denotes the Cartan generator of $\mathfrak{sl}_2$ acting on the finite-dimensional spin-$\frac12$ representation associated with the insertion at $z$. In the standard basis, $h^{(z)}=\operatorname{diag}(1,-1)$, so that $g_C=\operatorname{diag}(\psi(z)^{-1/2},\psi(z)^{1/2})$. This transformation normalizes the upper-right entry of the KZ connection to unity, allowing the second component of the two-component system to be eliminated in favor of a single scalar differential equation. At the level of chiral conformal field theory, this passage from the affine current-algebra description to the Virasoro description is understood through quantum Drinfeld–Sokolov (DS) reduction. An important feature of the present construction is that the Whittaker character $\psi(z)$ is not constant. Consequently, the reduction must account for its position dependence through appropriately twisted local trivializations of the reduced modules. In particular, the zeros and poles of $\psi$ determine where these local twists become singular and thereby influence the structure of the resulting Virasoro representations. For monomial characters, this construction is described by twisted DS reduction \cite{ArakawaFrenkel2019}, implemented on vertex-algebra modules through Li's $\Delta$-operator \cite{Li1997} .


 Decompose this twist into its singular and regular parts.  As a one-form,
 \eqref{equation:Whittaker meromorphic differential}
is represented at infinity by the coefficient
\begin{equation}
 \psi_{w}(w)=-u\,w^{-3}(1-qw),
 \qquad
 \partial_{w}\log\psi_{w}(w)
 =-\frac{3}{w}-\frac{q}{1-qw}.
\end{equation}
In the reduced Heisenberg realization, the local twist shifts the
Cartan field by $-\kappa\partial\log\psi$.  Consequently,
\begin{equation}
 \label{equation:singular and regular twist}
 -\kappa\partial_{w}\log\psi_{w}(w)
 =\frac{3\kappa}{w}+\frac{\kappa q}{1-qw}.
\end{equation}
The singular term $3\kappa/w$ is interpreted as a spectral-flow twist at infinity, labeled by $-3\check{\omega}$, where $\check{\omega}=h/2$ denotes the fundamental coweight of $\mathfrak{sl}_2$. The coefficient $-3$ reflects the third-order pole of the Whittaker differential at infinity. In the present conventions, this twist shifts the reduced Heisenberg field to $a^{\mathrm{sf}}(w)=a(w)+3\kappa/w$, and consequently shifts its zero mode by $a_0^{\mathrm{sf}}=a_0+3\kappa$. The second term in equation \eqref{equation:singular and regular twist} is holomorphic at $w=0$ and corresponds instead to a regular change of local trivialization. Likewise, near the simple zero $q$, the singular local shift is
 $-\kappa\partial_{z}\log\psi(z)=-\kappa/(z-q)$; it produces the
 level-two degenerate module at $q$.  These singular contributions enter
 the BPZ Ward identities \cite{ArakawaFrenkel2019,GaiottoLamyPoirier2013}.

We choose the reduced local trivialization in which the holomorphic
contribution is included in the scalar normalization, while the
infinity Ward modes retain the singular spectral-flow contribution.
In this trivialization the upper oper component $\Phi$ and the
Virasoro block $\mathcal{G}$ are related by
\begin{equation}
\Phi=Z_{\mathrm{irr}}(q)\,\mathcal{G},
 \qquad
 Z_{\mathrm{irr}}(q)
 =\exp\left(\frac{q^{3}}{3}+\frac{tq}{2}\right).
\end{equation}
Indeed, the regular first-order drift in the scalar KZ and quantum isomonodromy
connections is removed because
$2\partial_{q}\log Z_{\mathrm{irr}}(q)=2q^{2}+t$.
The scalar KZ and quantum isomonodromy equations below verify this normalization
simultaneously.  The singular terms in \eqref{equation:singular and regular twist}
remain in the reduced theory as the spectral-flow shift and the local
degenerate modules.  These local data yield the two BPZ equations derived below.  The
normalization is the irregular counterpart of the KZ--BPZ prefactor in
\cite{RibaultTeschner2005,GaiottoLamyPoirier2013}.

Finally, the irregular Heisenberg modes have an isomonodromic meaning.
The positive eigenvalues
$a_{3}=-2$, $a_{2}=0$, $a_{1}=-t$, and $a_{0}=2\theta$
fix the prescribed irregular type, its formal exponential factors and
formal powers.  Their
corresponding negative modes are quantum conjugate operators, following from
\begin{equation}
 [a_{n},a_{-m}]=2\kappa n\delta_{n,m},
 \qquad n,m\geq1.
\end{equation}
In particular, the present family is equipped with the horizontal
time equation
\begin{equation}
\partial_{t}|W\rangle
 =-\frac{a_{-1}}{2\kappa}|W\rangle.
\end{equation}
Accordingly, $a_{-1}$ acts as the quantum generator of the deformation
conjugate to the coefficient $a_{1}=-t$.  Its semiclassical limit generates the isomonodromic deformation.

\subsection{Oper-gauge transformation and scalar reduction}

 We now apply the oper-gauge transformation
\begin{equation}
	\label{equation:oper gauge transformation}
	g_{C}
	=
	\psi(z)^{-h^{(z)}/2}
	=
	\begin{pmatrix}
		\bigl(u(z-q)\bigr)^{-1/2}&0\\
		0&\sqrt{u(z-q)}
	\end{pmatrix}
\end{equation}
where
\begin{equation}
	\psi(x)=u(x-q).
\end{equation}
The upper-right entry $u(z-q)$ in \eqref{equation:physical evaluation KZ} is normalized directly to $1$.

Define
\begin{equation}
	\cF_{C}=g_{C}\cF.
\end{equation}
A direct calculation from \eqref{equation:physical evaluation KZ} gives
\begin{equation}
	\label{equation:oper gauge KZ}
	\kappa\partial_{z}\cF_{C}
	=
	\begin{pmatrix}
		z^{2}+\dfrac{t}{2}-\partial_{q}-\dfrac{\kappa-1}{2(z-q)}
		&
		1
		\\[3mm]
		2(z^{2}-q^{2})\partial_{q}
		-2u(z-q)\partial_{u}
		-2z
		&
		-z^{2}-\dfrac{t}{2}+\partial_{q}+\dfrac{\kappa+1}{2(z-q)}
	\end{pmatrix}
	\cF_{C}.
\end{equation}

Let $\Phi$ be the upper component of $\cF_{C}$.  Since the upper-right entry in
\eqref{equation:oper gauge KZ} is $1$, the lower component can be
eliminated directly using the first row.  This gives
\begin{equation}
	\label{equation:exact scalar KZ}
	\begin{aligned}
		0={}&
		\Bigg[
		\kappa^{2}\partial_{z}^{2}
		-
		\frac{\kappa}{z-q}\partial_{z}
		-
		\partial_{q}^{2}
		+
		\left(
		2q^{2}+t-\frac{\kappa}{z-q}
		\right)\partial_{q}
		+
		2u(z-q)\partial_{u}
		\\
		&\quad
		-
		\left(z^{2}+\frac{t}{2}\right)^{2}
		-
		(\kappa-2)z
		+
		\kappa q
		+
		\frac{\kappa(2q^{2}+t)}{2(z-q)}
		-
		\frac{3(\kappa^{2}-1)}{4(z-q)^{2}}
		\Bigg]\Phi .
	\end{aligned}
\end{equation}

Before the oper-gauge transformation, the diagonal global Cartan Ward identity is
\begin{equation}
	\cF\left(
	\left(
	h_{0}^{(\infty)}+h^{(z)}
	\right)v
	\right)
	=
	0,
\end{equation}
Here, $h_0^{(\infty)}$ denotes the Cartan zero mode acting on the irregular Wakimoto module at infinity, whereas $h^{(z)}$, as introduced before, is the Cartan generator acting on the spin-$\frac12$ representation inserted at $z$.

On the irregular cyclic state,
\begin{equation}
	h_{0}^{(\infty)}
	=
	a_{0}
	-
	2\gamma_{0}\beta_{0}
	-
	2\gamma_{-1}\beta_{1}
	-
	2\gamma_{-2}\beta_{2},
\end{equation}
we therefore obtain
\begin{equation}
	h_{0}^{(\infty)}|W\rangle
	=
	\left(
	2\theta+2
	-2uq\,\gamma_{-1}
	+2u\,\gamma_{-2}
	\right)|W\rangle.
\end{equation}
On conformal blocks, this gives
\begin{equation}
	\begin{aligned}
		\cF\left(h_{0}^{(\infty)}v\right)
		&=
		\left(2\theta+2\right)\cF(v)
		-
		2uq\,\cF(\gamma_{-1}v)
		+
		2u\,\cF(\gamma_{-2}v)
		\\
		&=
		\left(
		2\theta+2
		-2q\partial_{q}
		+2q\partial_{q}
		-2u\partial_{u}
		\right)\cF(v)
		\\
		&=
		\left(
		2\theta-2u\partial_{u}+2
		\right)\cF(v).
	\end{aligned}
\end{equation}
Hence, 
\begin{equation}
	\left[
	\left(
	2\theta-2u\partial_{u}+2
	\right)\mathbf{1}
	+
	h^{T}
	\right]\cF
	=
	0.
\end{equation}
Since $h^{T}=h$, the upper component satisfies
\begin{equation}
	\left(
	2\theta-2u\partial_{u}+3
	\right)\cF_{1}
	=
	0.
\end{equation}

The upper component in oper gauge is
\begin{equation}
	\Phi
	=
	\left(u(z-q)\right)^{-1/2}\cF_{1},
	\qquad
	\cF_{1}
	=
	\sqrt{u(z-q)}\,\Phi.
\end{equation}
The Cartan Ward identity becomes
\begin{equation}
	\label{equation:scalar Cartan Ward}
	\left(
	2\theta-2u\partial_{u}+2
	\right)\Phi
	=
	0,
\end{equation}
or equivalently
\begin{equation}
	2\theta=2u\partial_{u}-2
\end{equation}
on the scalar KZ block.  The irregular prefactor
\eqref{equation:irregular prefactor} is independent of $u$, so the same
homogeneity relation holds for $\mathcal{G}$.

The Virasoro block $\mathcal{G}$ is obtained from the component $\Phi$ after the
oper-gauge transformation by a further prefactor
\begin{equation}
	\label{equation:irregular prefactor}
	\Phi
	=
	Z_{\mathrm{irr}}(q)\,\mathcal{G}
	=
	\exp\left(\frac{q^{3}}{3}+\frac{tq}{2}\right)\mathcal{G}.
\end{equation}
Conjugating \eqref{equation:exact scalar KZ} by this factor gives
\begin{equation}
	\label{equation:physical combined scalar KZ}
	\begin{aligned}
		0={}&
		\Bigg[
		\kappa^{2}\partial_{z}^{2}
		-
		\partial_{q}^{2}
		-
		\frac{\kappa}{z-q}
		\left(
		\partial_{z}+\partial_{q}
		\right)
		+
		2u(z-q)\partial_{u}
		\\
		&\qquad
		-
		(z^{4}-q^{4})
		-
		t(z^{2}-q^{2})
		-
		(\kappa-2)(z-q)
		-
		\frac{3(\kappa^{2}-1)}{4(z-q)^{2}}
		\Bigg]\mathcal{G}.
	\end{aligned}
\end{equation}
This completes the affine route.  We now turn to the Virasoro modules and their BPZ equations.

\subsection{Quantum isomonodromy equation}
\label{subsection:PII original time connection}

We first derive the quantum isomonodromy equation.  The mode $a_{-1}$ can be rewritten entirely in terms of affine
modes before Drinfeld--Sokolov reduction.  From $h=a-2:\gamma\beta:$,
\begin{equation}
	\begin{aligned}
		a_{-1}|W\rangle
		={}&
		\left(
		h_{-1}
		+2\gamma_{-1}e_{0}
		+2uq\gamma_{-2}
		-2u\gamma_{-3}
		\right)|W\rangle.
	\end{aligned}
\end{equation}
The zero mode of the Wakimoto field $f$ acts as
\begin{equation}
f_{0}|W\rangle
	=
	\left(
	u\gamma_{-1}^{2}
	-t\gamma_{-1}
	-2\gamma_{-3}
	\right)|W\rangle.
\end{equation}
Using
\begin{equation}
	f_{2}|W\rangle=-2\gamma_{-1}|W\rangle,
	\qquad
	f_{1}|W\rangle=-2\gamma_{-2}|W\rangle,
\end{equation}
one obtains the affine identity on the cyclic state
\begin{equation}
	\label{equation:PII a minus one affine resolution}
		a_{-1}|W\rangle
		=
		\Bigg(
		h_{-1}
		-f_{2}e_{0}
		-uqf_{1}
		+uf_{0}
		-\frac{u^{2}}{4}f_{2}^{2}
		-\frac{ut}{2}f_{2}
		\Bigg)\,|W\rangle.
\end{equation}
The two orderings of $f_{2}e_{0}$ agree on the cyclic state, since
$[e_{0},f_{2}]|W\rangle=h_{2}|W\rangle=0$.

The quantum isomonodromy connection gives
\begin{equation}
 \partial_{t}|W\rangle=-\frac{a_{-1}}{2\kappa}|W\rangle,
 \qquad
 \cF(a_{-1}v)=-2\kappa\partial_{t}\cF(v).
\end{equation}
The global affine Ward identity at infinity gives, on the evaluation column,
\begin{equation}
	\cF(h_{-1}v)=-zh\,\cF(v),
	\qquad
	\cF(e_{0}v)=-f\,\cF(v),
	\qquad
	\cF(f_{0}v)=-e\,\cF(v),
\end{equation}
where the exchange of $e$ and $f$ in the last two formulas is the evaluation transpose.
Together with \eqref{equation:physical family insertion identities},
\begin{equation}
	\cF(f_{2}v)=-\frac{2}{u}\partial_{q}\cF(v),
	\qquad
	\cF(f_{1}v)=-2\left(\frac{q}{u}\partial_{q}-\partial_{u}\right)\cF(v),
\end{equation}
and
\begin{equation}
	\cF(f_{2}^{2}v)=\frac{4}{u^{2}}\partial_{q}^{2}\cF(v).
\end{equation}
They give the independent affine isomonodromy equation
\begin{equation}
	\label{equation:PII affine isomonodromy equation}
	2\kappa\partial_{t}\cF
		=
		\Bigg[
		\partial_{q}^{2}
		-\left(2q^{2}+t\right)\partial_{q}
		+2qu\partial_{u}
		+zh
		+\frac{2}{u}f\partial_{q}
		+ue
		\Bigg]\cF.
\end{equation}
For the spin-$\frac{1}{2}$ insertion at $z$, the same equation takes
the matrix form
\begin{equation}
	\label{equation:PII physical matrix isomonodromy equation}
	\begin{aligned}
		2\kappa\partial_{t}\cF
		={}&
		\Bigg\{
		\left[
		\partial_{q}^{2}
		-\left(2q^{2}+t\right)\partial_{q}
		+2qu\partial_{u}
		\right]\mathbf{1}
		+
		\begin{pmatrix}
			z&u\\[2mm]
			\dfrac{2}{u}\partial_{q}&-z
		\end{pmatrix}
		\Bigg\}\cF.
	\end{aligned}
\end{equation}
Equation
\eqref{equation:PII affine isomonodromy equation} is the quantum
isomonodromy equation complementary to the KZ equation.
The two differential operators commute; their compatibility is an
identity in $q,u,z,t$, before imposing the Cartan Ward constraint.\
\vspace{2 mm}

{\bf{Remark :}} We define the {\it quantum isomonodromy problem} here following the definition in \cite{Jimbo_2008}. Our construction provides an alternative towards defining it and below, we explain the consistency of both approaches. Given a linear system 
\begin{equation}
\label{eq:isomonodromyquantum}
\begin{split}
& \kappa \partial_z \Psi(z,t) = \widehat{A}\, \Psi(z,t) 
\\ & 
\kappa \partial_t \Psi(z,t) = (\widehat{B}+\widehat{H})\, \Psi(z,t)
\end{split}    
\end{equation}
where $\widehat{A},\widehat{B}$ and $\widehat{H}$ are three operator-valued matrices satisfying the compatibility equation 
\begin{equation}
\kappa\partial_t \widehat{A} - \kappa\partial_z \widehat{B} + [\widehat{A},\widehat{B}] + [\widehat{A},\widehat{H}] = 0.
\end{equation}
The definitions need to account for the ``quantum Hamiltonian'' as we demonstrate below for the Painlev\'e II example. For $\Psi = \mathcal{F}$, one has the equation \eqref{equation:physical evaluation KZ}, from which we read off 
\begin{equation}
\widehat{A} = \begin{pmatrix}
z^2+ \frac{t}{2} - \partial_q && u(z-q) \\
\frac{2(z+q)}{u}\partial_q - 2\partial_u && -z^2 -\frac{t}{2} + \partial_q
\end{pmatrix}
\end{equation}
which is the quantization of ${\bf{A}}$ through the replacements $\hat{p} = -\partial_q$ and $\hat{\theta} = u\partial_u$, with the classical product $pq$ ordered as $q\hat{p}$. Applying the same replacements to the second Lax matrix ${\bf{B}}$ and the Hamiltonian \eqref{eq:Hamiltoniancl}, with $pq^2$ ordered as $q^2\hat{p}$, gives
\begin{equation}
\begin{split}
&\widehat{B} = \begin{pmatrix}
\frac{z}{2} & \frac{u}{2} \\
\frac{1}{u}\partial_q & -\frac{z}{2}
\end{pmatrix}
\\ & 
\widehat{H}_{II} = \bigg[\frac12 \partial_q^2 - \bigg(q^2 + \frac{t}{2}\bigg)\partial_q + q u \partial_u\bigg] \, {\bf{1}}
\end{split}
\end{equation}
Then in the second equation in \eqref{eq:isomonodromyquantum}, identifying $\widehat{H} = \widehat{H}_{II}$, one obtains equation \eqref{equation:PII affine isomonodromy equation}, which was aptly called before as the {\it quantum isomonodromy equation}. This is also consistent with the claim in \cite{BanerjeeHaghighatLatifi2026}, where the notion of {\it quantum flatness} was introduced [in particular section 4, equations 4.94-4.103, for Painlev\'e II].

\vspace{2 mm}

The affine equation \eqref{equation:PII affine isomonodromy equation} can be reduced by the same
oper-gauge transformation as the KZ system.  Write
\begin{equation}
	\cF_{C}=
	\begin{pmatrix}
		\Phi\\
		\Xi
	\end{pmatrix}.
\end{equation}
Conjugating \eqref{equation:PII affine isomonodromy equation} by
\eqref{equation:oper gauge transformation} and taking the upper component gives
\begin{equation}
	\begin{aligned}
		2\kappa\partial_{t}\Phi
		={}&
		\Bigg[
		\partial_{q}^{2}
		-\left(
		2q^{2}+t+\frac{1}{z-q}
		\right)\partial_{q}
		+2qu\partial_{u}
		\\
		&
		+q+z
		+\frac{2q^{2}+t}{2(z-q)}
		-\frac{1}{4(z-q)^{2}}
		\Bigg]\Phi
		+\frac{1}{z-q}\Xi.
	\end{aligned}
\end{equation}
The first row of \eqref{equation:oper gauge KZ} is
\begin{equation}
	\Xi
	=
	\kappa\partial_{z}\Phi
	-
	\left(
	z^{2}+\frac{t}{2}
	-\partial_{q}
	-\frac{\kappa-1}{2(z-q)}
	\right)\Phi.
\end{equation}
Substitution cancels the extra first-order $q$-derivative and all polynomial scalar terms,
leaving
\begin{equation}
	\label{equation:PII oper gauge isomonodromy equation}
		2\kappa\partial_{t}\Phi
		=
		\Bigg[
		\partial_{q}^{2}
		-\left(2q^{2}+t\right)\partial_{q}
		+2qu\partial_{u}
		+\frac{\kappa}{z-q}\partial_{z}
		+\frac{2\kappa-3}{4(z-q)^{2}}
		\Bigg]\Phi.
\end{equation}
Now use the same scalar pairing factor as in
\eqref{equation:irregular prefactor},
\begin{equation}
	\Phi
	=
	\exp\left(\frac{q^{3}}{3}+\frac{tq}{2}\right)\mathcal{G}.
\end{equation}

Equation \eqref{equation:PII oper gauge isomonodromy equation} becomes
\begin{equation}
	\label{equation:PII scalar isomonodromy equation}
	\begin{aligned}
		2\kappa\partial_{t}\mathcal{G}
		=
		\Bigg[
		&\partial_{q}^{2}
		+\frac{\kappa}{z-q}\partial_{z}
		+2qu\partial_{u}
		-q^{4}-tq^{2}-\frac{t^{2}}{4}
		\\
		&
		-(\kappa-2)q
		+\frac{2\kappa-3}{4(z-q)^{2}}
		\Bigg]\mathcal{G}.
	\end{aligned}
\end{equation}

\subsection{Drinfeld--Sokolov reduction and Virasoro modules}
\label{subsection:PII DS local modules}

We now implement Drinfeld--Sokolov reduction \cite{DrinfeldSokolov1985}
by applying the oper-gauge transformation intrinsically to the local modules, with
\begin{equation}
	\psi(x)=u(x-q).
\end{equation}
At $z$, where $\psi(z)\neq 0$, it is a regular change of local trivialization.  Its simple zero at $q$ produces a singular local Heisenberg module after Drinfeld--Sokolov reduction, while at infinity it induces the corresponding spectral-flow shift of the irregular Wakimoto module.

To obtain the reduced stress tensor, work first in a local
trivialization in which the Whittaker character is normalized to
$e=\beta=1$.  In our Wakimoto conventions the Sugawara tensor is
$T_{\mathrm{Sug}}=:\beta\partial\gamma:
+\frac{1}{4\kappa}:a^{2}:-\frac{1}{2\kappa}\partial a$.
The Cartan improvement
$T_{\mathrm{imp}}=T_{\mathrm{Sug}}+\frac12\partial h$
makes $e$ a field of conformal weight zero, as required by the
constant constraint
\cite[Section~2.2, equation~(2.5) and
Theorems~2.2--2.3]{KacRoanWakimoto2003}.
Using $h=a-2:\gamma\beta:$, it gives
$T_{\mathrm{imp}}=-:\gamma\partial\beta:
+\frac{1}{4\kappa}:a^{2}:
+\frac{\kappa-1}{2\kappa}\partial a$.
Introduce fermionic DS ghosts $b_{\mathrm{gh}},c_{\mathrm{gh}}$,
of weights zero and one in this improved conformal structure, with
$T_{\mathrm{gh}}=:\partial b_{\mathrm{gh}}\,c_{\mathrm{gh}}:$
\cite[Example~1.3 and Theorem~2.2(b)]{KacRoanWakimoto2003}.
For the BRST differential
$d=[Q,\cdot]_{\mathrm{gr}}$, where
$Q=\operatorname{Res}_{x}c_{\mathrm{gh}}(x)(\beta(x)-1)$,
we have $d\gamma=c_{\mathrm{gh}}$ and
$db_{\mathrm{gh}}=\beta-1$.  The graded derivation rule therefore
yields
$-:\gamma\partial\beta:+T_{\mathrm{gh}}
=-d(:\gamma\partial b_{\mathrm{gh}}:)$.
Thus the combined $\beta\gamma$--ghost contribution is BRST-exact,
rather than identically zero, and the stress tensor surviving in
cohomology is
$T_{\mathrm{DS}}=\frac{1}{4\kappa}:a^{2}:
+\frac{\kappa-1}{2\kappa}\partial a$.
Expanding $a(x)=\sum_m a_m x^{-m-1}$ then gives the mode formula
below.  For the original nonconstant character, this Virasoro action
is transported by the local twist, so the same formula is evaluated
on the corresponding twisted Heisenberg modes.

Throughout this section we use the Drinfeld--Sokolov Heisenberg realization
\begin{equation}
	\label{equation:DS Virasoro modes}
	\kappa L_{n}
	=
	\frac{1}{4}
	\sum_{m\in\mathbb{Z}}:a_{m} a_{n-m}:
	-
	\frac{\kappa-1}{2}(n+1)a_{n} .
\end{equation}
The corresponding Virasoro central charge is
$1-6(\kappa-1)^{2}/\kappa$.

\paragraph*{Twisted infinity module}

The irregular Heisenberg data are
\begin{equation}
	a_{3}=-2,
	\qquad
	a_{2}=0,
	\qquad
	a_{1}=-t,
	\qquad
	a_{0}=2\theta .
\end{equation}
At infinity, choose the local trivialization for the
$-3\check{\omega}$ spectral-flow twist, with
$\check{\omega}=h/2$:
\begin{equation}
a^{\mathrm{sf}}(w)
	=
	a(w)
	+
	\frac{3\kappa}{w}.
\end{equation}
Equivalently,
\begin{equation}
	a_{n}^{\mathrm{sf}}=a_{n}
	\quad (n\geq 1),
	\qquad
	a_{0}^{\mathrm{sf}}=a_{0}+3\kappa,
	\qquad
	a_{-r}^{\mathrm{sf}}=a_{-r}
	\quad (r\geq 1).
\end{equation}
Substitution into \eqref{equation:DS Virasoro modes} gives the highest
irregular Virasoro modes
\begin{equation}
	\kappa L_{6}^{(\infty)}=1,
	\qquad
	\kappa L_{5}^{(\infty)}=0,
	\qquad
	\kappa L_{4}^{(\infty)}=t,
\end{equation}
together with
\begin{equation}
	\label{equation:twisted infinity L3}
	\left.
	\kappa L_{3}^{(\infty)}
	\right|_{\mathrm{sf}}
	=
	-2\theta+\kappa-4
\end{equation}
and
\begin{equation}
	\label{equation:twisted infinity L2}
	\left.
	\kappa L_{2}^{(\infty)}
	\right|_{\mathrm{sf}}
	=
	-a_{-1}
	+
	\frac{t^{2}}{4} .
\end{equation}
In the reduced conformal-block pairing defining $\mathcal{G}$, use
the quantum isomonodromy connection
\begin{equation}
	\nabla_{t}^{M}
	=
	\partial_{t}+\frac{a_{-1}}{2\kappa}.
\end{equation}
Horizontality gives
\begin{equation}
 \mathcal{G}(a_{-1}v)=-2\kappa\partial_{t}\mathcal{G}(v).
\end{equation}
The infinity modes acting on $\mathcal{G}$ are therefore
\begin{equation}
 \begin{aligned}
 \kappa L_{2}^{(\infty)}&=2\kappa\partial_{t}+\frac{t^{2}}{4},\qquad
 \kappa L_{3}^{(\infty)}=-2u\partial_{u}+\kappa-2,\\
 \kappa L_{4}^{(\infty)}&=t,\qquad
 \kappa L_{5}^{(\infty)}=0,\qquad
 \kappa L_{6}^{(\infty)}=1.
 \end{aligned}
\end{equation}

Here the global Cartan Ward identity
\eqref{equation:scalar Cartan Ward} gives
$2\theta=2u\partial_{u}-2$ in the $L_{3}$ coefficient.

\paragraph*{The degenerate module at $q$.}

Drinfeld--Sokolov reduction requires the position-dependent Whittaker
character $\psi(x)$ of the raising current to be normalized to unity.
This is achieved locally by a Cartan transformation that rescales the
Wakimoto fields $\beta$ and $\gamma$ inversely.  Its position dependence
shifts the affine Cartan current, while normal ordering of the
$\beta\gamma$ product supplies an additional correction, together
producing the Heisenberg shift $a_x\mapsto a_x+\delta a_x$, with
\begin{equation}
	\delta a_{x}(x)
	=
 -\kappa\partial_{x}\log\psi(x)
 =-\frac{\kappa}{x-q}.
\end{equation}
Here $\kappa=k+2$ and $\psi(x)=u(x-q)$.  At the simple zero $q$,
the transformation is singular and introduces a Heisenberg charge
$-\kappa$, which becomes the additional degenerate Virasoro insertion
after reduction.  At the regular insertion $z$, the transformation is
nonsingular, whereas the third-order pole of the Whittaker differential
at infinity produces the corresponding spectral-flow shift.  In
particular, the residue at $x=q$ fixes the new local Heisenberg charge:
\begin{equation}
	a_{0}|q\rangle
	=
	-\kappa|q\rangle,
	\qquad
	a_{n>0}|q\rangle=0.
\end{equation}
The standard Heisenberg weight formula from
\eqref{equation:DS Virasoro modes} then gives
\begin{equation}
	\label{equation:q conformal weight from twist}
	\Delta_{q}
	=
	\frac{3\kappa-2}{4}.
\end{equation}

The standard level-two Virasoro degeneracy at this charge gives
\begin{equation}
	\label{equation:q null vector}
	\left(
	L_{-1}^{2}-\kappa L_{-2}
	\right)|q\rangle=0.
\end{equation}
Thus both the conformal weight and the BPZ degeneracy at $q$ are consequences
of the local module created by the zero of the twisting function.

\paragraph*{The fundamental module at $z$.}

At $z$ the twisting function is nonzero:
\begin{equation}
	\psi(z)=u(z-q)\neq0.
\end{equation}
Since $\psi(z)\neq 0$, the factor $\psi(z)^{h^{(z)}/2}$ defining the oper-gauge transformation gives a regular change of local trivialization of the fundamental module.  On the conformal-block evaluation column this is represented by the inverse factor $\psi(z)^{-h^{(z)}/2}$ used above.  After DS reduction, the surviving component of the spin-$\frac{1}{2}$ module has
\begin{equation}
	a_{0}|z\rangle
	=
	|z\rangle,
	\qquad
	a_{n>0}|z\rangle=0.
\end{equation}
The same Heisenberg weight formula gives
\begin{equation}
	\label{equation:z conformal weight from spin module}
	\Delta_{z}
	=
	\frac{3\kappa^{-1}-2}{4}.
\end{equation}

The corresponding level-two degeneracy gives
\begin{equation}
	\label{equation:z null vector}
	\left(
	L_{-1}^{2}-\kappa^{-1}L_{-2}
	\right)|z\rangle=0.
\end{equation}

\subsection{BPZ equations and matching of the joint system}
\label{sec:BPZ operators}

With the infinity Ward polynomial and the two local degenerate modules fixed, the two BPZ equations follow directly from the two local null vectors and
the standard Virasoro Ward identity for a conformal block.  Since
$L_{-1}^{(q)}=\partial_{q}$ and $L_{-1}^{(z)}=\partial_{z}$, contour deformation
gives
\begin{equation}
	L_{-2}^{(q)}\mathcal{G}
	=
	\left(
	P(q)
	-
	\frac{1}{z-q}\partial_{z}
	+
	\frac{\Delta_{z}}{(z-q)^{2}}
	\right)\mathcal{G}
\end{equation}
and
\begin{equation}
	L_{-2}^{(z)}\mathcal{G}
	=
	\left(
	P(z)
	+
	\frac{1}{z-q}\partial_{q}
	+
	\frac{\Delta_{q}}{(z-q)^{2}}
	\right)\mathcal{G}.
\end{equation}
The polynomial part of the infinity Ward identity is
$P_{\kappa}(x)=\sum_{n=2}^{6}x^{n-2}L_{n}^{(\infty)}$.  Using the modes above gives
\begin{equation}
		\label{equation:irregular Ward polynomial}
		P_{\kappa}(x)
		=
  2\partial_{t}+\frac{t^{2}}{4\kappa}
  +\frac{x}{\kappa}\left(-2u\partial_{u}+\kappa-2\right)
  +\frac{t}{\kappa}x^{2}+\frac{x^{4}}{\kappa}.
\end{equation}
Together with the null relation \eqref{equation:q null vector}, this gives
\begin{equation}
	\label{equation:q BPZ from twisted module}
	\left[
	\partial_{q}^{2}
	-
	\kappa
	\left(
	\frac{\Delta_{z}}{(z-q)^{2}}
	-
	\frac{1}{z-q}\partial_{z}
	+
	P(q)
	\right)
	\right]\mathcal{G}
	=
	0,
\end{equation}
while \eqref{equation:z null vector} gives
\begin{equation}
	\label{equation:z BPZ from twisted module}
	\left[
	\partial_{z}^{2}
	-
	\kappa^{-1}
	\left(
	\frac{\Delta_{q}}{(z-q)^{2}}
	+
	\frac{1}{z-q}\partial_{q}
	+
	P(z)
	\right)
	\right]\mathcal{G}
	=
	0.
\end{equation}
The conformal weights entering these equations are precisely the local
$L_{0}$ eigenvalues obtained in
\eqref{equation:q conformal weight from twist} and
\eqref{equation:z conformal weight from spin module}.

Equation \eqref{equation:PII scalar isomonodromy equation} is exactly the $q$-BPZ equation
\eqref{equation:q BPZ from twisted module}, rewritten as an evolution equation in $t$.
Thus the affine family connection and the local null-vector equation
give the same scalar time evolution in this normalization.

Multiplying \eqref{equation:z BPZ from twisted module} by $\kappa^{2}$ and
subtracting \eqref{equation:q BPZ from twisted module} gives
\begin{equation}
	\begin{aligned}
		0
		={}&
		\Bigg[
		\kappa^{2}\partial_{z}^{2}
		-
		\partial_{q}^{2}
		-
		\frac{\kappa}{z-q}
		\left(
		\partial_{z}+\partial_{q}
		\right)
		-
		\kappa\left(P(z)-P(q)\right)
		\\
		&\qquad
		-
		\frac{\kappa(\Delta_{q}-\Delta_{z})}{(z-q)^{2}}
		\Bigg]\mathcal{G}.
	\end{aligned}
\end{equation}

The scalar KZ operator $\mathcal{K}_{\mathrm{II}}$ appearing in
\eqref{equation:physical combined scalar KZ} satisfies
\begin{equation}
\mathcal{K}_{\mathrm{II}}
	=
	\kappa^{2}\mathcal{E}_{z}-\mathcal{E}_{q},
\end{equation}
where $\mathcal{E}_{q}$ and $\mathcal{E}_{z}$ denote the operators in
\eqref{equation:q BPZ from twisted module} and
\eqref{equation:z BPZ from twisted module}.  For $\kappa\neq0$,
the scalar systems are equivalent:
\begin{equation}
	\left\{
	\begin{aligned}
		&\text{scalar KZ equation},\\
		&\text{scalar isomonodromy equation}
	\end{aligned}
	\right.
	\quad\Longleftrightarrow\quad
	\left\{
	\begin{aligned}
		&\mathrm{BPZ}_{z},\\
		&\mathrm{BPZ}_{q}.
	\end{aligned}
	\right.
\end{equation}

\section{Matrix integrals for the irregular KZ--BPZ correspondence}
\label{section:PII integral solutions}

The differential equations derived in the preceding section establish the irregular KZ–BPZ correspondence at the level of affine and Virasoro conformal blocks. To make this correspondence more explicit, it is useful to construct integral representations of their solutions. The Wakimoto free-field realization provides a natural framework for this purpose: by inserting screening operators into the correlation functions, the affine conformal blocks can be expressed as multidimensional contour integrals of hypergeometric type \cite{SchechtmanVarchenko1990,Kohno2014}. In the presence of an irregular Whittaker state, the prescribed higher current modes generate an exponential potential in the screening variables, leading to a complex matrix-integral, or $\beta$-ensemble, representation. Such integrals provide a concrete link between current-algebra representations, quantum differential equations, and the associated integrable structures, extending the screening-integral approach to irregular conformal blocks discussed in \cite{Rumanov2014,HaghighatLiuReshetikhin2023,GukovHaghighatLiuReshetikhin2024,Haghighat2025}.

In the present Painlev\'e II construction, the rank-three irregular state gives rise to a cubic potential, while the fundamental insertion at $z$ and the zero $q$ of the Whittaker differential determine additional factors in the integrand. An important feature of this representation is that the same integral satisfies differential equations in both $z$ and $q$, together with the deformation equation in the irregular time $t$. These relations follow from Schwinger–Dyson identities and allow us to reconstruct the full two-component irregular KZ system as well as its quantum isomonodromic evolution. Furthermore, the choice of integration cycle is an essential part of the construction: different admissible twisted rapid-decay cycles can yield distinct solutions, providing a geometric setting for studying their analytic continuation and possible Stokes phenomena. The integral representation also furnishes a useful starting point for the semiclassical and Nekrasov–Shatashvili limits, where saddle-point equations, Bethe roots, and spectral curves emerge. We now develop this construction explicitly for a single fundamental insertion.

\subsection{Matrix integral representation and differential equations}
\label{subsection:PII screening integral}
We write
\begin{equation}
	\label{equation:PII A function}
	V(x):=\frac{2}{3}x^{3}+tx.
\end{equation}
Thus $\exp\left(\frac{V(q)}{2}\right)=Z_{\mathrm{irr}}(q)$ is the
irregular prefactor in \eqref{equation:irregular prefactor}.  We use
$V$ for the scalar cubic potential, reserving $A(z)$ for the
matrix-valued KZ operator of the preceding affine construction.  Throughout this section,
$\mathcal{I}_{N}$ denotes an undressed integral; the conformal-block
column is denoted by $\cF$, as in Section~\ref{section:PII affine BPZ}.

For the rank-three irregular insertion at infinity and the spin-$\frac{1}{2}$
insertion at $z$, consider
\begin{equation}
	\label{equation:PII matrix integral}
	\begin{aligned}
		\mathcal{I}_{N}(q,t;z)
		={}&\frac{1}{N!}\int_{\Gamma}
		\prod_{a=1}^{N}\dd x_{a}
		\prod_{a<b}(x_{a}-x_{b})^{\frac{2}{\kappa}}
		\\
		&\times\prod_{a=1}^{N}(x_{a}-q)(z-x_{a})^{-\frac{1}{\kappa}}
		\exp\left[-\frac{1}{\kappa}\sum_{a=1}^{N}V(x_{a})\right].
	\end{aligned}
\end{equation}
 Here ``matrix integral'' denotes the eigenvalue representation of a
complex $\beta$ ensemble, with $\beta=2/\kappa$.  It is constructed
by the Wakimoto screening prescription.  The screening
number is $N\in\mathbb{Z}_{\geq0}$.  The cycle $\Gamma$
is a twisted rapid-decay cycle in the configuration space excluding
$x_{a}=x_{b}$ and $x_{a}=z$.  The local coefficient system records the
branches of the Vandermonde and $(z-x_{a})$ powers.  At infinity the
cycle ends in sectors with
$\operatorname{Re}\left(2x_{a}^{3}/(3\kappa)\right)>0$.
Transport $\Gamma$ horizontally as $z,t$ vary, so parameter
differentiation acts on the period through the displayed integrand.
The total derivatives below have vanishing twisted boundary terms.
The matrix size is related to the affine
Cartan parameter via charge conservation by
\begin{equation}
	\label{equation:PII N theta relation}
	N=\theta+\frac{3}{2}.
\end{equation}

The integral \eqref{equation:PII matrix integral} satisfies the scalar equation
\begin{equation}
	\label{equation:PII matrix integral scalar equation}
	\begin{aligned}
		0={}&
		\Bigg[
		\kappa^{2}\partial_{z}^{2}
		-\partial_{q}^{2}
		+\left(
		\kappa(2z^{2}+t)
		-\frac{\kappa(\kappa+1)}{z-q}
		\right)\partial_{z}
		\\
		&\qquad
		+\left(
		2q^{2}+t
		-\frac{\kappa+1}{z-q}
		\right)\partial_{q}
		+2N(z-q)
		\Bigg]\mathcal{I}_{N}.
	\end{aligned}
\end{equation}
\paragraph{Differential equations.} The integration-by-parts calculation in Appendix~\ref{appendix:PII matrix integral checks} gives the pair
\begin{equation}
	\label{equation:PII matrix integral z equation}
	\begin{aligned}
		0={}&
		\Bigg[
		\kappa^{2}\partial_{z}^{2}
		+\left(
		\kappa(2z^{2}+t)
		-\frac{\kappa^{2}}{z-q}
		\right)\partial_{z}
		-\frac{\kappa}{z-q}\partial_{q}
		\\
		&\qquad
		-2\kappa\partial_{t}
		+2Nz
		\Bigg]\mathcal{I}_{N},
	\end{aligned}
\end{equation}
and
\begin{equation}
	\label{equation:PII matrix integral q equation}
	\begin{aligned}
		0={}&
		\Bigg[
		\partial_{q}^{2}
		+\frac{\kappa}{z-q}\partial_{z}
		+\left(
		\frac{1}{z-q}-2q^{2}-t
		\right)\partial_{q}
		\\
		&\qquad
		-2\kappa\partial_{t}
		+2Nq
		\Bigg]\mathcal{I}_{N}.
	\end{aligned}
\end{equation}
Subtracting \eqref{equation:PII matrix integral q equation} from
\eqref{equation:PII matrix integral z equation} reproduces
\eqref{equation:PII matrix integral scalar equation}.

\subsection{Reconstruction of the irregular KZ equation}
\label{subsection:PII integral reconstruction}

The upper component of the evaluation-column KZ equation
\eqref{equation:physical evaluation KZ} is obtained by the dressing
\begin{equation}
	\label{equation:PII matrix integral top component}
	\cF_{1}
	=
	u^{N}
	\exp\left(\frac{V(z)}{2\kappa}\right)
	\mathcal{I}_{N}.
\end{equation}
By \eqref{equation:PII N theta relation}, the global Cartan Ward identity is
\begin{equation}
	u\partial_{u}\cF_{1}
	=N\cF_{1}
	=\left(
	\theta+\frac{3}{2}
	\right)\cF_{1}.
\end{equation}

For later comparison with the KZ system, the two Schwinger--Dyson equations
take the following form after this dressing:
\begin{equation}
\begin{aligned}
		0={}&
		\Bigg[
		\kappa^{2}\partial_{z}^{2}
		-\frac{\kappa^{2}}{z-q}\partial_{z}
		-\frac{\kappa}{z-q}\partial_{q}
		-2\kappa\partial_{t}
		-\left(
		z^{2}+\frac{t}{2}
		\right)^{2}
		\\
		&\qquad
		+\frac{\kappa\left(z^{2}+\frac{t}{2}\right)}{z-q}
		+(2N-2\kappa+1)z
		\Bigg]\cF_{1},
	\end{aligned}
\end{equation}
and
\begin{equation}
\begin{aligned}
		0={}&
		\Bigg[
		\partial_{q}^{2}
		+\frac{\kappa}{z-q}\partial_{z}
		+\left(
		\frac{1}{z-q}-2q^{2}-t
		\right)\partial_{q}
		-2\kappa\partial_{t}
		+2Nq
		\\
		&\qquad
		-\frac{zq+\frac{t}{2}}{z-q}
		\Bigg]\cF_{1}.
	\end{aligned}
\end{equation}

Define the lower component by
\begin{equation}
\cF_{2}
	=
	\frac{u^{N-1}}{z-q}
	\exp\left(\frac{V(z)}{2\kappa}\right)
	\left(
	\kappa\partial_{z}+\partial_{q}
	\right)
	\mathcal{I}_{N}.
\end{equation}
Using \eqref{equation:PII matrix integral derivative identities}, this can also be written as
\begin{equation}
	\label{equation:PII matrix integral lower component resolvent}
	\begin{aligned}
		\cF_{2}
		&=
		\frac{u^{N-1}}{z-q}
		\exp\left(\frac{V(z)}{2\kappa}\right)
		\left\langle
		R(q)-R(z)
		\right\rangle
		\\
		&=
		u^{N-1}
		\exp\left(\frac{V(z)}{2\kappa}\right)
		\left\langle
		\sum_{a=1}^{N}
		\frac{1}{(q-x_{a})(z-x_{a})}
		\right\rangle.
	\end{aligned}
\end{equation}
The column $\cF=(\cF_{1},\cF_{2})^{T}$ therefore satisfies
the irregular KZ equation \eqref{equation:physical evaluation KZ}.
Thus the matrix integral gives an explicit solution of the position equation.

\subsection{Quantum isomonodromy from the matrix integral}
The Schwinger--Dyson identities
\eqref{equation:PII matrix integral z equation} and
\eqref{equation:PII matrix integral q equation} also give the
$t$-equation of Section~\ref{subsection:PII original time connection}.  As $\partial_{t}V(z)=z$, differentiating the first
component of \eqref{equation:PII matrix integral top component} gives
\begin{equation}
	2\kappa\partial_{t}\cF_{1}
	=
	u^{N}\exp\left(\frac{V(z)}{2\kappa}\right)
	\left(2\kappa\partial_{t}+z\right)\mathcal{I}_{N}.
\end{equation}
Equation \eqref{equation:PII matrix integral q equation} therefore gives
\begin{equation}
2\kappa\partial_{t}\cF_{1}
	=
	\left[
	\partial_{q}^{2}-(2q^{2}+t)\partial_{q}
	+2qu\partial_{u}+z
	\right]\cF_{1}
	+u\cF_{2}.
\end{equation}
To obtain the second row, apply $\kappa\partial_{z}+\partial_{q}$ to
\eqref{equation:PII matrix integral q equation} and use
\eqref{equation:PII matrix integral z equation} to eliminate the
remaining second-order combination.  Written entirely in terms of
the undressed integral, the resulting identity is
\begin{equation}
	\label{equation:PII matrix integral lower isomonodromy identity}
	\begin{aligned}
		2\kappa\partial_{t}
		\left(\frac{(\kappa\partial_{z}+\partial_{q})\mathcal{I}_{N}}{z-q}\right)
		={}&\partial_{q}^{2}
		\left(\frac{(\kappa\partial_{z}+\partial_{q})\mathcal{I}_{N}}{z-q}\right)
		\\
		&-(2q^{2}+t)\partial_{q}
		\left(\frac{(\kappa\partial_{z}+\partial_{q})\mathcal{I}_{N}}{z-q}\right)
		\\
		&+(2q(N-1)-2z)
		\frac{(\kappa\partial_{z}+\partial_{q})\mathcal{I}_{N}}{z-q}
		+2\partial_{q}\mathcal{I}_{N}.
	\end{aligned}
\end{equation}
The factor $\exp\left(\frac{V(z)}{2\kappa}\right)$ contributes $z$
to $2\kappa\partial_{t}$ when the lower component is dressed.
Using $u\partial_{u}\cF_{2}=(N-1)\cF_{2}$ in
\eqref{equation:PII matrix integral lower isomonodromy identity}, we find
\begin{equation}
2\kappa\partial_{t}\cF_{2}
	=
	\left[
	\partial_{q}^{2}-(2q^{2}+t)\partial_{q}
	+2qu\partial_{u}-z
	\right]\cF_{2}
	+\frac{2}{u}\partial_{q}\cF_{1}.
\end{equation}
Combining the two rows gives the quantum isomonodromy equation
\eqref{equation:PII physical matrix isomonodromy equation}, with
the conventions of
Sections~\ref{subsection:PII original affine construction}
and~\ref{subsection:PII original time connection}.

Finally, the oper and Virasoro blocks of Section~\ref{section:PII affine BPZ} are recovered by
\begin{equation}
	\label{equation:PII integral oper Virasoro dictionary}
	\begin{aligned}
		\Phi
		&=
		\bigl(u(z-q)\bigr)^{-1/2}\cF_{1}
		=
		u^{N-\frac{1}{2}}(z-q)^{-\frac{1}{2}}
		\exp\left(\frac{V(z)}{2\kappa}\right)\mathcal{I}_{N},
		\\
		\mathcal{G}
		&=
		Z_{\mathrm{irr}}(q)^{-1}\Phi
		=
		u^{N-\frac{1}{2}}(z-q)^{-\frac{1}{2}}
		\exp\left(\frac{V(z)}{2\kappa}-\frac{V(q)}{2}\right)
		\mathcal{I}_{N}.
	\end{aligned}
\end{equation}
The $q$-Schwinger--Dyson equation
\eqref{equation:PII matrix integral q equation}, conjugated by the
prefactor in \eqref{equation:PII integral oper Virasoro dictionary},
becomes precisely the scalar quantum isomonodromy equation
\eqref{equation:PII scalar isomonodromy equation}; the prefactor
cancels the first-order $\partial_{q}$ term.
Likewise, the difference of the two Schwinger--Dyson equations,
followed by the same dressing, gives
\eqref{equation:physical combined scalar KZ}.  Hence the integral
reproduces both members of the KZ--isomonodromy/BPZ system of Section~\ref{section:PII affine BPZ}.

\section{Limits of the matrix integral}
\label{section:PII limits}

The matrix integral \eqref{equation:PII matrix integral} admits two
degenerate limits, which are the subject of this section.  In the
Nekrasov--Shatashvili limit $\kappa\to0$ at fixed $N$, the integral is
dominated by a Bethe saddle determined by a Yang--Yang function.  In
the classical limit $\epsilon_{1}\to0$ with $\epsilon_{2}$ fixed, one
has $\kappa\to\infty$ and $N\to\infty$ with
$N/\kappa=-\epsilon_{1}N/\epsilon_{2}\to-\Theta/\epsilon_{2}$ finite,
and the same integral, in this double scaling, reduces to a
Hamilton--Jacobi problem whose characteristics obey Painlev\'e II.
The two limits are therefore two degenerations of one
$\beta$-ensemble, and they provide complementary descriptions of the
same irregular KZ--BPZ system: isomonodromic dynamics with an
$\epsilon_{2}$ oper on one side, and Bethe--Gaudin spectral geometry
with an $\epsilon_{1}$ oper on the other.  Further limits of both end at the same
Seiberg--Witten curve, a coincidence that is analysed in
Section~\ref{section:PII planar SW geometry} and explained by quantum
Langlands duality in Section~\ref{section:PII Langlands interpretation}.

\subsection{Classical limit and Painlev\'e II}
\label{section:PII classical WZW limit}

The irregular KZ–BPZ correspondence established above provides a quantum description of the Painlev\'e II isomonodromic system, together with explicit matrix-integral solutions. A natural question is how the classical Painlev\'e dynamics emerges from this quantum construction. We address this by taking the limit $\epsilon_1\to0$ with $\epsilon_2$ fixed, while appropriately scaling the irregular parameters and representation labels. In this limit, the affine conformal block develops a common classical exponential whose action satisfies a Hamilton–Jacobi equation. Its characteristic equations reproduce the Painlev\'e II Hamiltonian dynamics, while the reduced KZ equation becomes a scalar oper with an apparent singularity at the dynamical coordinate. The compatibility between these structures identifies the motion of the apparent singularity with the Painlevé-II isomonodromic flow. A subsequent $\epsilon_2\to0$ limit recovers the associated classical spectral curve. Thus, the classical limit provides a direct bridge from the quantum affine and Virasoro descriptions to the Hamiltonian, oper and Seiberg–Witten geometry of Painlev\'e II.

In this subsection, we take the classical limit $\epsilon_{1}\to0$ with
$\epsilon_{2}$ fixed, following
Frenkel--Gukov--Teschner~\cite[Appendix~D, especially
(D.4)--(D.7) and (D.12)--(D.13)]{FrenkelGukovTeschner2015}.
Scaling the representation labels and the irregular coordinates in
the equations of Sections~\ref{section:PII affine BPZ} and~\ref{section:PII integral solutions} gives the classical Hamilton--Jacobi
equation and the scalar Painlev\'e II oper.  A subsequent
$\epsilon_{2}\to0$ limit gives
the classical spectral curve.

\subsubsection{Classical scaling}

We take one spin-$\frac{1}{2}$ field at $z$, with the two-component column
$\cF$ in \eqref{equation:physical evaluation KZ}.  The coordinate
$q$ parametrizes the irregular-state family.

For an internal WZW representation, the conventions of
\cite[Eq.~(D.2)]{FrenkelGukovTeschner2015} are
\begin{equation}
	j_{e}=-\frac{1}{2}+\mathrm{i}\frac{a_{e}}{\epsilon_{1}},
	\qquad
	\kappa=k+2=-\frac{\epsilon_{2}}{\epsilon_{1}}.
\end{equation}
Our irregular state has $a_{0}|W\rangle=2\theta|W\rangle$.
To parametrize its unshifted Cartan weight analogously, we take
\begin{equation}
	\label{equation:PII classical WZW theta weight}
	\theta=-\frac{1}{2}+\mathrm{i}\frac{a_{e}}{\epsilon_{1}},
	\qquad
	\Theta:=\lim_{\epsilon_{1}\to0}\epsilon_{1}\theta=\mathrm{i}a_{e},
	\qquad \epsilon_{2}\ne0\ \text{fixed}.
\end{equation}
This identifies the unshifted Cartan parameters; finite zero-mode shifts
leave the scaled weight $\Theta$ unchanged.
For the one-field matrix integral the exact neutrality relation is
\begin{equation}
	\label{equation:PII classical WZW neutrality}
	N=\theta+\frac{3}{2}
	=1+\mathrm{i}\frac{a_{e}}{\epsilon_{1}},
	\qquad \epsilon_{1}N\longrightarrow\Theta.
\end{equation}
Thus neutrality sends $N$ to infinity with
$\epsilon_{1}N\longrightarrow\Theta$, while the finite-point field
retains spin $\frac{1}{2}$.

\paragraph*{Rescaling the irregular KZ equation.}

In the original coordinates $q,z,t$, the KZ system
\eqref{equation:physical evaluation KZ} reads
\begin{equation}
	-\epsilon_{2}\partial_{z}\cF
	=\epsilon_{1}
	\begin{pmatrix}
		z^{2}+\frac{t}{2}-\partial_{q} & u(z-q)\\[1mm]
		\frac{2(z+q)}{u}\partial_{q}-2\partial_{u}&
		-z^{2}-\frac{t}{2}+\partial_{q}
	\end{pmatrix}\cF.
\end{equation}
The weight scaling \eqref{equation:PII classical WZW theta weight}
keeps the Cartan term $\epsilon_{1}u\partial_{u}$ finite on the
chosen weight sector.  Balancing the translation term
$\kappa\partial_{z}$ with the irregular polynomial $z^{2}$ fixes
the coordinate powers $1/3$ and $2/3$:
\begin{equation}
	\label{equation:PII classical WZW cubic balance}
	z=\epsilon_{1}^{-1/3}Z,\qquad
	q=\epsilon_{1}^{-1/3}Q,\qquad
	t=\epsilon_{1}^{-2/3}T,\qquad
	x_{a}=\epsilon_{1}^{-1/3}\xi_{a}.
\end{equation}
The $x_{a}$ rescaling applies to the integral representation, and $u$
remains unchanged.

To balance the two components, introduce the column
\begin{equation}
	\label{equation:PII classical WZW rescaled column}
	\widehat{\cF}
	:=\begin{pmatrix}\cF_{1}\\ \epsilon_{1}^{1/3}u\cF_{2}\end{pmatrix}.
\end{equation}
A direct change of variables in the exact equations
\eqref{equation:physical evaluation KZ} and
\eqref{equation:PII physical matrix isomonodromy equation} yields
\begin{equation}
	\label{equation:PII classical WZW exact Z connection}
	-\epsilon_{2}\partial_{Z}\widehat{\cF}
	=\begin{pmatrix}
		Z^{2}+\frac{T}{2}-\epsilon_{1}\partial_{Q} & Z-Q\\[2mm]
		2\epsilon_{1}\bigl((Z+Q)\partial_{Q}-u\partial_{u}\bigr)&
		-Z^{2}-\frac{T}{2}+\epsilon_{1}\partial_{Q}
	\end{pmatrix}\widehat{\cF},
\end{equation}
\begin{equation}
	\begin{aligned}
		-2\epsilon_{2}\partial_{T}\widehat{\cF}
		={}&\left[
		\epsilon_{1}\partial_{Q}^{2}-(2Q^{2}+T)\partial_{Q}
		+2Qu\partial_{u}\right]\widehat{\cF}\\
		&+\begin{pmatrix}
			Z&1\\[1mm]2\epsilon_{1}\partial_{Q}&-Z-2Q
		\end{pmatrix}\widehat{\cF}.
	\end{aligned}
\end{equation}
The extra $-2Q$ in the lower diagonal entry of the second equation
comes from conjugating $u\partial_{u}$ by the factor $u$ in the second
component of \eqref{equation:PII classical WZW rescaled column}.
These are the rescaled KZ and isomonodromy equations at finite
$\epsilon_{1}$.

\subsubsection{Classical action and Painlev\'e II}
\label{subsection:PII WZW asymptotics}

Following \cite[Eq.~(D.12)]{FrenkelGukovTeschner2015}, factor the
common classical exponential from the irregular block:
\begin{equation}
	\label{equation:PII classical WZW WKB ansatz}
	\widehat{\cF}
	=u\exp\left(\frac{\mathcal{W}(Z,Q,T,u;\epsilon_{2})}{\epsilon_{1}}\right)
	\left[\begin{pmatrix}\psi_{1}(Z;Q,T)\\\psi_{2}(Z;Q,T)\end{pmatrix}
	+o(1)\right].
\end{equation}
The explicit factor $u$ records the finite term in
$N=\Theta/\epsilon_{1}+1$; the remaining leading column is independent of $u$. 
Our sign convention is $\mathcal{W}=-Y^{\mathrm{WZ}}$ relative to
\cite[Eq.~(D.12)]{FrenkelGukovTeschner2015}.
The Cartan Ward identity gives
$u\partial_{u}\widehat{\cF}=N\widehat{\cF}$, hence
\begin{equation}
	u\partial_{u}\mathcal{W}=\Theta,
	\qquad
	P:=\partial_{Q}\mathcal{W}.
\end{equation}
Initially we allow the common classical action to depend on $Z$.
After extracting the exponential, the right-hand side of
\eqref{equation:PII classical WZW exact Z connection} is finite:
$\epsilon_{1}\partial_{Q}$ contributes $P$, and the Cartan term
contributes $\Theta$.  Substituting $\widehat{\cF}$ into
\eqref{equation:PII classical WZW exact Z connection} and dividing out
the common factor $u\,e^{\mathcal{W}/\epsilon_{1}}$, one gets
\begin{equation}
\label{eq: leadingeps1zero}
-\frac{\epsilon_2}{\epsilon_1}(\partial_Z {\cal{W}}) \begin{pmatrix} \psi_1 \\ \psi_2 \end{pmatrix}
-\epsilon_2\begin{pmatrix}\partial_Z\psi_1 \\ \partial_Z\psi_2
\end{pmatrix}
= \widehat{A}_{\mathrm{cl}} \begin{pmatrix} \psi_1 \\ \psi_2 \end{pmatrix}+o(1)
\end{equation}
where 
\begin{equation}
\widehat{A}_{\mathrm{cl}}
= \begin{pmatrix}
Z^2 + \frac{T}{2} - P && Z-Q \\ 2[(Z+Q)P-\Theta] && -Z^2 - \frac{T}{2} + P
\end{pmatrix}
\end{equation}
The left-hand side in \eqref{eq: leadingeps1zero} has a term of
order $\epsilon_{1}^{-1}$ proportional to
$-\epsilon_{2}\partial_{Z}\mathcal{W}$.  Since $\epsilon_{2}$ is
fixed and nonzero, its coefficient must vanish, so
$\partial_{Z}\mathcal{W}=0$.  Then $\mathcal{W}$ is independent of $Z$ and we therefore suppress the $Z$
argument of $\mathcal{W}$ below.

The leading term of the exact $T$ equation gives a Hamilton--Jacobi
identity for the common classical action,
\begin{equation}
	\label{equation:PII classical WZW Hamilton Jacobi}
	-2\epsilon_{2}\partial_{T}\mathcal{W}
	=P^{2}-(2Q^{2}+T)P+2Q\Theta.
\end{equation}
Upon differentiating with respect to $Q$, the corresponding
characteristics satisfy
\begin{equation}
	\label{equation:PII classical WZW characteristics}
	\frac{\dd Q}{\dd T}
	=\frac{2P-2Q^{2}-T}{2\epsilon_{2}},
	\qquad
	\frac{\dd P}{\dd T}
	=\frac{4QP-2\Theta}{2\epsilon_{2}}.
\end{equation}
These are the Hamilton--Jacobi characteristics in the present
time convention.
Eliminating $P$ gives Painlev\'e II in :
\begin{equation}
	\label{equation:PII classical WZW Painleve equation}
	\epsilon_{2}^{2}\frac{\dd^{2}Q}{\dd T^{2}}
	=2Q^{3}+TQ-\Theta-\frac{\epsilon_{2}}{2},
\end{equation}
At $\epsilon_{2}=-1$, the identifications $q=Q$, $t=T$,
$\theta=\Theta$ and $p=-P$ recover the standard Hamiltonian
convention of \eqref{eq:Laxpaircl} and the Painlev\'e II equation
\eqref{eq:PIIscl}.  Thus the standard momentum is the negative of
the action derivative $P=\partial_{Q}\mathcal{W}$.

\paragraph{Scalar oper and apparent singularity.}\phantomsection
\label{subsection:PII WZW scalar oper}

Apply the classical scaling and
\eqref{equation:PII classical WZW WKB ansatz} directly to the scalar
KZ equation \eqref{equation:exact scalar KZ}.  In the oper gauge of
Section~\ref{section:PII affine BPZ}, the leading scalar wave function is
\begin{equation}
	\phi_{\mathrm{cl}}(Z)=(Z-Q)^{-\frac{1}{2}}\psi_{1}(Z).
\end{equation}
The leading scalar equation is
\begin{equation}
	\label{equation:PII classical WZW scalar oper}
	\left[\epsilon_{2}^{2}\partial_{Z}^{2}-U(Z)\right]\phi_{\mathrm{cl}}(Z)=0,
\end{equation}
where
\begin{equation}
	\label{equation:PII classical WZW oper potential}
	\begin{aligned}
		U(Z)={}&\left(Z^{2}+\frac{T}{2}\right)^{2}
		-(2\Theta+\epsilon_{2})Z
		+P^{2}-(2Q^{2}+T)P+2Q\Theta+\epsilon_{2}Q\\
		&+\frac{\epsilon_{2}\left(Q^{2}+\frac{T}{2}-P\right)}{Z-Q}
		+\frac{3\epsilon_{2}^{2}}{4(Z-Q)^{2}}.
	\end{aligned}
\end{equation}
In view of \eqref{equation:PII classical WZW Hamilton Jacobi}, the
constant term in this potential can alternatively be written using
$-2\epsilon_{2}\partial_{T}\mathcal{W}$.  The coefficient of the double
pole is $3\epsilon_{2}^{2}/4$, so the two local exponents of the scalar
equation at $Z=Q$ are $-1/2$ and $3/2$.  The regular term in the
local expansion is $\left(Q^{2}+\frac{T}{2}-P\right)^{2}$, which
satisfies the no-logarithm condition for the displayed simple and
double poles.  Thus the singularity is apparent, with scalar
monodromy $-\mathbf{1}$ and trivial projective monodromy.  In the convention
$\left(\epsilon_{2}^{2}\partial_{Z}^{2}+t(Z)\right)\phi_{\mathrm{cl}}=0$, we have
$t(Z)=-U(Z)$, matching the local apparent-pole structure described in
\cite[Eq.~(D.15)]{FrenkelGukovTeschner2015}.

\paragraph{Semiclassical spectral curve}

Taking the subsequent $\epsilon_{2}\to0$ limit of
\eqref{equation:PII classical WZW scalar oper} gives the spectral curve
\begin{equation}
	\label{equation:PII classical WZW subsequent spectral curve}
	y^{2}=
	\left(Z^{2}+\frac{T}{2}\right)^{2}-2\Theta Z
	+P^{2}-(2Q^{2}+T)P+2Q\Theta.
\end{equation}
This passage from the
$\epsilon_{2}$-oper to its spectral curve follows the order
of limits in \cite[Appendix~D.4]{FrenkelGukovTeschner2015},
applied to the rank-three irregular scalar equation.

{\it Thus we see here that the semiclassical limit of the irregular KZ–isomonodromy system produces a Hamilton–Jacobi problem in the dynamical variable $Q$, whose characteristics obey Painlev\'e II, together with an $\epsilon_2$ oper in the spectral coordinate $Z$. The apparent singularity at $Z=Q$ has trivial projective monodromy, and the Hamilton–Jacobi evolution governs its isomonodromic motion.}

\subsection{NS limit}
\label{section:PII NS comparison}

In this subsection, we take the Nekrasov--Shatashvili (NS) limit,
$\epsilon_{2}\to0$ with nonzero $\epsilon_{1}$, in the
parameter convention of \cite{NekrasovShatashvili2009}.
We first work in the original coordinates $q,z,t$, with one
spin-$\frac{1}{2}$ field at $z$
and $N$ screening integrations.  The saddle
determines the NS action, a common eigenvector of the position
and quantum isomonodromy Hamiltonians, and its scalar $q$-oper.
Counting the resulting Bethe states gives a complete set of simultaneous
eigenstates in the finite polynomial sector for generic spectral parameters.
After the finite-$N$ analysis, we take the subsequent
$\epsilon_{1}\to0$ limit and compare its spectral curve with
the classical-limit construction of
Subsection~\ref{section:PII classical WZW limit}.

The significance of the NS limit is that it reveals the quantum integrable structure underlying the irregular KZ–BPZ correspondence. Unlike the classical limit of Subsection~\ref{section:PII classical WZW limit}, which recovers the nonlinear Painlev\'e II dynamics and its isomonodromic Lax system, the NS limit turns the screening integrals into a finite-dimensional Bethe eigenvalue problem governed by a Yang–Yang function. The critical points of this function determine simultaneous eigenstates of the commuting KZ and quantum isomonodromy Hamiltonians, while the Bethe roots construct a scalar oper whose accessory parameters encode their joint spectrum. The resulting Bethe states form a complete eigenbasis in generic finite-screening sectors, and their large-$N$ limit reconstructs the associated spectral curve. Thus, the classical and NS limits expose two complementary aspects of the same quantum system: isomonodromic dynamics and Gaudin-type spectral geometry. Their relationship, made precise through quantum Langlands duality, will be developed in Section \ref{section:PII Langlands interpretation}.

\subsubsection{Yang--Yang function and the Bethe equations}

Recall the parameter relation and the exact neutrality condition
\begin{equation}
	\kappa=k+2=-\frac{\epsilon_{2}}{\epsilon_{1}},
	\qquad N=\theta+\frac{3}{2}.
\end{equation}
The NS scaling is
\begin{equation}
	\label{equation:PII NS actual limit}
	\epsilon_{2}\longrightarrow0,
	\qquad \epsilon_{1}\ne0\ \text{fixed},
	\qquad \kappa\longrightarrow0,
\end{equation}
with all remaining variables held fixed.  The weight scaling
\eqref{equation:PII classical WZW theta weight} enters the subsequent
$\epsilon_{1}\to0$ limit.

For the integral \eqref{equation:PII matrix integral}, define its
Yang--Yang function using the cubic potential of
\eqref{equation:PII A function}:
\begin{equation}
	\label{equation:PII NS Yang Yang}
	\mathcal{Y}_{N}(x_{1},\ldots,x_{N};z,t)
	=\sum_{a=1}^{N}V(x_{a})
	+\sum_{a=1}^{N}\log(z-x_{a})
	-2\sum_{a<b}\log(x_{a}-x_{b}).
\end{equation}
Thus the exact undressed period in equation \eqref{equation:PII matrix integral} may be written as
\begin{equation}
	\mathcal{I}_{N}
	=\frac{1}{N!}\int_{\Gamma}
	\exp\left(-\frac{\mathcal{Y}_{N}}{\kappa}\right)
	\prod_{a=1}^{N}(x_{a}-q)\prod_{a=1}^{N}\dd x_{a}.
\end{equation}
At fixed $N$, the factor $\prod_{a=1}^{N}(x_{a}-q)$ contributes at
order $\kappa^{0}$ to the logarithm of the integrand and enters the
saddle amplitude.  Including it in the exponent shifts
$\mathcal{Y}_{N}$ by $-\kappa\sum_{a=1}^{N}\log(x_{a}-q)$.

The NS saddle $x_{1}^{*},\ldots,x_{N}^{*}$ is determined by
\begin{equation}
	\label{equation:PII NS Bethe equations}
	2(x_{a}^{*})^{2}+t
	-\frac{1}{z-x_{a}^{*}}
	-2\sum_{b\ne a}\frac{1}{x_{a}^{*}-x_{b}^{*}}=0,
	\qquad a=1,\ldots,N.
\end{equation}
Stationary phase gives, up to a common factor independent of $q$,
\begin{equation}
	\label{equation:PII NS integral stationary phase}
	\mathcal{I}_{N}
	\sim C_{N}(\kappa;z,t)
	\exp\left(-\frac{\mathcal{Y}_{N}(x^{*};z,t)}{\kappa}\right)
	\prod_{a=1}^{N}(q-x_{a}^{*}),
\end{equation}
For the unit steepest-descent cycle through the nondegenerate saddle
$x^{*}$, the prefactor $C_N$ is given by
\begin{equation}
	C_{N}(\kappa;z,t)
	:=\frac{(-1)^{N}}{N!}
	\frac{(2\pi\kappa)^{N/2}}
	{\sqrt{\det\left(
			\left.\partial_{x_{a}}\partial_{x_{b}}
			\mathcal{Y}_{N}(x;z,t)\right|_{x=x^{*}}
			\right)}}.
\end{equation}
The factor $(-1)^{N}$ converts
$\prod_{a}(x_{a}^{*}-q)$ to $\prod_{a}(q-x_{a}^{*})$.
The orientation of the steepest-descent cycle fixes the square-root
branch; for a general cycle, its constant coefficient is
absorbed into $C_{N}$.  Higher saddle corrections are of relative
order $O(\kappa)$ and need not be independent of $q$.  In particular,
for fixed $z,t$ and a nondegenerate saddle,
$\epsilon_{2}\log C_{N}\to0$, so $C_{N}$ does not contribute to
$\mathcal{W}_{\mathrm{NS}}$.

The WZW column of Section~\ref{section:PII integral solutions} is
\begin{equation}
	\label{equation:PII NS exact column}
	\begin{aligned}
		\cF_{1}&=u^{N}\exp\left(\frac{V(z)}{2\kappa}\right)\mathcal{I}_{N},\\
		\cF_{2}&=u^{N-1}\exp\left(\frac{V(z)}{2\kappa}\right)
		\left\langle\sum_{a=1}^{N}
		\frac{1}{(q-x_{a})(z-x_{a})}\right\rangle.
	\end{aligned}
\end{equation}
The second line uses the same unnormalized integral bracket as
Section~\ref{section:PII integral solutions}.  All nonzero components
share the leading $\epsilon_{2}^{-1}$ exponential.  With the normalization of
\eqref{equation:PII NS exact column}, this exponent is
\begin{equation}
	\mathcal{W}_{\mathrm{NS}}(z,t)
	:=\lim_{\epsilon_{2}\to0}\epsilon_{2}\log\cF_{1}
	=\epsilon_{1}\left(
	\mathcal{Y}_{N}(x^{*};z,t)-\frac{V(z)}{2}\right).
\end{equation}
Using $-1/\kappa=\epsilon_{1}/\epsilon_{2}$ gives the action
$\mathcal{W}_{\mathrm{NS}}(z,t)$.

Define the monic polynomial
\begin{equation}
	P_{N}(q):=\prod_{a=1}^{N}(q-x_{a}^{*}).
\end{equation}
The leading column, with the common saddle normalization removed,
is
\begin{equation}
	\label{equation:PII NS Bethe vector}
	v=\begin{pmatrix}
		u^{N}P_{N}(q)\\[1mm]
		\displaystyle u^{N-1}P_{N}(q)
		\sum_{a=1}^{N}\frac{1}{(q-x_{a}^{*})(z-x_{a}^{*})}
	\end{pmatrix},
	\qquad
	\cF\sim\exp\left(\frac{\mathcal{W}_{\mathrm{NS}}}{\epsilon_{2}}\right)v.
\end{equation}
Thus the component amplitudes retain the light-field data through
$P_{N}(q)$ and the resolvent insertion.

Let $H_{1}$ and $H_{t}$ denote the operators on the right-hand
sides of \eqref{equation:physical evaluation KZ} and
\eqref{equation:PII physical matrix isomonodromy equation}.
They are independent of $\kappa$, and
$\partial_{z}H_{t}=2\partial_{t}H_{1}$, so flatness gives
$[H_{1},H_{t}]=0$ at fixed $z,t$.  Since
$\mathcal{W}_{\mathrm{NS}}$ is independent of $q,u$, the leading
KZ and quantum isomonodromy equations give the joint eigenvalue problem
\begin{equation}
	H_{1}v=E_{1}v,
	\qquad H_{t}v=E_{t}v,
\end{equation}
Differentiating the critical value, the derivatives of the Bethe
roots drop out by the Bethe equations.  This gives
\begin{equation}
	\label{equation:PII NS derivatives superpotential}
	\begin{aligned}
		E_{1}
		&=-\frac{1}{\epsilon_{1}}\partial_{z}\mathcal{W}_{\mathrm{NS}}
		=z^{2}+\frac{t}{2}-\sum_{a=1}^{N}\frac{1}{z-x_{a}^{*}},\\
		E_{t}
		&=-\frac{2}{\epsilon_{1}}\partial_{t}\mathcal{W}_{\mathrm{NS}}
		=z-2\sum_{a=1}^{N}x_{a}^{*}.
	\end{aligned}
\end{equation}
The factors $-1/\epsilon_{1}$ and $-2/\epsilon_{1}$ follow from
$\kappa=-\epsilon_{2}/\epsilon_{1}$ and the normalization
$2\kappa\partial_{t}\cF=H_{t}\cF$.

\subsubsection{Scalar oper}
\label{subsection:PII NS scalar oper}

The scalar normalization relating the oper and Virasoro blocks is
given in Section~\ref{section:PII integral solutions} by
$\mathcal{G}=u^{N-\frac{1}{2}}(z-q)^{-\frac{1}{2}}
\exp\left({V(z)}/{(2\kappa)}-{V(q)}/{2}\right)\mathcal{I}_{N}$.
For the saddle $x^{*}$, define the corresponding NS wave function by
\begin{equation}
	\label{equation:PII NS scalar wavefunction}
	\chi(q;x^{*},z,t):=
	\exp\left(-\frac{V(q)}{2}\right)
	(z-q)^{-\frac{1}{2}}
	\prod_{a=1}^{N}(q-x_{a}^{*}).
\end{equation}
Then the exact block has the saddle asymptotic
\begin{equation}
	\mathcal{G}(q;z,t)
	\sim u^{N-\frac{1}{2}}C_{N}(\kappa;z,t)
	\exp\left(\frac{\mathcal{W}_{\mathrm{NS}}(z,t)}{\epsilon_{2}}\right)
	\chi(q;x^{*},z,t).
\end{equation}
We use $\mathcal{G}$ for the exact finite-$\epsilon_{2}$ block and
$\chi$ for its leading NS wave function after extracting the
Cartan power, common exponential and Gaussian factor.  The same
notation is used for the dual theory below.  The wave function $\chi$ depends on the selected
Bethe roots $x^{*}$ and on $q,z,t$; in particular, $q$ is the spectral
coordinate.

The scalar oper also follows from the exact $q$-Schwinger--Dyson
equation; the intermediate resolvent calculation is
\eqref{equation:PII q loop equation} in
Appendix~\ref{appendix:PII matrix integral checks}.

The exact $q$-Schwinger--Dyson identity
\eqref{equation:PII matrix integral q equation} reads
\begin{equation}
	\label{equation:PII NS input exact q equation}
	\begin{aligned}
		0={}&\left[\partial_{q}^{2}
		+\frac{\kappa}{z-q}\partial_{z}
		+\left(\frac{1}{z-q}-2q^{2}-t\right)\partial_{q}
		-2\kappa\partial_{t}+2Nq\right]\mathcal{I}_{N}.
	\end{aligned}
\end{equation}
At the chosen saddle,
$\kappa\partial_{z}\log\mathcal{I}_{N}\to
-\sum_{a=1}^{N}(z-x_{a}^{*})^{-1}$ and
$\kappa\partial_{t}\log\mathcal{I}_{N}\to
-\sum_{a=1}^{N}x_{a}^{*}$.
Substitution of \eqref{equation:PII NS integral stationary phase} into
\eqref{equation:PII NS input exact q equation} therefore gives
\begin{equation}
	\label{equation:PII NS polynomial ODE}
	\begin{aligned}
		0={}&P_{N}''(q)
		+\left(\frac{1}{z-q}-2q^{2}-t\right)P_{N}'(q)\\
		&+\left(2Nq+2\sum_{a=1}^{N}x_{a}^{*}
		-\frac{1}{z-q}\sum_{a=1}^{N}\frac{1}{z-x_{a}^{*}}\right)P_{N}(q).
	\end{aligned}
\end{equation}

Some remarks are in order about the equation \eqref{equation:PII NS polynomial ODE}. 
\begin{itemize}
\item First, it provides a differential equation formulation of the finite Bethe system: the Bethe equations  \eqref{equation:PII NS Bethe equations} are precisely the conditions ensuring the existence of a monic polynomial solution $P_N(q)$ of degree $N$. Indeed, evaluating the differential equation at the zeros of $P_N$ reproduces the Bethe equations, while the converse follows from the cancellation of the polynomial remainder.

\item Secondly, the equation retains the geometric information associated with the fundamental insertion at $z$. Although its coefficients have simple poles at $q=z$, the singular contribution is removable when acting on the Bethe polynomial, since $P_N'(z)/P_N(z)$ equals the Bethe resolvent at $z$. Evaluating the resulting polynomial identity at this point yields the finite-root Hamilton–Jacobi relation \eqref{equation:PII NS z Hamilton Jacobi}, which will also ensure the absence of logarithmic monodromy at the corresponding apparent singularity of the scalar oper.

\item Finally, the dependence on the Bethe configuration enters through the first symmetric sum of the roots and their resolvent evaluated at $z$. These quantities determine the joint eigenvalues $E_t$ and $E_1$, respectively. Upon removing the first-derivative term by the scalar normalization , equation \eqref{equation:PII NS polynomial ODE} becomes the Schr\"odinger-type oper equation \eqref{equation:PII NS unscaled quantum oper}, whose additional parameters encode the joint spectrum of the irregular KZ and quantum isomonodromy operators. 

\end{itemize}


Conjugating \eqref{equation:PII NS polynomial ODE} by the prefactor in
\eqref{equation:PII NS scalar wavefunction} removes its first
derivative and yields the NS spectral equation for the saddle wave
function,
\begin{equation}
	\label{equation:PII NS unscaled quantum oper}
	\left[\partial_{q}^{2}-U_{\mathrm{NS}}(q)\right]
	\chi(q;x^{*},z,t)=0,
\end{equation}
where
\begin{equation}
	\label{equation:PII NS unscaled oper potential}
	\begin{aligned}
		U_{\mathrm{NS}}(q)={}&
		\left(q^{2}+\frac{t}{2}\right)^{2}
		-(2N+1)q+E_{t}
		+\frac{E_{1}}{q-z}+\frac{3}{4(q-z)^{2}}.
	\end{aligned}
\end{equation}
The residue $E_{1}$ and the constant term $E_{t}$ therefore
encode the joint spectrum obtained from the NS action.  The local
double pole $3/[4(q-z)^{2}]$ is the oper gauge contribution
of the single spin-$\frac{1}{2}$ insertion.

\paragraph{Apparent singularity}
The other exact loop equation,
\eqref{equation:PII matrix integral z equation}, gives a complementary
leading Hamilton--Jacobi relation.  Writing
$\sum_{a=1}^{N}(z-x_{a}^{*})^{-1}$ for the saddle resolvent, it is
\begin{equation}
	\label{equation:PII NS z Hamilton Jacobi}
	\left(\sum_{a=1}^{N}\frac{1}{z-x_{a}^{*}}\right)^{2}
	-(2z^{2}+t)\sum_{a=1}^{N}\frac{1}{z-x_{a}^{*}}
	+2Nz+2\sum_{a=1}^{N}x_{a}^{*}=0.
\end{equation}
Equivalently, using \eqref{equation:PII NS derivatives superpotential},
\begin{equation}
	\label{equation:PII NS z spectral relation}
	\left(\frac{\partial_{z}\mathcal{W}_{\mathrm{NS}}}{\epsilon_{1}}\right)^{2}
	=\left(z^{2}+\frac{t}{2}\right)^{2}
	-2Nz-2\sum_{a=1}^{N}x_{a}^{*}.
\end{equation}
{\it Thus the critical-level WZW equations give a Hamilton--Jacobi
problem in $z$ and a quantum oper in $q$ determined by finitely many
Bethe roots.}
To see the local meaning of
\eqref{equation:PII NS z spectral relation}, expand the oper potential
at $q=z$:
\begin{equation}
	U_{\mathrm{NS}}(q)
	=\frac{3}{4(q-z)^{2}}+\frac{E_{1}}{q-z}
	+\left[\left(z^{2}+\frac{t}{2}\right)^{2}
	-(2N+1)z+E_{t}\right]+O(q-z).
\end{equation}
The Frobenius ansatz
$\chi=(q-z)^{\rho}\sum_{n\geq0}c_{n}(q-z)^{n}$, with $c_{0}\ne0$,
gives the indicial equation
\begin{equation}
	\rho(\rho-1)-\frac{3}{4}=0,
	\qquad \rho=-\frac{1}{2},\ \frac{3}{2}.
\end{equation}
The exponent difference is two.  On the resonant
$\rho=-1/2$ branch, the first recurrence gives
$c_{1}=-E_{1}c_{0}$, while the next recurrence is compatible without
a logarithm precisely when
\begin{equation}
	E_{1}^{2}
	=\left(z^{2}+\frac{t}{2}\right)^{2}-(2N+1)z+E_{t}.
\end{equation}
Using \eqref{equation:PII NS derivatives superpotential}, this condition
is exactly \eqref{equation:PII NS z spectral relation}.  Hence $q=z$
is an apparent singularity; both local monodromy eigenvalues are
$-1$, so the projective monodromy is trivial.

\subsubsection{Completeness of Bethe vectors}
\label{subsection:PII NS joint spectrum completeness}

The charge condition
$N=\theta+\frac{3}{2}\in\mathbb{Z}_{\geq 0}$ selects the finite
polynomial sector
\begin{equation}
	\label{equation:PII NS polynomial sector}
	\mathcal{V}_{N}
	=u^{N}\mathbb{C}[q]_{\leq N}
	\oplus u^{N-1}\mathbb{C}[q]_{\leq N-1},
	\qquad \dim\mathcal{V}_{N}=2N+1.
\end{equation}
The upper spin component contains $N+1$ polynomial coefficients and
the lower component contains $N$; the latter is absent for $N=0$.
For a generic fixed irregular time $t$ and generic spectral variable
$z$, the Bethe equations \eqref{equation:PII NS Bethe equations}
give $2N+1$ independent simultaneous eigenstates of $H_{1}$ and
$H_{t}$ in this sector.  Each solution has distinct roots away from
$z$, and permuting the roots does not produce a new state.

The two Hamiltonians preserve $\mathcal{V}_{N}$.  The proof and
state count are given in
Appendix~\ref{appendix:PII Bethe completeness proof}.
At exceptional parameter values, levels and eigenvectors can
coalesce into nontrivial Jordan blocks; the examples in
Appendix~\ref{appendix:PII Bethe exceptional points} explain the
genericity condition.

Completeness here counts both spin components.  At fixed $z,t$,
the scalar wave functions \eqref{equation:PII NS scalar wavefunction}
share the same gauge factor and contain a polynomial with at most
$N+1$ independent coefficients.  Their joint spectral information
is also recorded by the accessory parameters $E_{1},E_{t}$ in
\eqref{equation:PII NS unscaled oper potential}.

\subsubsection{Large-\texorpdfstring{$N$}{N} spectral curve}
\label{subsection:PII NS subsequent classical limit}

We take the large-$N$ limit of the Bethe system.
The weight parametrization
\eqref{equation:PII classical WZW theta weight} and the
neutrality condition give
\begin{equation}
	\theta=-\frac{1}{2}+\mathrm{i}\frac{a_{e}}{\epsilon_{1}},
	\qquad N=1+\mathrm{i}\frac{a_{e}}{\epsilon_{1}},
	\qquad \epsilon_{1}N\longrightarrow\Theta=\mathrm{i}a_{e}.
\end{equation}
Balancing the quartic term and the term $Nq$ in the oper potential
gives the cubic scaling already obtained from the KZ equation:
\begin{equation}
	q=\epsilon_{1}^{-\frac{1}{3}}Q,
	\qquad z=\epsilon_{1}^{-\frac{1}{3}}Z,
	\qquad t=\epsilon_{1}^{-\frac{2}{3}}T,
	\qquad x_{a}^{*}=\epsilon_{1}^{-\frac{1}{3}}\xi_{a}^{*}.
\end{equation}
With $\zeta$ as the spectral variable, define the finite-root
resolvent, first moment and rescaled screening number by
\begin{equation}
	\Omega_{N}(\zeta):=\epsilon_{1}\sum_{a=1}^{N}\frac{1}{\zeta-\xi_{a}^{*}},
	\qquad M_{1}:=\epsilon_{1}\sum_{a=1}^{N}\xi_{a}^{*},
	\qquad \mathcal{N}:=\epsilon_{1}N.
\end{equation}
The Bethe equations \eqref{equation:PII NS Bethe equations} become
\begin{equation}
	2(\xi_{a}^{*})^{2}+T
	-\frac{\epsilon_{1}}{Z-\xi_{a}^{*}}
	-2\epsilon_{1}\sum_{\substack{b=1\\b\neq a}}^{N}
	\frac{1}{\xi_{a}^{*}-\xi_{b}^{*}}=0.
\end{equation}
They imply the exact finite-root Riccati relation
\begin{equation}
	\label{equation:PII NS scaled Riccati}
	\begin{aligned}
		0={}&\Omega_{N}(\zeta)^{2}
		+\epsilon_{1}\partial_{\zeta}\Omega_{N}(\zeta)
		-(2\zeta^{2}+T)\Omega_{N}(\zeta)+2\mathcal{N}\zeta+2M_{1}\\
		&-\epsilon_{1}\frac{\Omega_{N}(\zeta)-\Omega_{N}(Z)}{\zeta-Z}.
	\end{aligned}
\end{equation}
This identity holds before taking the large-$N$ limit.

Up to a $Q$-independent normalization, the saddle wave function
\eqref{equation:PII NS scalar wavefunction} becomes
\begin{equation}
	\chi(Q;\xi^{*},Z,T)
	=\exp\left[-\frac{1}{\epsilon_{1}}
	\left(\frac{Q^{3}}{3}+\frac{TQ}{2}\right)\right]
	(Z-Q)^{-\frac{1}{2}}\prod_{a=1}^{N}(Q-\xi_{a}^{*}).
\end{equation}
Since $\partial_{q}^{2}=\epsilon_{1}^{\frac{2}{3}}\partial_{Q}^{2}$,
the rescaled oper is
\begin{equation}
	\left[\epsilon_{1}^{2}\partial_{Q}^{2}
	-\mathcal{U}_{\mathrm{NS}}(Q)\right]
	\chi(Q;\xi^{*},Z,T)=0,
\end{equation}
with potential obtained from
\eqref{equation:PII NS unscaled oper potential}:
\begin{equation}
	\label{equation:PII NS scaled oper}
	\begin{aligned}
		\mathcal{U}_{\mathrm{NS}}(Q)
		={}&\left(Q^{2}+\frac{T}{2}\right)^{2}
		-(2N+1)\epsilon_{1}Q+\epsilon_{1}Z-2M_{1}\\
		&+\epsilon_{1}
		\frac{Z^{2}+\frac{T}{2}-\Omega_{N}(Z)}{Q-Z}
		+\frac{3\epsilon_{1}^{2}}{4(Q-Z)^{2}}.
	\end{aligned}
\end{equation}
The rescaled oper has quantum parameter $\epsilon_{1}$.

In the large-$N$ limit at fixed $\Theta$, write
$\Omega_{N}(\zeta)\longrightarrow\Omega^{(0)}(\zeta)$ and
$M_{1}\longrightarrow M_{1}^{(0)}$ for the formal leading limits.
The leading resolvent equation from
\eqref{equation:PII NS scaled Riccati} is
\begin{equation}
	\label{equation:PII NS limiting Riccati}
	\left(\Omega^{(0)}(\zeta)\right)^{2}
	-(2\zeta^{2}+T)\Omega^{(0)}(\zeta)+2\Theta\zeta+2M_{1}^{(0)}=0.
\end{equation}
The spectral curve is
\begin{equation}
	y^{2}=\left(\zeta^{2}+\frac{T}{2}\right)^{2}-2\Theta\zeta-2M_{1}^{(0)},
	\qquad y(\zeta)=\zeta^{2}+\frac{T}{2}-\Omega^{(0)}(\zeta).
\end{equation}

{\it Hence the classical and NS limits provide complementary realizations of the same irregular KZ–BPZ system: one produces Painlev\'e II dynamics with an $\epsilon_2$-oper, while the other produces Bethe/Gaudin spectral geometry with an $\epsilon_1$-oper; as we shall elaborate in what follows, quantum Langlands duality exchanges these two descriptions.}

\paragraph{Classical and NS limits under Langlands duality}
\label{subsection:PII NS comparison classical WZW}

Langlands duality exchanges the
spectral and isomonodromic coordinates together with the two
deformation parameters.  In the balanced coordinates the exchange is
\begin{equation}
	Q^{\vee}=-Z,\qquad Z^{\vee}=-Q,\qquad T^{\vee}=T.
\end{equation}
It identifies the original classical limit with the dual NS limit,
and the original NS limit with the dual classical limit.  The exact
coordinate and charge maps, including the finite quantum shifts,
are recorded in
Subsection~\ref{subsection:PII finite coupling dictionary}, with their
derivation in Appendix~\ref{appendix:PII finite coupling verification}.

The original NS saddle determines the $q$-oper
\eqref{equation:PII NS unscaled quantum oper}, with one apparent
singularity at $q=z$.  After the cubic rescaling its potential is
\eqref{equation:PII NS scaled oper}.  The leading spectral curve is
\begin{equation}
	\label{equation:PII NS comparison NS curve}
	y^{2}=\left(\zeta^{2}+\frac{T}{2}\right)^{2}-2\Theta\zeta-2M_{1}^{(0)}.
\end{equation}
Setting $P=\Omega^{(0)}(Q)$ and evaluating
\eqref{equation:PII NS limiting Riccati} at $Q$ gives
\begin{equation}
	P^{2}-(2Q^{2}+T)P+2Q\Theta=-2M_{1}^{(0)}.
\end{equation}
Substitution into the classical spectral curve
\eqref{equation:PII classical WZW subsequent spectral curve}
reproduces \eqref{equation:PII NS comparison NS curve}.
The classical and NS limits give the same spectral polynomial.

{\bf Remark :}
Together Subsections~\ref{section:PII classical WZW limit} and~\ref{section:PII NS comparison} establish the following :
\begin{equation}
\label{eq:omega_limits}
\begin{tikzcd}[
	column sep=5.5em,
	row sep=3.2em,
	every arrow/.append style={-{Straight Barb[length=2.1mm,width=1.5mm]}},
	every label/.append style={font=\small},
]
	(\epsilon_1,\epsilon_2)
		\arrow[r, "\epsilon_1\to0"]
		\arrow[d, "\epsilon_2\to0"']
	&
	\text{Painlev\'e II}+\epsilon_2\text{-oper}
		\arrow[d, "\epsilon_2\to0"]
	\\
	\text{Bethe/Gaudin}+\epsilon_1\text{-oper}
		\arrow[r, "\epsilon_1\to0"']
	&
	\text{Seiberg--Witten curve}
\end{tikzcd}
\end{equation}

The two limiting procedures therefore reveal complementary aspects of the same underlying integrable structure: the Painlev\'e II isomonodromic dynamics and the irregular Gaudin spectral problem. Although both lead to the same Seiberg–Witten geometry upon removing the remaining deformation parameter, this agreement alone does not establish a correspondence between the two quantum descriptions. A natural question is whether their relationship can be understood before taking these limits. In the following section \ref{section:PII Langlands interpretation}, we show that quantum Langlands duality provides precisely such a connection. The transformation $\kappa\leftrightarrow\kappa^{-1}$ exchanges the spectral insertion and the Whittaker zero, intertwines the corresponding BPZ and affine systems, and maps the classical limit of one theory to the NS limit of its dual. The large-$N$ limit of the NS saddle, together with the Seiberg--Witten prepotential and differential that it yields, is developed in Section~\ref{section:PII planar SW geometry}.

\section{Quantum Langlands duality}
\label{section:PII Langlands interpretation}

The exact quantum Langlands map exchanges the fundamental insertion
with the simple zero of the Whittaker differential and relates the
joint affine systems at the dual level $\kappa^{\vee}=\kappa^{-1}$.
It interchanges the classical and NS limits of
Subsections~\ref{section:PII classical WZW limit}
and~\ref{section:PII NS comparison}.  We first record the
correspondence dictionary, then apply the original finite-$N$ NS
data to reconstruct dual Painlev\'e II trajectories and Lax wave
functions.  Their characteristic determinants provide the
finite-screening input for
Section~\ref{section:PII planar SW geometry}.

The correspondence realizes the coupling inversion in
\cite{FrenkelGukovTeschner2015,ArakawaFrenkel2019,GiribetNakayamaNicolas2008}.
The dual affine construction, DS reduction, screening integrals,
and step-by-step covariance and reconstruction calculations are
collected in Appendix~\ref{appendix:PII finite coupling verification}.

\subsection{Exact quantum Langlands dictionary}
\label{subsection:PII finite coupling dictionary}
Choose a branch of $a^{3}=\kappa^{-1}$\footnote{The irregular Ward polynomial in equation \eqref{equation:irregular Ward polynomial} is given by 
\begin{equation*}
P_\kappa(x) = 2\partial_t + \frac{t^2}{4\kappa} + \frac{tx^2 + x^4}{\kappa} + \frac{x}{\kappa}(-2u\partial_u + \kappa -2)
\end{equation*}
The operator $\mathcal{T}_{\mathrm{BPZ}}$ exchanges the variables $q^\vee$ and $z^\vee$, as defined in the equation \eqref{equation:PII Langlands transformed block}. Then from equation \eqref{equation:PII Langlands BPZ dictionary}, one imposes 
\begin{equation*}
 P_{\kappa^\vee}(a x) = a^{-2} P_\kappa (x).
 \end{equation*} This forces this the choice $a^3 = \kappa^{-1}$ upon us and this is also consistent with the BPZ covariance relations in equation \eqref{equation:PII Langlands BPZ dictionary}. More details can be found in appendix \ref{subsection:PII finite coupling affine exchange}.}.  The coordinates, dual level
and Cartan charges are related by
\begin{equation}
	\label{equation:PII Langlands finite coordinate map}
	\begin{gathered}
		\kappa^{\vee}=\kappa^{-1},\qquad
		q^{\vee}=az,\qquad z^{\vee}=aq,\qquad
		t^{\vee}=a^{2}t,\\
		\theta^{\vee}+\frac{5}{2}
		=\frac{\theta+\frac{5}{2}}{\kappa},\qquad
		u=(u^{\vee})^{\frac{1}{\kappa}}.
	\end{gathered}
\end{equation}
The reduced Virasoro data are
\begin{equation}
	\begin{gathered}
		c(\kappa)=13-6\left(\kappa+\kappa^{-1}\right)
		=c(\kappa^{\vee}),\\
		\Delta_{q}=\frac{3\kappa-2}{4},\qquad
		\Delta_{z}=\frac{3\kappa^{-1}-2}{4},\qquad
		\Delta^{\vee}_{q^{\vee}}=\Delta_{z},\qquad
		\Delta^{\vee}_{z^{\vee}}=\Delta_{q}.
	\end{gathered}
\end{equation}
We have 
\begin{itemize}
    \item the original theory at level $\kappa$ 
    , with a conformal block $\mathcal{G} (q;z,t,u)$ and 
     \item a Langlands dual theory at level $\kappa^\vee = \kappa^{-1}$, with a conformal block $\mathcal{G}^\vee(q^\vee; z^\vee,t^\vee,u^\vee)$.
\end{itemize}
There are two BPZ solution spaces : 
\begin{equation}
S_\kappa = \big\{\mathcal{G} : \mathcal{E}_q \mathcal{G} = \mathcal{E}_z \mathcal{G} = 0, \,\,u\partial_u \mathcal{G} = (\theta+1)\mathcal{G}\big\}.
\end{equation}
Then along with the Langlands-dual BPZ operators, as defined in equation \eqref{equation:PII independently derived dual BPZ}, the dual solution space is
\begin{equation}
S_{\kappa^\vee} = \{\mathcal{G}^\vee : \mathcal{E}^\vee_{q^\vee} \mathcal{G}^\vee = \mathcal{E}^\vee_{z^\vee} \mathcal{G}^\vee= 0, \,\,u^\vee\partial_{u^\vee}\mathcal{G}^\vee =(\theta^\vee+1)\mathcal{G}^\vee \}. 
\end{equation}
Define an operator 
\begin{equation}
\label{eq:intertwiner}
\mathcal{T}_{\mathrm{BPZ}} : S_\kappa \rightarrow \, S_{\kappa^\vee}, \quad 
\mathcal{G} \mapsto \mathcal{G}^\vee,
\end{equation}
such that (with the argument order $(q;z,t,u)$ explicit),  \footnote{The prefactor of $(u^\vee)^{\frac32(\kappa^{-1}-1)}$ is fixed using equation \eqref{equation:PII Langlands finite coordinate map}.}
\begin{equation}
	\label{equation:PII Langlands transformed block}    \mathcal{G}^\vee(q^{\vee};z^{\vee},t^{\vee},u^{\vee})
		=\,\, (u^{\vee})^{\frac{3}{2}(\kappa^{-1}-1)}
		\mathcal{G}\left(\frac{z^{\vee}}{a};
		\frac{q^{\vee}}{a},
		\frac{t^{\vee}}{a^{2}},
		(u^{\vee})^{\frac{1}{\kappa}}\right).
\end{equation}
The BPZ-operator dictionary is
\begin{equation}
	\label{equation:PII Langlands BPZ dictionary}
	\begin{aligned}
		\mathcal{E}_{q^{\vee}}^{\vee}\mathcal{T}_{\mathrm{BPZ}}
		&=a^{-2}\mathcal{T}_{\mathrm{BPZ}}\mathcal{E}_{z},\\
		\mathcal{E}_{z^{\vee}}^{\vee}\mathcal{T}_{\mathrm{BPZ}}
		&=a^{-2}\mathcal{T}_{\mathrm{BPZ}}\mathcal{E}_{q}.
	\end{aligned}
\end{equation}
Writing $\mathscr{D}_{\kappa}$ and $\mathscr{L}_{\kappa}$ for
scalar reduction and affine reconstruction, the corresponding map
of joint affine solutions is
\begin{equation}
	\label{equation:PII Langlands affine intertwiner}
	\mathcal{T}_{\mathrm{aff}}
	=\mathscr{L}_{\kappa^{\vee}}
	\mathcal{T}_{\mathrm{BPZ}}\mathscr{D}_{\kappa},
	\qquad
	\mathscr{D}_{\kappa^{\vee}}\mathcal{T}_{\mathrm{aff}}
	=\mathcal{T}_{\mathrm{BPZ}}\mathscr{D}_{\kappa}.
\end{equation}
The component formulas for these mutually inverse reduction and
reconstruction maps are given in
Appendix~\ref{appendix:PII affine reconstruction verification}.

For the screening realizations, neutrality and the Vandermonde
exponents give the dictionary
\begin{equation}
	\label{equation:PII Langlands dual screening number}
	\begin{gathered}
		N=\theta+\frac{3}{2},\qquad
		N^{\vee}=\theta^{\vee}+\frac{3}{2},\qquad
		N^{\vee}+1=\frac{N+1}{\kappa},\\
		\beta=\frac{2}{\kappa},\qquad
		\beta^{\vee}=\frac{2}{\kappa^{\vee}}
		=2\kappa=\frac{4}{\beta}.
	\end{gathered}
\end{equation}
Screening integrals require nonnegative integer screening numbers;
their cycles select solution branches related by
\eqref{equation:PII Langlands transformed block}, as discussed in
Appendix~\ref{subsection:PII dual screening representation}.

The deformation parameters are exchanged as
$\epsilon_{1}^{\vee}=\epsilon_{2}$ and
$\epsilon_{2}^{\vee}=\epsilon_{1}$.  Thus
\begin{equation}
	\label{equation:PII Langlands limit dictionary}
	\begin{array}{ccl}
		\epsilon_{2}\to0\ \text{(original NS)}
		&\longleftrightarrow&\epsilon_{1}^{\vee}\to0\ \text{(dual classical)},\\ \qquad
		\epsilon_{1}\to0\ \text{(original classical)}
		&\longleftrightarrow&\epsilon_{2}^{\vee}\to0\ \text{(dual NS)}.
	\end{array}
\end{equation}
In balanced coordinates the dictionary reads
\begin{equation}
	Q^{\vee}=-Z,\qquad Z^{\vee}=-Q,\qquad T^{\vee}=T.
\end{equation}

We denote by $\alpha^\vee$ the parameter of the dual Painlev\'e II equation in the standard normalization $w''=2w^3+tw+\alpha^\vee$. As we shall derive in subsection \ref{subsection:PII Langlands limiting opers}, the finite-screening sector with $N$ Bethe roots corresponds to the half-integer value $\alpha^\vee=N+\frac12$.
For the original finite-$N$ NS-to-dual-classical limit, the finite
dual Painlev\'e coordinates and charge are
\begin{equation}
	\label{equation:PII Langlands unscaled coordinate exchange}
	w=-z,\qquad \lambda=-q,\qquad t^{\mathrm{PII}}=t,
\end{equation}
\begin{equation}
	\label{equation:PII Langlands reverse classical charge}
	\epsilon_{1}^{\vee}\theta^{\vee}
	\longrightarrow-\epsilon_{1}(N+1),\qquad
	\alpha^{\vee}=N+\frac{1}{2}.
\end{equation}
Here $w$ is the isomonodromic coordinate and $\lambda$ is the
spectral coordinate.  The momentum, action and Lax wave function
are identified in the next subsection; the limiting coordinate
and charge calculations are in
Appendix~\ref{subsection:PII Langlands exchanged limits}.

\subsection{From finite-\texorpdfstring{$N$}{N} NS data to dual Painlev\'e II}
\label{subsection:PII Langlands limiting opers}

We focus on the original NS-to-dual-classical row of the duality,
keeping the original screening number $N$ finite.  The original
coordinates $q,z,t$ and the original NS Bethe roots $x_a^*$ are those
of Subsection~\ref{section:PII NS comparison}.  The dual variables are $w,p$ and the spectral coordinate is
$\lambda$.  The coupling and charge data are
\begin{equation}
	\label{equation:PII Langlands NS to dual charge}
	\epsilon_2\to0,\quad \epsilon_1^{\vee}=\epsilon_2\to0,
	\quad\epsilon_2^{\vee}=\epsilon_1\ne0,\qquad
	\alpha^{\vee}=N+\frac12.
\end{equation}
\paragraph*{Bethe roots, action and the Painlev\'e momentum.}
The finite Bethe equations and their monic polynomial are
\begin{equation}
	\label{equation:PII Langlands Bethe regularity in PII}
	\begin{gathered}
		2(x_a^*)^2+t-\frac{1}{z-x_a^*}
		-2\sum_{b\ne a}\frac{1}{x_a^*-x_b^*}=0,\\
		P_N(q)=\prod_{a=1}^N(q-x_a^*),\qquad
		R_N(q)=\frac{P'_N(q)}{P_N(q)},\qquad
		s_N=\sum_{a=1}^Nx_a^*.
	\end{gathered}
\end{equation}
The exact NS-envelope derivatives from
\eqref{equation:PII NS derivatives superpotential} read
\begin{equation}
	\label{equation:PII Langlands NS finite action derivatives}
	\frac{\partial_z\mathcal W_{\mathrm{NS}}}{\epsilon_1}
	=R_N(z)-z^2-\frac t2,\qquad
	\frac{\partial_t\mathcal W_{\mathrm{NS}}}{\epsilon_1}
	=s_N-\frac z2.
\end{equation}
For the dual position $w=-z$ define the dimensionless characteristic
action (the Cartan term is suppressed)
\begin{equation}
	\label{equation:PII Langlands limiting action exchange inverse}
	\mathcal W^{\vee}(w,t)
	=\frac{1}{\epsilon_1}\mathcal W_{\mathrm{NS}}(-w,t)
	+\frac{w^3}{3}+\frac{tw}{2}.
\end{equation}
Taking the derivative with respect to $w$ on a chosen NS saddle gives
\begin{equation}
	\label{equation:PII Langlands special oper identification}
	p:=\partial_w\mathcal W^{\vee}
	=2w^2+t-R_N(-w)
	=2w^2+t+\sum_{a=1}^N\frac{1}{w+x_a^*},
	\qquad w=-z.
\end{equation}
The finite-root resolvent determines the dual momentum at $w=-z$.
The accessory identity becomes
\begin{equation}
	\label{equation:PII Langlands special accessory identity}
	-2s_N=p^2-(2w^2+t)p-2Nw.
\end{equation}
This identifies the NS accessory coefficient with the dual Painlev\'e
Hamiltonian data.

\paragraph*{The dual Lax wave function and its apparent pole.}
The normalized original NS scalar wave function remains
\begin{equation}
	\label{equation:PII Langlands NS finite polynomial wave}
	\chi_N(q;z,t)=e^{-q^3/3-tq/2}(z-q)^{-1/2}P_N(q).
\end{equation}
Under the interchange $q=-\lambda$, $z=-w$, it gives a
particular solution of the dual scalar Painlev\'e II Lax equation:
\begin{equation}
	\label{equation:PII Langlands limiting oper correspondence}
	\begin{gathered}
		\phi_{\mathrm{cl}}^{\vee}(\lambda,t)
		=C(t)e^{\lambda^3/3+t\lambda/2}
		(\lambda-w)^{-1/2}\prod_{a=1}^N(\lambda+x_a^*),\\
		\bigl[\partial_{\lambda}^2-U^{\vee}(\lambda)\bigr]\phi_{\mathrm{cl}}^{\vee}=0,
		\qquad U^{\vee}(\lambda)=U_{\mathrm{NS}}(-\lambda),
	\end{gathered}
\end{equation}
where $U_{\mathrm{NS}}$ is given in
\eqref{equation:PII NS unscaled oper potential}, explicitly:
\begin{equation}
	\label{equation:PII Langlands unscaled dual oper explicit}
	\begin{aligned}
		U^{\vee}(\lambda)={}&
		\left(\lambda^2+\frac t2\right)^2+(2N+1)\lambda
		-w-2s_N\\
		&-\frac{p-w^2-t/2}{\lambda-w}
		+\frac{3}{4(\lambda-w)^2}.
	\end{aligned}
\end{equation}
The point $\lambda=w$ is the scalar apparent singularity; the
zeros $\lambda=-x_a^*$ are not singularities of the connection.
Indeed the Miura logarithmic derivative is
\begin{equation}
	\label{equation:PII Langlands Miura factorization}
	\begin{aligned}
		\partial_\lambda\log\phi_{\mathrm{cl}}^{\vee}
		&=\lambda^2+\frac t2+\sum_{a=1}^N\frac{1}{\lambda+x_a^*}
		-\frac{1}{2(\lambda-w)},\\
		U^{\vee}&=\bigl(\partial_\lambda\log\phi_{\mathrm{cl}}^{\vee}\bigr)^2
		+\partial_\lambda^2\log\phi_{\mathrm{cl}}^{\vee}.
	\end{aligned}
\end{equation}
Cancellation of the residues at the polynomial zeros reproduces
\begin{equation}
	\label{equation:PII Langlands reflected finite Bethe equations}
	2(x_a^*)^2+t+\frac{1}{w+x_a^*}
	-2\sum_{b\ne a}\frac{1}{x_a^*-x_b^*}=0.
\end{equation}
The NS Bethe polynomial therefore supplies the dual Lax wave
function with apparent point $\lambda=w$.

\paragraph*{Time evolution and the dual Painlev\'e equation.}
The scalar time equation fixes the previously free normalization:
\begin{equation}
	\label{equation:PII Langlands unscaled time equation}
	2\,\partial_t\phi_{\mathrm{cl}}^{\vee}
	=\frac{1}{\lambda-w}\partial_\lambda\phi_{\mathrm{cl}}^{\vee}
	+\frac{\phi_{\mathrm{cl}}^{\vee}}{2(\lambda-w)^2},
	\qquad \frac{C'(t)}{C(t)}=\frac{w(t)}2.
\end{equation}
Along an isomonodromic characteristic the moving Bethe zeros obey
\begin{equation}
	\label{equation:PII Langlands Bethe zero dynamics}
	(x_a^*)'=-\frac{1}{2(w+x_a^*)}
	=(x_a^*)^2+\frac t2
	-\sum_{b\ne a}\frac{1}{x_a^*-x_b^*}.
\end{equation}
Together with the resolvent dictionary this gives
\begin{equation}
	\label{equation:PII Langlands dual PII Hamilton system}
	\begin{aligned}
		w'&=p-w^2-\frac t2,\\
		p'&=2wp+N+1,\\
		w''&=2w^3+tw+N+\frac12.
	\end{aligned}
\end{equation}
Equivalently the original insertion evolves as $z(t)=-w(t)$.
In the standard Hamiltonian convention of \eqref{eq:Laxpaircl},
the same trajectory is described by
\begin{equation*}
	q_{\mathrm{std}}=w,\qquad
	p_{\mathrm{std}}=p-2w^{2}-t,\qquad
	\theta_{\mathrm{std}}=-N.
\end{equation*}
These identifications bring the first two equations of
\eqref{equation:PII Langlands dual PII Hamilton system} into the
standard Hamiltonian form and give
$\alpha_{\mathrm{std}}=\frac{1}{2}-\theta_{\mathrm{std}}=N+\frac{1}{2}$.
In particular, the dual momentum $p$ used here differs from the
standard momentum by $2w^{2}+t$.
At fixed $(z,t)$, the $2N+1$ Bethe branches give the initial data
for the dual Painlev\'e trajectories.

\paragraph*{Finite-dimensional spectral data.}
The original critical Gaudin operators preserve the
$(2N+1)$-dimensional polynomial sector
\begin{equation}
	\mathcal{V}_{N}=\mathbb{C}[q]_{\le N}\oplus\mathbb{C}[q]_{\le N-1}.
\end{equation}
After extracting the Cartan powers, the operators are
\begin{equation}
	\label{equation:PII Langlands finite critical position operator}
	H_{1}(z,t)=\begin{pmatrix}
		z^{2}+t/2-\partial_{q}&z-q\\
		2((z+q)\partial_{q}-N)&-z^{2}-t/2+\partial_{q}
	\end{pmatrix},
\end{equation}
and
\begin{equation}
	\label{equation:PII Langlands unscaled finite time operator}
	H_{t}(z,t)=\bigl[\partial_{q}^{2}-(2q^{2}+t)\partial_{q}\bigr]\mathbf{1}
	+\begin{pmatrix}2Nq+z&1\\2\partial_{q}&2(N-1)q-z\end{pmatrix}.
\end{equation}
On a Bethe eigenvector, $H_{1}$ has eigenvalue
$-\partial_{z}\mathcal{W}_{\mathrm{NS}}/\epsilon_{1}
=p-w^{2}-t/2$.  Consequently, at any fixed $(w,t)$ the
possible initial momenta are the roots of the finite determinant
\begin{equation}
	\label{equation:PII Langlands general invariant determinant}
	D_{N}(w,p,t):=
	\det_{\mathcal{V}_{N}}\!\left[
	\left(p-w^{2}-\frac{t}{2}\right)\mathbf{1}-H_{1}(-w,t)
	\right]=0.
\end{equation}
It has degree $2N+1$ in $p$ and polynomial coefficients in $w,t$.
The finite Gaudin operators also obey
\begin{equation}
	\label{equation:PII Langlands finite critical quadratic identity}
	H_{1}^{2}-H_{t}
	=\left[\left(z^{2}+\frac{t}{2}\right)^{2}-(2N+1)z\right]\mathbf{1},
	\qquad\partial_{z} H_{t}=2\partial_{t} H_{1}.
\end{equation}
The Painlev\'e vector field transports the characteristic polynomial:
\begin{equation}
	\label{equation:PII Langlands general invariant polynomial flow}
	\left[\partial_{t}+
	\left(p-w^{2}-\frac{t}{2}\right)\partial_{w}
	+(2wp+N+1)\partial_{p}\right]D_{N}=-2wD_{N}.
\end{equation}
The zero locus $D_{N}=0$ parametrizes admissible Painlev\'e initial data.

At $N=0$ and $N=1$, respectively,
\begin{equation}
	\label{equation:PII Langlands empty invariant curve}
	D_{0}=p-2w^{2}-t,
\end{equation}
\begin{equation}
	\label{equation:PII Langlands one root invariant curve}
	D_{1}=p(2w^{2}+t-p)^{2}+4w(2w^{2}+t-p)+2.
\end{equation}
The empty sector reduces to $w'=w^{2}+t/2$.
The one-root curve equivalently follows by eliminating $x_{1}^*$ from
$2(x_{1}^*)^{2}+t+1/(w+x_{1}^*)=0$ and
$p=2w^{2}+t+1/(w+x_{1}^*)$.

Increasing the screening number by one induces the B\"acklund step
\begin{equation}
	\label{equation:PII Langlands screening Backlund shift}
	w_{N+1}=-w_{N}-\frac{N+1}{p_{N}},\qquad
	p_{N+1}=2w_{N+1}^{2}+t-p_{N},
\end{equation}
where $p_{N}\ne0$.  On the NS Gaudin columns, the polynomial
Darboux transformation reads
\begin{equation}
	\label{equation:PII Langlands polynomial Darboux shift}
	v_{N+1}(q)\ \propto\
	\begin{pmatrix}q-w_{N}&1/2\\p_{N}&0\end{pmatrix}v_{N}(q),
\end{equation}
with each vector evaluated on its corresponding moving insertion
$z_{N}=-w_{N}$.

\subsection{Free-field screening duality and T-duality origin of the Langlands exchange}
\label{sec:hecke-separation-langlands}

To understand why the level inversion $\kappa^{\vee}=\kappa^{-1}$ simultaneously exchanges the fundamental insertion at $z$ with the Whittaker zero at $q$ and swaps the classical and Nekrasov--Shatashvili limits, it is instructive to recall how the Langlands dual algebra ${}^{L}\mathfrak{g}$ emerges from the Wakimoto realization in the conformal field theory approach to geometric Langlands \cite[Sections~8.5--8.6]{Frenkel2005LanglandsLectures} (see also \cite{FeiginFrenkel1992,Frenkel2005Wakimoto}).

\paragraph{Wakimoto screening and the Virasoro kernel.}
In the Wakimoto free-field realization of $\widehat{\mathfrak{sl}}_{2}$ at noncritical shifted level $\kappa=k+2\neq 0$, the affine currents are embedded into the tensor product of the first-order $(\beta,\gamma)$ system and the Heisenberg chiral algebra $\pi_{0}$ generated by $a(x)=\sqrt{2\kappa}\,\partial_{x}\varphi(x)$, with $\varphi(x)\varphi(y)\sim +\log(x-y)$. For generic $\kappa$, the image of the affine vertex algebra inside the free-field algebra is cut out as the kernel of the Wakimoto screening operator $\int \beta(x)V_{-}(x)\,\dd x$, whose purely bosonic factor is the vertex operator
\begin{equation}
	\label{equation:PII Frenkel magnetic screening}
	V_{-}(x)=\;:e^{-\sqrt{2/\kappa}\,\varphi(x)}:\;.
\end{equation}
Upon passing to Drinfeld--Sokolov reduction (or restricting to the $\beta\gamma$-vacuum sector), the reduced Virasoro algebra $\mathcal{W}_{\kappa}(\mathfrak{sl}_{2})$ is realized inside $\pi_{0}$ as the kernel of the bosonic screening charge $\int V_{-}(x)\,\dd x$. Indeed, the DS stress tensor
\begin{equation}
	\label{equation:PII Frenkel Virasoro}
	T_{\kappa}(x)
	=\frac{1}{4\kappa}:a(x)^{2}:+\frac{\kappa-1}{2\kappa}\partial_{x}a(x)
	=\frac{1}{2}:(\partial_{x}\varphi)^{2}:+\frac{\kappa-1}{\sqrt{2\kappa}}\,\partial_{x}^{2}\varphi,
	\qquad
	c(\kappa)=1-\frac{6(\kappa-1)^{2}}{\kappa},
\end{equation}
is precisely characterized by the property that $V_{-}(x)$ has conformal weight~$1$, so that $\int V_{-}(x)\,\dd x$ commutes with $T_{\kappa}(x)$.

The key observation of Feigin and Frenkel \cite{FeiginFrenkel1992,Frenkel2005LanglandsLectures} is that $T_{\kappa}(x)$ is manifestly invariant under the involution $\sqrt{\kappa}\mapsto -\sqrt{\kappa}^{-1}$ (i.e.\ $\kappa\mapsto\kappa^{\vee}=\kappa^{-1}$, or $a(x)\mapsto a^{\vee}(x)=-a(x)/\kappa$), and hence there is a \emph{second} vertex operator of conformal weight~$1$ with respect to the very same stress tensor $T_{\kappa}(x)$:
\begin{equation}
	\label{equation:PII Frenkel electric screening}
	V_{+}(x)=\;:e^{\sqrt{2\kappa}\,\varphi(x)}:\;.
\end{equation}
For irrational $\kappa$, the kernels of the two screening operators inside $\pi_{0}$ coincide and both equal the Virasoro vertex algebra generated by $T_{\kappa}(x)$:
\begin{equation}
	\label{equation:PII Frenkel kernel duality}
	\mathcal{W}_{\kappa}(\mathfrak{sl}_{2})
	=\ker_{\pi_{0}}\int V_{-}(x)\,\dd x
	=\ker_{\pi_{0}}\int V_{+}(x)\,\dd x
	\simeq \mathcal{W}_{\kappa^{\vee}}({}^{L}\mathfrak{sl}_{2}),
	\qquad
	\kappa^{\vee}=\kappa^{-1}.
\end{equation}

\paragraph{T-duality, weights versus coweights, and the $q\leftrightarrow z$ exchange.}
As explained in \cite[Section~8.6]{Frenkel2005LanglandsLectures}, the coincidence of the two screening kernels has a transparent physical interpretation in terms of target-space T-duality. Compactifying the full boson $\varphi(x,\bar{x})$ on a circle of radius $R=\sqrt{\kappa}$ (for a general simple Lie algebra $\mathfrak{g}$, on the torus $\mathfrak{h}_{\mathbb{R}}^{*}/\sqrt{\kappa}^{-1}P$, where $P$ is the weight lattice), the operator $V_{+}(x)$ is the chiral component of the \emph{electric} vertex operator carrying unit momentum and zero winding, labeled by the simple coroots $\check{\alpha}_{j}\in\check{P}$, whereas $V_{-}(x)$ is the chiral component of the \emph{magnetic} vertex operator carrying zero momentum and unit winding, labeled by the simple roots $\alpha_{j}\in P$. Target-space T-duality $R\leftrightarrow 1/R$ ($\sqrt{\kappa}\leftrightarrow 1/\sqrt{\kappa}$) exchanges the weight lattice $P$ of $\mathfrak{g}$ with the coweight lattice $\check{P}={}^{L}P$ of the Langlands dual algebra ${}^{L}\mathfrak{g}$ and interchanges electric and magnetic screenings. Consequently, deforming the free boson by the magnetic screening of $\mathfrak{g}$ at level $\kappa$ produces the exact same chiral algebra as deforming the T-dual boson by the magnetic screening of ${}^{L}\mathfrak{g}$ at level $\kappa^{\vee}=\kappa^{-1}$.

In our Painlev\'e~II construction, this weight--coweight T-duality directly explains the exchange of the two punctures in \eqref{equation:PII Langlands finite coordinate map}--\eqref{equation:PII Langlands BPZ dictionary}:
\begin{enumerate}
	\item The fundamental affine insertion at $z$ is a highest-weight module labeled by the fundamental \emph{weight} $\omega=\frac{1}{2}\alpha\in P$; in the reduced Heisenberg realization it carries charge $a_{0}=1$ and couples to the electric/light degenerate vertex operator, with null vector $(L_{-1}^{2}-\kappa^{-1}L_{-2})|z\rangle=0$.
	\item The simple zero of the Whittaker differential at $q$ induces a spectral-flow / Li twist labeled by the fundamental \emph{coweight} $-\check{\omega}=-\frac{1}{2}h\in\check{P}$; in the reduced Heisenberg realization it carries charge $a_{0}=-\kappa$ and couples to the magnetic/heavy degenerate vertex operator, with null vector $(L_{-1}^{2}-\kappa L_{-2})|q\rangle=0$.
\end{enumerate}
Under the T-duality map $a^{\vee}(x)=-a(x)/\kappa$ of \eqref{equation:PII Langlands reduced Heisenberg exchange}, the Heisenberg charges $(1,-\kappa)$ are mapped to $(-\kappa^{-1},1)=(-\kappa^{\vee},1)$: the fundamental weight insertion at $z$ becomes the coweight Whittaker zero at $q^{\vee}=az$, and the coweight Whittaker zero at $q$ becomes the fundamental weight insertion at $z^{\vee}=aq$ \cite{ArakawaFrenkel2019}.

\paragraph{Critical-level limit and the exchange of NS and classical regimes.}
Finally, the two screening operators govern the opposite asymptotic regimes $\kappa\to 0$ and $\kappa\to\infty$ \cite[Section~8.5]{Frenkel2005LanglandsLectures}. In the critical-level limit $\kappa\to 0$ ($k\to -2$), the rescaled field $\tilde{a}(x)=a(x)$ generates a commutative chiral algebra $\tilde{\pi}_{0}$, the magnetic screening $\int V_{-}(x)\,\dd x$ regularizes to a derivation whose kernel is the Feigin--Frenkel center $\mathfrak{z}(\widehat{\mathfrak{sl}}_{2})\subset V_{-2}(\mathfrak{sl}_{2})$, while the electric screening $\int V_{+}(x)\,\dd x$ reduces to the classical limit of $\kappa T_{\kappa}(x)$, namely the Miura oper potential $\frac{1}{4}\tilde{a}(x)^{2}-\frac{1}{2}\partial_{x}\tilde{a}(x)\in\operatorname{Fun}\operatorname{Op}_{{}^{L}\mathfrak{sl}_{2}}$. Thus the finite-$\kappa$ duality \eqref{equation:PII Frenkel kernel duality} degenerates as $\kappa\to 0$ into the isomorphism between the critical-level Gaudin/Bethe algebra ($\epsilon_{2}\to 0$, our NS limit) and classical ${}^{L}\mathfrak{sl}_{2}$-opers ($\epsilon_{1}^{\vee}\to 0$, our dual classical WZW/Painlev\'e II limit). This is the exact mechanism behind the crossed commutative diagram \eqref{eq:omega_limits}.

\section{Seiberg--Witten geometry in the Large $N$ Limit}
\label{section:PII planar SW geometry}

We compute the leading $\epsilon_{1}$ expansion of the NS action
$\mathcal{W}_{\mathrm{NS}}$ and of the wave action $S$ from the
large-$N$ limit of the Bethe saddle.  The saddle determines a
two-cut spectral curve, whose periods and on-shell action give the
Seiberg--Witten prepotential and differential.  We first fix the
scaling and the normalization of the expansion, then evaluate the
leading coefficients.

\subsection{NS actions and the Large $N$ Limit}
\label{subsection:PII chapter7 Large Limit}

Our starting point is the reduced Virasoro conformal block
$\mathcal{G}(q;z,t)=u^{N-\frac{1}{2}}(z-q)^{-\frac{1}{2}}
\exp{\left(\frac{V(z;t)}{2\kappa}-\frac{V(q;t)}{2}\right)}
\mathcal{I}_{N}(q,t;z)$ of \eqref{equation:PII integral oper Virasoro dictionary},
obtained by quantum Drinfeld--Sokolov reduction
(\S\ref{subsection:PII DS local modules}) of the
$\widehat{\mathfrak{sl}}_{2}$ Wakimoto--Whittaker block with screening
integral $\mathcal{I}_{N}$ in \eqref{equation:PII matrix integral} and
$V(x;t)=\frac{2}{3}x^{3}+tx$.  Physically, $\mathcal{G}$ describes the
$H_{1}$ Argyres--Douglas theory at infinity probed by two level-two
degenerate surface defects: one at $z$ of weight
$\Delta_{z}=(3\kappa^{-1}-2)/4$ descending from the fundamental
insertion, and one at $q$ of weight $\Delta_{q}=(3\kappa-2)/4$ created
by the zero of the Whittaker differential $\psi(x)=u(x-q)$.  In the
integrand \eqref{equation:PII matrix integral}, these enter as
$\prod_{a}(z-x_{a})^{-1/\kappa}$ and $\prod_{a}(x_{a}-q)^{1}$, with
$\kappa=-\epsilon_{2}/\epsilon_{1}$.

The NS limit is taken first: $\epsilon_{2}\to0$ ($\kappa\to0$) at fixed
$\epsilon_{1}\ne0$ and $N\in\mathbb{Z}_{\geq0}$
(\S\ref{section:PII NS comparison}).  Because of their different
$\kappa$-scalings, the defect at $z$ is \emph{heavy}
($\Delta_{z}\sim -3\epsilon_{1}/(4\epsilon_{2})$) and enters the
saddle-point functional at order $\epsilon_{2}^{-1}$, whereas the
defect at $q$ is \emph{light} ($\Delta_{q}\to-1/2$) and does not shift
the Bethe roots $x_{a}^{*}$ of \eqref{equation:PII NS Bethe equations},
entering instead at $\mathcal{O}(\epsilon_{2}^{0})$.  Continuing all
logarithms from a nondegenerate saddle $x^{*}$ gives the factorization
\eqref{equation:PII NS scalar wavefunction}:
\begin{equation}
	\label{equation:PII chapter7 W chi definitions}
	\begin{aligned}
		\mathcal{G}(q;z,t)
		&\sim u^{N-\frac{1}{2}}C_{N}(\kappa;z,t)\,
		\exp{\left(\frac{\mathcal{W}_{\mathrm{NS}}(z,t)}{\epsilon_{2}}\right)}
		\chi_{N}(q;z,t),\\
		\mathcal{W}_{\mathrm{NS}}(z,t)
		&=\epsilon_{1}\left(\mathcal{Y}_{N}(x^{*};z,t)
		-\frac{V(z;t)}{2}\right),\\
		\mathcal{Y}_{N}(x;z,t)
		&=\sum_{a=1}^{N}V(x_{a};t)
		+\sum_{a=1}^{N}\log{(z-x_{a})}
		-2\sum_{a<b}\log{(x_{a}-x_{b})},\\
		\chi_{N}(q;z,t)
		&=\exp{\left(-\frac{V(q;t)}{2}\right)}
		(z-q)^{-1/2}P_{N}(q),\qquad
		P_{N}(q)=\prod_{a=1}^{N}(q-x_{a}^{*}).
	\end{aligned}
\end{equation}
Here $\mathcal{W}_{\mathrm{NS}}(z,t)=\lim_{\epsilon_{2}\to0}
\epsilon_{2}\log{\mathcal{G}}$ is the NS action carrying the heavy
defect $z$, while the monic polynomial $P_{N}(q)$ fixes the
normalization of the light-defect wave function $\chi_{N}(q;z,t)$.

To pass to the Seiberg--Witten curve, we take the correlated
\emph{Large-$N$ Limit} balancing $V(x;t)$ and the mass parameter
$\Theta=\epsilon_{1}(N-1)$:
\begin{equation}
	\label{equation:PII chapter7 scaled variables}
	\begin{gathered}
		\epsilon_{1}\to0,\qquad
		N=\frac{\Theta}{\epsilon_{1}}+1\to\infty,\\
		q=\epsilon_{1}^{-1/3}Q,\qquad
		z=\epsilon_{1}^{-1/3}Z,\qquad
		t=\epsilon_{1}^{-2/3}T,\qquad
		x_{a}^{*}=\epsilon_{1}^{-1/3}\xi_{a}^{*}.
	\end{gathered}
\end{equation}
The macroscopic coordinates $\Theta,Q,Z,T$ and the fillings are held
fixed on a two-cut saddle branch $C_{+},C_{-}$ away from colliding
branch points, with $Q\neq Z$ outside the cuts and
\begin{equation}
	\label{equation:PII chapter7 leading fillings}
	\epsilon_{1}N_{\pm}\to\nu_{\pm},\qquad
	N_{+}+N_{-}=N,\qquad \nu_{+}+\nu_{-}=\Theta.
\end{equation}
Since $V(x;t)=\epsilon_{1}^{-1}V(\xi;T)$ with
$V(Q;T)=\frac{2}{3}Q^{3}+TQ$, the critical value in these coordinates is
\begin{equation}
	\label{equation:PII chapter7 scaled action}
	\begin{aligned}
		\mathcal{W}_{\mathrm{NS}}-\mathcal{W}_{\mathrm{scale}}
		={}&\sum_{a=1}^{N}V(\xi_{a}^{*};T)-\frac{V(Z;T)}{2}\\
		&+\epsilon_{1}\left[
		\sum_{a=1}^{N}\log{(Z-\xi_{a}^{*})}
		-2\sum_{a<b}\log{(\xi_{a}^{*}-\xi_{b}^{*})}\right],\\
		\mathcal{W}_{\mathrm{scale}}
		={}&\left(\frac{\Theta^{2}}{3\epsilon_{1}}
		-\frac{\epsilon_{1}}{3}\right)\log{\epsilon_{1}},
	\end{aligned}
\end{equation}
where $\mathcal{W}_{\mathrm{scale}}$ is independent of $(Q,Z,T)$ and of
the filling split at fixed $\Theta$.  Up to the coordinate scaling
factor $\epsilon_{1}^{-(N-1/2)/3}$, the light-defect wave function takes
the WKB form
\begin{equation}
	\label{equation:PII chapter7 S definition}
	\begin{aligned}
		\chi_{N}(\epsilon_{1}^{-1/3}Q;z,t)
		&=\epsilon_{1}^{-(N-1/2)/3}
		\exp{\left(\frac{S(Q;Z,T;\epsilon_{1})}{\epsilon_{1}}\right)}
		\sim\exp{\left(\frac{S(Q;Z,T;\epsilon_{1})}{\epsilon_{1}}\right)},\\
		S(Q;Z,T;\epsilon_{1})
		&=-\frac{V(Q;T)}{2}
		+\epsilon_{1}\sum_{a=1}^{N}\log{(Q-\xi_{a}^{*})}
		-\frac{\epsilon_{1}}{2}\log{(Z-Q)}.
	\end{aligned}
\end{equation}
Expanding both actions in powers of $\epsilon_{1}$ (with the
$\mathcal{O}(\epsilon_{1}\log\epsilon_{1})$ subtraction derived in
Appendix~\ref{appendix:PII action normalization}),
\begin{equation}
	\label{equation:PII chapter7 action expansion convention}
	\begin{aligned}
		\mathcal{W}_{\mathrm{NS}}-\mathcal{W}_{\mathrm{scale}}
		-\left(\Theta+\frac{7\epsilon_{1}}{6}\right)\log{\epsilon_{1}}
		&\sim\sum_{j\geq0}\epsilon_{1}^{j-1}\mathcal{W}_{j}(Z,T),\\
		S(Q;Z,T;\epsilon_{1})&\sim\sum_{j\geq0}\epsilon_{1}^{j}S_{j}(Q;Z,T),
	\end{aligned}
\end{equation}
the leading coefficients are
\begin{equation}
	\label{equation:PII chapter7 leading action}
	\mathcal{W}_{0}(T)
	:=\lim_{\epsilon_{1}\to0}\epsilon_{1}
	\left(\mathcal{W}_{\mathrm{NS}}-\mathcal{W}_{\mathrm{scale}}\right),
	\qquad S_{0}(Q;T):=\lim_{\epsilon_{1}\to0}S.
\end{equation}
In $\mathcal{W}_{\mathrm{NS}}$, the bulk Coulomb gas
$\sum_{a}V(\xi_{a}^{*})-2\epsilon_{1}\sum_{a<b}\log{(\xi_{a}^{*}-\xi_{b}^{*})}$
is of order $\epsilon_{1}^{-1}\mathcal{W}_{0}(T)$, whereas the
heavy-defect terms $-\frac{1}{2}V(Z;T)+\epsilon_{1}\sum_{a}\log{(Z-\xi_{a}^{*})}$
enter only at $\mathcal{O}(\epsilon_{1}^{0})$: thus $\mathcal{W}_{0}(T)$ is
a purely closed-string bulk quantity independent of $Z$, while
$\mathcal{W}_{j\geq1}(Z,T)$ combine closed-string pieces with open
integrals bounded by $Z$ (which decouple as $Z\to\infty$).  Conversely,
$S(Q;Z,T;\epsilon_{1})$ is an open WKB integral bounded by the light
defect $Q$: its leading term $S_{0}(Q;T)$ depends only on $Q$, while
the coupling $-\frac{\epsilon_{1}}{2}\log{(Z-Q)}$ and the $Z$-shift of
$\xi_{a}^{*}$ enter from $S_{1}$ onward.

\subsection{Formal expansion and special coordinates}
\label{subsection:PII planar SW proof}

Introduce the resolvent and first moment, with spectral variable
$\zeta$:
\begin{equation}
	\label{equation:PII chapter7 finite resolvent}
	\Omega_{N}(\zeta):=\epsilon_{1}\sum_{a=1}^{N}
	\frac{1}{\zeta-\xi_{a}^{*}},\qquad
	M_{1}:=\epsilon_{1}\sum_{a=1}^{N}\xi_{a}^{*}.
\end{equation}
One can compute the exact finite $N$ resolvent identity implied by the Bethe equations as 
\begin{equation}
\label{eq:exactresolventBethe}
\Omega_N^2 - (2\zeta^2+T) \Omega_N + 2\epsilon_1N\zeta + 2M_1 + \epsilon_1\partial_\zeta \Omega_N 
+ \epsilon_1 \frac{\Omega_N(\zeta) - \Omega_N(Z)}{Z-\zeta} = 0.
\end{equation}

The formal limits $\Omega^{(0)}$ and $M_{1}^{(0)}$ are
determined by the continuum density $\rho$:
\begin{equation}
	\label{equation:PII chapter7 density and fillings}
	\begin{gathered}
		\epsilon_{1}\sum_{a=1}^{N}\delta_{\xi_{a}^{*}}
		\longrightarrow\rho(\zeta)\,\dd\zeta,\qquad
		\int_{C_{\pm}}\rho(\zeta)\,\dd\zeta=\nu_{\pm},\\
		\Omega^{(0)}(\zeta)=\int_{C_{+}\cup C_{-}}
		\frac{\rho(\xi)}{\zeta-\xi}\,\dd\xi,\qquad
		M_{1}^{(0)}=\int_{C_{+}\cup C_{-}}\xi\rho(\xi)\,\dd\xi.
	\end{gathered}
\end{equation}
The leading Bethe equation follows from equation \eqref{eq:exactresolventBethe}
\begin{equation}
	\label{equation:PII chapter7 leading loop}
	\left(\Omega^{(0)}(\zeta)\right)^{2}
	-(2\zeta^{2}+T)\Omega^{(0)}(\zeta)+2\Theta\zeta+2M_{1}^{(0)}=0.
\end{equation}
Consequently, with the sheet fixed at infinity,
\begin{equation}
	\label{equation:PII chapter7 y curve}
	\begin{split}
		&y(\zeta):=\zeta^{2}+\frac{T}{2}-\Omega^{(0)}(\zeta) \, \, = \,\,
		\zeta^{2}+\frac{T}{2}-\frac{\Theta}{\zeta} +O\left(\zeta^{-2}\right).\\
	& y(\zeta)^{2}=\left(\zeta^{2}+\frac{T}{2}\right)^{2}
		-2\Theta\zeta-2M_{1}^{(0)}
    \end{split}        
\end{equation}
The remaining modulus $M_{1}^{(0)}$ is fixed by the filling period. Notice that the insertion coordinate $Z$ has disappeared from the leading curve. Its effect starts showing up from the subleading orders. 

From the equations \eqref{equation:PII chapter7 scaled action} and \eqref{equation:PII chapter7 leading action} let us derive the NS action as a variational functional. 
\begin{equation}
\begin{split}
& \epsilon_1 \sum_{a=1}^N V(\xi_a^*;T) \rightarrow \int_{C_{+}\cup C_{-}} V(\zeta;T) \rho(\zeta) d\zeta ,
\\ & 
-\frac{\epsilon_1 V(Z;T)}{2} + \epsilon_1^2 \sum_{a=1}^N \log(Z-\xi^*_a) 
\sim \,\, O(\epsilon_1) ,
\\ & 
-2\epsilon_1^2 \sum_{a<b} \log(\xi_a^*-\xi_b^*) \rightarrow - \int_{C_{+}\cup C_{-}}\int_{C_{+}\cup C_{-}} \rho(\zeta)\rho(\xi) \log(\zeta-\xi) d\zeta d\xi.
\end{split}
\end{equation}
The bulk coefficient is then the constrained critical value
\begin{equation}
	\label{equation:PII chapter7 planar YY functional}
	\mathcal{W}_{0}
	=\operatorname{crit}_{\int_{C_{\pm}}\rho=\nu_{\pm}}
	\left[
	\int V(\zeta;T)\rho(\zeta)\,\dd\zeta
	-\iint\rho(\zeta)\rho(\xi)\log{(\zeta-\xi)}\,\dd\zeta\,\dd\xi
	\right].
\end{equation}
Both integrals run over the two cuts, with the logarithmic
continuation inherited from the finite saddle.  The insertion term $-\frac{\epsilon_1 V(Z;T)}{2} + \epsilon_1^2 \sum_{a=1}^N \log(Z-\xi^*_a)$
from \eqref{equation:PII chapter7 scaled action} in equation \eqref{equation:PII chapter7 leading action} is subleading,
so $\mathcal{W}_{0}$ is independent of $Z$.
Computing the variation of $\mathcal{W}_0$ in equation \eqref{equation:PII chapter7 planar YY functional} under $\rho \rightarrow \rho + d\rho$, with Lagrangian multipliers $\mu_\pm$, the equilibrium potential and its two constant values are computed as 
\begin{equation}
	\label{equation:PII chapter7 chemical potential}
	\Phi_{\mathrm{eq}}(\zeta):=V(\zeta;T)-2\int\rho(\xi)\log{(\zeta-\xi)}\,\dd\xi,
	\qquad \Phi_{\mathrm{eq}}\big|_{C_{\pm}} :=\mu_{\pm}.
\end{equation}
The constrained variation and the cubic virial identity give
an evaluation of the critical value:
\begin{equation}
	\label{equation:PII chapter7 absolute W0}
	\begin{aligned}
		\mathcal{W}_{0}
		&=\frac{1}{2}\int V(\zeta;T)\rho(\zeta)\,\dd\zeta
		+\frac{\nu_{+}\mu_{+}+\nu_{-}\mu_{-}}{2}\\
		&=\frac{\Theta^{2}}{6}+\frac{T^{3}}{24}
		-\frac{T u_{\mathrm{CB}}}{6}
		+\frac{\nu_{+}\mu_{+}+\nu_{-}\mu_{-}}{2},\qquad
		u_{\mathrm{CB}}:=\frac{T^{2}}{4}-2M_{1}^{(0)}.
	\end{aligned}
\end{equation}
Similarly, the wave action follows directly from
\eqref{equation:PII chapter7 S definition}:
\begin{equation}
	\label{equation:PII chapter7 leading wave}
	\begin{aligned}
		S_{0}(Q)
		&=-\frac{\Phi_{\mathrm{eq}}(Q)}{2}\\
		&=-\frac{Q^{3}}{3}-\frac{TQ}{2}+\Theta\log{Q}
		+\int_{\infty}^{Q}\left(
		\Omega^{(0)}(\zeta)-\frac{\Theta}{\zeta}\right)\dd\zeta,
		\qquad \dd S_{0}=-y(Q)\,\dd Q.
	\end{aligned}
\end{equation}
The primitive is fixed by the monic normalization, with paths
on the chosen sheet.

These results identify the SW data as
\begin{equation}
	\label{equation:PII chapter7 SW data}
	\Sigma:\quad y^{2}=\zeta^{4}+T\zeta^{2}-2\Theta\zeta+u_{\mathrm{CB}},
	\qquad \lambda_{\mathrm{SW}}:=y(\zeta)\,\dd\zeta.
\end{equation}
In the local wave coordinate, the pullback of this differential
at $\zeta=Q$ is $y(Q)\,\dd Q=-\dd S_{0}(Q)$;
the SW periods, by contrast, are integrals on $\Sigma$.
For distinct branch points this is the genus-one curve of the
$H_{1}$ Argyres--Douglas theory \cite{GrassiGu2019} and $u_{\mathrm{CB}}$ is the Coulomb branch parameter .
Let $A$ surround $C_{+}$ counterclockwise on the physical sheet.
Let $B$ run from $C_{-}$ to $C_{+}$ on that sheet and return
on the second sheet, using the continuation in $\Phi_{\mathrm{eq}}$.
Then
\begin{equation}
	\label{equation:PII chapter7 a period}
	a:=\frac{1}{2\pi\mathrm{i}}\oint_{A}\lambda_{\mathrm{SW}}
	=-\nu_{+},\qquad
	a_{D}:=\frac{1}{2\pi\mathrm{i}}\oint_{B}\lambda_{\mathrm{SW}}.
\end{equation}
Since $\dd\Phi_{\mathrm{eq}}=2\lambda_{\mathrm{SW}}$, the on-shell
variations are
\begin{equation}
	\label{equation:PII chapter7 B relation}
	\left.\partial_{\nu_{+}}\mathcal{W}_{0}\right|_{\Theta,T}
	=\mu_{+}-\mu_{-}=\oint_{B}\lambda_{\mathrm{SW}},
	\qquad
	\left.\partial_{T}\mathcal{W}_{0}\right|_{\nu_{+},\Theta}
	=M_{1}^{(0)}=\frac{T^{2}}{8}-\frac{u_{\mathrm{CB}}}{2}.
\end{equation}
For the convention $\partial_{a}\mathcal{F}_{\mathrm{SW}}=a_{D}$,
our action normalization therefore gives
\begin{equation}
	\label{equation:PII chapter7 W0 SW identification}
	\mathcal{W}_{0}(a,T,\Theta)
	=-2\pi\mathrm{i}\,\mathcal{F}_{\mathrm{SW}}(a,T,\Theta)
	+C(T,\Theta),
\end{equation}
where $C(T,\Theta)$ is an integration constant.
The absolute normalization of $\mathcal{W}_{0}$ is fixed by
the critical value in \eqref{equation:PII chapter7 planar YY functional}.

Therefore in this subsection, we realize that both pieces of Seiberg–Witten special geometry emerge directly from the large-$N$ Bethe system: the spectral differential is recovered from the logarithmic derivative of the Bethe polynomial, while its periods and prepotential arise from the filling fractions and the critical Yang–Yang action.

\subsection{Exact finite-\texorpdfstring{$N$}{N} reconstruction}
\label{subsection:PII exact finite reconstruction}

The monic data of Subsection~\ref{subsection:PII chapter7 Large Limit}
also determine the action at finite $N$ through the scaled
determinant of Appendix~\ref{appendix:PII action normalization}.
Write $\widehat{\tau}_{N}(T):=\tau_{N}(\epsilon_{1}^{-2/3}T)$.
Fixing $N,\epsilon_{1},Z,T$, the projective seed $c_{+}/c_{-}$ is
determined by the insertion condition
\begin{equation}
\label{equation:PII exact seed condition}
\epsilon_{1}\left(
\widehat{\tau}_{N}\partial_{T}\widehat{\tau}_{N+1}
-\widehat{\tau}_{N+1}\partial_{T}\widehat{\tau}_{N}\right)
=Z\widehat{\tau}_{N}\widehat{\tau}_{N+1}.
\end{equation}
The two determinants are homogeneous of degrees $N$ and $N+1$ in
$(c_{+},c_{-})$, so this is generically an equation of degree
$2N+1$ for their ratio.  Branches with a vanishing adjacent tau
denominator and the required fillings and logarithms are discarded.

At fixed seed the exact finite-$N$ action is
\begin{equation}
\label{equation:PII exact scaled action}
\begin{aligned}
\mathcal{W}_{\mathrm{NS}}
={}&\frac{\Theta^{2}}{3\epsilon_{1}}-\frac{\epsilon_{1}}{3}
+\frac{Z}{3}\left(\epsilon_{1}\partial_{T}Z-T\right)
-(\Theta+\epsilon_{1})\log{(-2)}\\
&+\epsilon_{1}\left[
(N+1)\log{\widehat{\tau}_{N}}
-N\log{\widehat{\tau}_{N+1}}
-\frac{2T}{3}\partial_{T}\log{\widehat{\tau}_{N}}\right],
\\
Z={}&\epsilon_{1}\partial_{T}
\log{\frac{\widehat{\tau}_{N+1}}{\widehat{\tau}_{N}}},
\qquad
\partial_{T}Z=\epsilon_{1}\partial_{T}^{2}
\log{\frac{\widehat{\tau}_{N+1}}{\widehat{\tau}_{N}}}.
\end{aligned}
\end{equation}
The displayed equation for $Z$ is equivalent to
\eqref{equation:PII exact seed condition}, and the action is
invariant under a common rescaling of the seed.  The monic
polynomial built from the same hierarchy gives the exact
finite-$N$ wave action
\begin{equation}
\label{equation:PII exact wave action}
S=-\frac{V(Q;T)}{2}
+\epsilon_{1}\log{\left[
\frac{\epsilon_{1}^{N/3}
P_{N}(\epsilon_{1}^{-1/3}Q)}{(Z-Q)^{1/2}}\right]}.
\end{equation}
The derivation of \eqref{equation:PII exact scaled action} from the
finite Bethe equations is given in
Appendix~\ref{appendix:PII quotient identity}.  Its adjacent-size
quotient is what fixes the logarithmic normalization of
Subsection~\ref{subsection:PII chapter7 Large Limit}.

\section*{Acknowledgements}
We thank Nicolai Reshetikhin, Mauricio Romo and Yinan Wang for helpful
discussions. BH would also like to thank Tom\'a\v{s} Proch\'azka for inviting him to
Prague during the summer and for discussions that clarified several
concepts underlying this work. He also thanks the Institute of Physics
of the Czech Academy of Sciences (FZU) for its hospitality during the
visit.

\paragraph{AI statement.}
AI-assisted tools were used for algebraic checks, notation auditing,
and LaTeX editing. The authors retain full responsibility for the
mathematical content and final manuscript.

\appendix

\section{Matrix-integral and Schwinger--Dyson identities}
\label{appendix:PII matrix integral checks}

We verify the matrix-integral equations by Schwinger--Dyson identities.

Let $\mathcal{M}$ denote the part of the integrand in
\eqref{equation:PII matrix integral} that excludes the differentials.  Then
\begin{equation}
	\label{equation:PII log derivative matrix integrand}
	\partial_{x_{a}}\log\mathcal{M}
	=
	\frac{2}{\kappa}
	\sum_{b\neq a}\frac{1}{x_{a}-x_{b}}
	+\frac{1}{x_{a}-q}
	+\frac{1}{\kappa(z-x_{a})}
	-\frac{1}{\kappa}(2x_{a}^{2}+t).
\end{equation}
Introduce the resolvents
\begin{equation}
	R(\zeta)=\sum_{a=1}^{N}\frac{1}{\zeta-x_{a}},
	\qquad
	R_{2}(\zeta)=\sum_{a=1}^{N}\frac{1}{(\zeta-x_{a})^{2}}.
\end{equation}
The elementary resolvent identity is
\begin{equation}
	\label{equation:PII resolvent pair identity}
	\sum_{a=1}^{N}
	\sum_{b\neq a}
	\frac{1}{\zeta-x_{a}}
	\frac{1}{x_{a}-x_{b}}
	=
	\frac{1}{2}
	\left(
	R(\zeta)^{2}-R_{2}(\zeta)
	\right).
\end{equation}

Writing $\langle\cdots\rangle$ for the unnormalized period with the
same $1/N!$ prefactor as $\mathcal{I}_{N}$ and the displayed insertion,
integration by parts gives
\begin{equation}
	0
	=
	\int_{\Gamma}
	\prod_{b=1}^{N}\dd x_{b}
	\sum_{a=1}^{N}
	\partial_{x_{a}}
	\left(
	\frac{1}{\zeta-x_{a}}\mathcal{M}
	\right).
\end{equation}
Using \eqref{equation:PII log derivative matrix integrand} and
\eqref{equation:PII resolvent pair identity}, this becomes
\begin{equation}
	\begin{aligned}
		0
		={}&
		\Bigg\langle
		R(\zeta)^{2}
		+(\kappa-1)R_{2}(\zeta)
		+\frac{\kappa}{\zeta-q}
		\left(
		R(\zeta)-R(q)
		\right)
		\\
		&\qquad
		-\frac{1}{\zeta-z}
		\left(
		R(\zeta)-R(z)
		\right)
		-(2\zeta^{2}+t)R(\zeta)
		+2N\zeta
		+2\sum_{a=1}^{N}x_{a}
		\Bigg\rangle.
	\end{aligned}
\end{equation}
The two distinguished limits are therefore
\begin{equation}
	\label{equation:PII z loop equation}
	\begin{aligned}
		0
		={}&
		\Bigg\langle
		R(z)^{2}
		+\kappa R_{2}(z)
		+\frac{\kappa}{z-q}
		\left(
		R(z)-R(q)
		\right)
		\\
		&\qquad
		-(2z^{2}+t)R(z)
		+2Nz
		+2\sum_{a=1}^{N}x_{a}
		\Bigg\rangle,
	\end{aligned}
\end{equation}
and
\begin{equation}
	\label{equation:PII q loop equation}
	\begin{aligned}
		0
		={}&
		\Bigg\langle
		R(q)^{2}
		-R_{2}(q)
		-\frac{1}{q-z}
		\left(
		R(q)-R(z)
		\right)
		\\
		&\qquad
		-(2q^{2}+t)R(q)
		+2Nq
		+2\sum_{a=1}^{N}x_{a}
		\Bigg\rangle.
	\end{aligned}
\end{equation}

Differentiation of the matrix integral gives
\begin{equation}
	\label{equation:PII matrix integral derivative identities}
	\kappa\partial_{z}\mathcal{I}_{N}
	=-\langle R(z)\rangle,
	\qquad
	\partial_{q}\mathcal{I}_{N}
	=\langle R(q)\rangle,
\end{equation}
\begin{equation}
	\kappa^{2}\partial_{z}^{2}\mathcal{I}_{N}
	=
	\left\langle
	R(z)^{2}+\kappa R_{2}(z)
	\right\rangle,
	\qquad
	\partial_{q}^{2}\mathcal{I}_{N}
	=
	\left\langle
	R(q)^{2}-R_{2}(q)
	\right\rangle,
\end{equation}
and
\begin{equation}
	\left\langle
	\sum_{a=1}^{N}x_{a}
	\right\rangle
	=-\kappa\partial_{t}\mathcal{I}_{N}.
\end{equation}

\section{Derivation of quantum Langlands duality}
\label{appendix:PII finite coupling verification}
\label{appendix:PII Wakimoto computations}

This appendix derives the dictionary of
Subsection~\ref{subsection:PII finite coupling dictionary}.
The dual Wakimoto modes, affine connections and DS reduction are
followed by the BPZ covariance and affine reconstruction.  The dual
screening integral and its loop equations are treated together,
before the neutrality and semiclassical calculations fix the charge
and limiting-action correspondences.

\subsection{Dual affine construction and DS reduction}
\label{subsection:PII dual affine construction}

We construct the dual affine system underlying the dictionary in
Subsection~\ref{subsection:PII finite coupling dictionary}.  For
$\slTwo$, the dual level is $\kappa^{\vee}=\kappa^{-1}$.
Place an unflowed spin-$\frac{1}{2}$ module at $z^{\vee}$ and
take the Whittaker character
\begin{equation}
	\psi^{\vee}(x)=u^{\vee}(x-q^{\vee}).
\end{equation}
Its one-form has a simple zero at $q^{\vee}$ and a third-order
pole at infinity.  We derive the affine connections before their
Virasoro reduction.

\paragraph*{The irregular module.}
In this subsection, $\beta,\gamma,a$ and $e,h,f$ denote the
oscillators and currents of the dual representation.  Their
commutators and Wakimoto formulas are those of Section~\ref{subsection:PII original affine construction} with
$\kappa$ replaced by $\kappa^{\vee}$; the affine level is
$\kappa^{\vee}-2$.  Take a Fock reference vector
$|2\theta^{\vee}\rangle$ annihilated by
$\beta_{n\geq1}$, $\gamma_{n\geq0}$ and $a_{n\geq1}$, with
$a_{0}=2\theta^{\vee}$.  In its formal completion set
\begin{equation}
	|W^{\vee}\rangle
	=\exp\left[
	u^{\vee}q^{\vee}\gamma_{-1}-u^{\vee}\gamma_{-2}
	-\frac{t^{\vee}}{2\kappa^{\vee}}a_{-1}
	-\frac{1}{3\kappa^{\vee}}a_{-3}
	\right]|2\theta^{\vee}\rangle.
\end{equation}
The oscillator commutators give
\begin{equation}
	\begin{gathered}
		\beta_{1}=u^{\vee}q^{\vee},\qquad
		\beta_{2}=-u^{\vee},\qquad \beta_{n\geq3}=0,
		\qquad \gamma_{n\geq0}=0,\\
		a_{0}=2\theta^{\vee},\qquad a_{1}=-t^{\vee},\qquad
		a_{2}=0,\qquad a_{3}=-2,\qquad a_{n\geq4}=0
	\end{gathered}
\end{equation}
on the cyclic vector.  The contragredient pairing defines the
corresponding irregular covector.  Inverting $w=1/x$ sends the
prescribed raising-current part to
$-x^{-2}(u^{\vee}q^{\vee}x^{2}-u^{\vee}x^{3})=\psi^{\vee}(x)$.

\paragraph*{Current modes and descendants.}\phantomsection
\label{appendix:PII dual Wakimoto modes}
The full affine action, including all descendant modes, is
\begin{equation}
	\label{equation:PII dual complete current modes}
	\begin{aligned}
		e_{n}&=\beta_{n},\\
		h_{n}&=a_{n}-2\sum_{r\in\mathbb{Z}}:\gamma_{r}\beta_{n-r}:,\\
		f_{n}&=-\sum_{r,s\in\mathbb{Z}}:\gamma_{r}\gamma_{s}\beta_{n-r-s}:
		+\sum_{r\in\mathbb{Z}}\gamma_{r}a_{n-r}
		-(\kappa^{\vee}-2)n\gamma_{n}.
	\end{aligned}
\end{equation}
In the zero-mode polarization
$\gamma_{0}\beta_{0}|W^{\vee}\rangle=-|W^{\vee}\rangle$,
the nonnegative current modes act as
\begin{equation}
	\begin{aligned}
		e_{1}&=u^{\vee}q^{\vee},&
		e_{2}&=-u^{\vee},&
		e_{n\geq3}&=0,\\
		h_{1}&=-t^{\vee}+2u^{\vee}\gamma_{-1},&
		h_{2}&=0,&
		h_{3}&=-2,\\
		f_{1}&=-2\gamma_{-2},&
		f_{2}&=-2\gamma_{-1},&
		f_{n\geq3}&=0,\\
		h_{0}&=2\theta^{\vee}+2
		-2u^{\vee}q^{\vee}\gamma_{-1}+2u^{\vee}\gamma_{-2},\\
		f_{0}&=u^{\vee}\gamma_{-1}^{2}
		-t^{\vee}\gamma_{-1}-2\gamma_{-3},
		\qquad h_{n\geq4}=0.
	\end{aligned}
\end{equation}
Here and below such mode identities are understood on the cyclic
vector.  Thus $e_{1},e_{2},h_{3}$ have prescribed scalar eigenvalues,
whereas $e_{0},h_{1},f_{0},f_{1},f_{2}$ and the unfixed negative
modes generate descendants.  Their action is retained through
\eqref{equation:PII dual complete current modes}.

The coherent family also fixes the insertion identities:
\begin{equation}
	\label{equation:PII dual horizontal family}
	\begin{aligned}
		\partial_{q^{\vee}}|W^{\vee}\rangle
		&=u^{\vee}\gamma_{-1}|W^{\vee}\rangle,\\
		\partial_{u^{\vee}}|W^{\vee}\rangle
		&=(q^{\vee}\gamma_{-1}-\gamma_{-2})|W^{\vee}\rangle,\\
		\partial_{t^{\vee}}|W^{\vee}\rangle
		&=-\frac{a_{-1}}{2\kappa^{\vee}}|W^{\vee}\rangle.
	\end{aligned}
\end{equation}

As in Section~\ref{subsection:PII DS local modules}, the singular part of the local twist shifts the reduced Heisenberg field by
$a^{\mathrm{sf}}(w)=a(w)+3\kappa^{\vee}/w$.  The resulting infinity stress tensor is
\begin{equation}
	\label{equation:PII dual stress separation}
	\begin{aligned}
		T_{\mathrm{red}}^{\vee}[a^{\mathrm{sf}}]
		&=T_{\mathrm{irr}}^{\vee}+T_{\mathrm{sf}}^{\vee},\\
		T_{\mathrm{irr}}^{\vee}
		&=\frac{1}{4\kappa^{\vee}}:a^{2}:
		+\frac{\kappa^{\vee}-1}{2\kappa^{\vee}}\partial a,\\
		T_{\mathrm{sf}}^{\vee}
		&=\frac{3}{2w}a(w)+\frac{3\kappa^{\vee}+6}{4w^{2}}.
	\end{aligned}
\end{equation}
This is the dual counterpart of the twist used in the original affine theory; the regular part is absorbed into the scalar normalization.  In particular, the singular shift fixes the $L_{3}^{(\infty)}$ coefficient used below.

\paragraph*{Sugawara irregular KZ and quantum isomonodromy connections.}
Evaluate the affine conformal block on the two spin basis vectors at
$z^{\vee}$ to obtain $\cF^{\vee}$.  Moving the Sugawara translation
contour to infinity gives
\begin{equation}
	\kappa^{\vee}\partial_{z^{\vee}}\cF^{\vee}
	=-\sum_{n\geq1}(z^{\vee})^{n-1}
	\left(\frac{h_{n}}{2}h^{T}+e_{n}f^{T}+f_{n}e^{T}\right)\cF^{\vee}.
\end{equation}
The explicit dual KZ equation is therefore
\begin{equation}
	\label{equation:PII dual affine KZ}
	\kappa^{\vee}\partial_{z^{\vee}}\cF^{\vee}
	=
	\begin{pmatrix}
		(z^{\vee})^{2}+\frac{t^{\vee}}{2}-\partial_{q^{\vee}}
		&u^{\vee}(z^{\vee}-q^{\vee})\\[2mm]
		\frac{2(z^{\vee}+q^{\vee})}{u^{\vee}}\partial_{q^{\vee}}
		-2\partial_{u^{\vee}}
		&-(z^{\vee})^{2}-\frac{t^{\vee}}{2}+\partial_{q^{\vee}}
	\end{pmatrix}\cF^{\vee}.
\end{equation}
Its components are the finite-spin evaluations of the affine block.
The independent time generator also has an affine expression:
\begin{equation}
	a_{-1}=h_{-1}-f_{2}e_{0}-u^{\vee}q^{\vee}f_{1}
	+u^{\vee}f_{0}
	-\frac{(u^{\vee})^{2}}{4}f_{2}^{2}
	-\frac{u^{\vee}t^{\vee}}{2}f_{2}.
\end{equation}
Using the finite-point Ward identities in
\eqref{equation:PII dual horizontal family} yields
\begin{equation}
	\label{equation:PII dual affine time}
	\begin{aligned}
		2\kappa^{\vee}\partial_{t^{\vee}}\cF^{\vee}
		=\Bigg\{&
		\left[\partial_{q^{\vee}}^{2}
		-(2(q^{\vee})^{2}+t^{\vee})\partial_{q^{\vee}}
		+2q^{\vee}u^{\vee}\partial_{u^{\vee}}\right]\mathbf{1}\\
		&+\begin{pmatrix}
			z^{\vee}&u^{\vee}\\[1mm]
			\frac{2}{u^{\vee}}\partial_{q^{\vee}}&-z^{\vee}
		\end{pmatrix}\Bigg\}\cF^{\vee}.
	\end{aligned}
\end{equation}
The quadratic derivative in the irregular coordinate is the image
of $f_{2}^{2}$.  The irregular KZ and quantum isomonodromy equations are first order in
their respective deformation variables.  Their differential
operators commute before imposing the Cartan constraint
\begin{equation}
	\left[(2\theta^{\vee}+2-2u^{\vee}\partial_{u^{\vee}})\mathbf{1}
	+h^{T}\right]\cF^{\vee}=0.
\end{equation}

\paragraph*{Reduction and the two dual BPZ equations.}
The scalar normalization is
\begin{equation}
	\begin{gathered}
		\cF_{1}^{\vee}
		=\sqrt{u^{\vee}(z^{\vee}-q^{\vee})}\,
		Z_{\mathrm{irr}}^{\vee}(q^{\vee})\mathcal{G}^{\vee},\\
		Z_{\mathrm{irr}}^{\vee}(q^{\vee})
		=\exp\left(\frac{(q^{\vee})^{3}}{3}
		+\frac{t^{\vee}q^{\vee}}{2}\right),\qquad
		u^{\vee}\partial_{u^{\vee}}\mathcal{G}^{\vee}
		=(\theta^{\vee}+1)\mathcal{G}^{\vee}.
	\end{gathered}
\end{equation}
At the zero $q^{\vee}$ the reduced local Heisenberg charge is
$-\kappa^{\vee}$; the surviving spin component at $z^{\vee}$ has
charge $1$.  Inserting these charges in the reduced mode formula
of Section~\ref{subsection:PII DS local modules} gives
\begin{equation}
	\begin{aligned}
		\Delta_{q^{\vee}}^{\vee}&=\frac{3\kappa^{\vee}-2}{4},&
		(L_{-1}^{2}-\kappa^{\vee}L_{-2})|q^{\vee}\rangle&=0,\\
		\Delta_{z^{\vee}}^{\vee}&=\frac{3(\kappa^{\vee})^{-1}-2}{4},&
		(L_{-1}^{2}-(\kappa^{\vee})^{-1}L_{-2})|z^{\vee}\rangle&=0.
	\end{aligned}
\end{equation}
The infinity tensor \eqref{equation:PII dual stress separation}
gives the complete positive-mode data relevant to contour deformation:
\begin{equation}
	\begin{gathered}
		\kappa^{\vee}L_{6}^{(\infty)}=1,\qquad
		\kappa^{\vee}L_{5}^{(\infty)}=0,\qquad
		\kappa^{\vee}L_{4}^{(\infty)}=t^{\vee},\\
		\kappa^{\vee}L_{3}^{(\infty)}
		=-2\theta^{\vee}+\kappa^{\vee}-4,\qquad
		\kappa^{\vee}L_{2}^{(\infty)}
		=-a_{-1}+\frac{(t^{\vee})^{2}}{4},\qquad
		L_{n>6}^{(\infty)}=0.
	\end{gathered}
\end{equation}
In particular the singular flow contributes $-3\kappa^{\vee}$ to
$\kappa^{\vee}L_{3}^{(\infty)}$ and leaves the displayed
$L_{2},L_{4},L_{5},L_{6}$ coefficients unchanged.  Horizontality
represents $a_{-1}$ by $-2\kappa^{\vee}\partial_{t^{\vee}}$ in the
reduced pairing.  The resulting polynomial is
\begin{equation}
	\begin{aligned}
		P_{\kappa^{\vee}}(x)
		={}&2\partial_{t^{\vee}}+\frac{(t^{\vee})^{2}}{4\kappa^{\vee}}
		+\frac{x}{\kappa^{\vee}}
		(-2u^{\vee}\partial_{u^{\vee}}+\kappa^{\vee}-2)\\
		&+\frac{t^{\vee}}{\kappa^{\vee}}x^{2}
		+\frac{x^{4}}{\kappa^{\vee}}.
	\end{aligned}
\end{equation}
The powers $x^{0},\ldots,x^{4}$ correspond respectively to
$L_{2},\ldots,L_{6}$; the cubic coefficient vanishes.
The two local null vectors now give
\begin{equation}
	\label{equation:PII independently derived dual BPZ}
	\begin{aligned}
		\mathcal{E}_{q^{\vee}}^{\vee}
		&=\partial_{q^{\vee}}^{2}-\kappa^{\vee}
		\left[P_{\kappa^{\vee}}(q^{\vee})
		-\frac{\partial_{z^{\vee}}}{z^{\vee}-q^{\vee}}
		+\frac{\Delta_{z^{\vee}}^{\vee}}{(z^{\vee}-q^{\vee})^{2}}\right],\\
		\mathcal{E}_{z^{\vee}}^{\vee}
		&=\partial_{z^{\vee}}^{2}-(\kappa^{\vee})^{-1}
		\left[P_{\kappa^{\vee}}(z^{\vee})
		+\frac{\partial_{q^{\vee}}}{z^{\vee}-q^{\vee}}
		+\frac{\Delta_{q^{\vee}}^{\vee}}{(z^{\vee}-q^{\vee})^{2}}\right],\\
		\mathcal{E}_{q^{\vee}}^{\vee}\mathcal{G}^{\vee}
		&=\mathcal{E}_{z^{\vee}}^{\vee}\mathcal{G}^{\vee}=0.
	\end{aligned}
\end{equation}
Eliminating the lower spin component in
\eqref{equation:PII dual affine KZ} gives
$\mathcal{K}_{\mathrm{II}}^{\vee}
=(\kappa^{\vee})^{2}\mathcal{E}_{z^{\vee}}^{\vee}
-\mathcal{E}_{q^{\vee}}^{\vee}$.
Reducing the independent quantum isomonodromy connection gives
$\mathcal{E}_{q^{\vee}}^{\vee}\mathcal{G}^{\vee}=0$.
Thus the two BPZ equations follow from the constructed affine
module and its irregular KZ and quantum isomonodromy connections.

\subsection{BPZ covariance at the dual level}
\label{subsection:PII finite coupling affine exchange}

The original and dual affine modules produce the two families of
degenerate Virasoro fields.  Their exchange is visible in the
original null-vector data:
\begin{equation}
	\begin{aligned}
		&(L_{-1}^{2}-\kappa L_{-2})|q\rangle=0,
		&&\Delta_{q}=\frac{3\kappa-2}{4},\\
		&(L_{-1}^{2}-\kappa^{-1}L_{-2})|z\rangle=0,
		&&\Delta_{z}=\frac{3\kappa^{-1}-2}{4}.
	\end{aligned}
\end{equation}
The reduced Virasoro data give
\begin{equation}
	c=13-6\left(\kappa+\frac{1}{\kappa}\right),
	\qquad
	\Delta_{q^{\vee}}^{\vee}=\Delta_{z},
	\qquad \Delta_{z^{\vee}}^{\vee}=\Delta_{q}.
\end{equation}
The central charge is invariant under coupling inversion, while the
two null-vector coefficients and conformal weights are exchanged.

The representation-theoretic origin of the exchange is visible in the
reduced Heisenberg tensor.  In a common coordinate,
the identification
\begin{equation}
	\label{equation:PII Langlands reduced Heisenberg exchange}
	a^{\vee}=-\frac{a}{\kappa},
	\qquad
	T_{\mathrm{red}}^{\vee}[a^{\vee}]=T_{\mathrm{red}}[a]
\end{equation}
rescales the Heisenberg commutator to level $\kappa^{-1}$ and
preserves the reduced Virasoro field.  It sends the charge $1$ of
the fundamental module to $-\kappa^{-1}$, and the charge $-\kappa$
at the Whittaker zero to $1$.  Thus finite-weight data and the
coweight specifying the DS twist exchange their roles.  This is the
weight--coweight interchange of twisted DS modules in
\cite{ArakawaFrenkel2019}.
The irregular Ward polynomial fixes the coordinate rescaling and
the charge at infinity in the calculation below.

To extend this exchange to the differential equations, write the
single-insertion operators of Section~\ref{section:PII affine BPZ} in full:
\begin{equation}
	\begin{aligned}
		\mathcal{E}_{q}
		&=\partial_{q}^{2}-\kappa\left[
		P_{\kappa}(q)-\frac{1}{z-q}\partial_{z}
		+\frac{\Delta_{z}}{(z-q)^{2}}\right],\\
		\mathcal{E}_{z}
		&=\partial_{z}^{2}-\kappa^{-1}\left[
		P_{\kappa}(z)+\frac{1}{z-q}\partial_{q}
		+\frac{\Delta_{q}}{(z-q)^{2}}\right],\\
		P_{\kappa}(x)
		&=2\partial_{t}+\frac{1}{\kappa}
		\left(\frac{t^{2}}{4}+tx^{2}+x^{4}\right)
		+\frac{x}{\kappa}(-2u\partial_{u}+\kappa-2).
	\end{aligned}
\end{equation}
Here $P_{\kappa}$ is the infinity Ward polynomial
\eqref{equation:irregular Ward polynomial}.  Its quartic term fixes
$a^{3}=\kappa^{-1}$; the remaining terms determine the charge map.

Under the coordinate map
\eqref{equation:PII Langlands finite coordinate map}, the chain rule is
\begin{equation}
	\partial_{q^{\vee}}=a^{-1}\partial_{z},\qquad
	\partial_{z^{\vee}}=a^{-1}\partial_{q},\qquad
	\partial_{t^{\vee}}=a^{-2}\partial_{t},\qquad
	z^{\vee}-q^{\vee}=-a(z-q).
\end{equation}
The transformed block
\eqref{equation:PII Langlands transformed block} satisfies
\begin{equation}
	\begin{aligned}
		u^{\vee}\partial_{u^{\vee}}\mathcal{G}^{\vee}
		={}&(u^{\vee})^{\frac{3}{2}(\kappa^{-1}-1)}
		\left[\frac{1}{\kappa}u\partial_{u}
		+\frac{3}{2}(\kappa^{-1}-1)\right]\\
		&\times\mathcal{G}\!\left(
		\frac{z^{\vee}}{a},\frac{q^{\vee}}{a},
		\frac{t^{\vee}}{a^{2}},(u^{\vee})^{1/\kappa}\right).
	\end{aligned}
\end{equation}
On the scalar Cartan sector
$u\partial_{u}\mathcal{G}=(\theta+1)\mathcal{G}$, the degree on the
right-hand side is
\begin{equation}
	\frac{\theta+1}{\kappa}
	+\frac{3}{2}(\kappa^{-1}-1)
	=\frac{\theta+\frac{5}{2}}{\kappa}-\frac{3}{2}
	=\theta^{\vee}+1.
\end{equation}
This gives the charge relation in
\eqref{equation:PII Langlands finite coordinate map}.

It remains to verify the infinity Ward polynomial.  Its dual linear
term becomes
\begin{equation}
	\begin{aligned}
		\frac{ax}{\kappa^{\vee}}
		(-2u^{\vee}\partial_{u^{\vee}}+\kappa^{\vee}-2)
		&=a\kappa x\left[
		-\frac{2}{\kappa}\left(u\partial_{u}+\frac{3}{2}\right)
		+3+\frac{1}{\kappa}-2\right]\\
		&=ax(-2u\partial_{u}+\kappa-2).
	\end{aligned}
\end{equation}
Because $a^{3}=\kappa^{-1}$, one has
$\kappa a^{4}=a=a^{-2}/\kappa$.  Hence
\begin{equation}
	\label{equation:PII Langlands Ward covariance}
	\begin{aligned}
		P_{\kappa^{\vee}}(ax)
		&=2a^{-2}\partial_{t}
		+\kappa a^{4}\left(\frac{t^{2}}{4}+tx^{2}+x^{4}\right)
		+ax(-2u\partial_{u}+\kappa-2)\\
		&=a^{-2}P_{\kappa}(x),
	\end{aligned}
\end{equation}
The first-order pole terms transform as
\begin{equation}
	-\frac{1}{z^{\vee}-q^{\vee}}\partial_{z^{\vee}}
	=\frac{a^{-2}}{z-q}\partial_{q},
	\qquad
	\frac{1}{z^{\vee}-q^{\vee}}\partial_{q^{\vee}}
	=-\frac{a^{-2}}{z-q}\partial_{z}.
\end{equation}
Combining these identities with the exchanged conformal weights gives
\begin{equation}
	\begin{aligned}
		\mathcal{E}_{q^{\vee}}^{\vee}\mathcal{G}^{\vee}
		&=a^{-2}(u^{\vee})^{\frac{3}{2}(\kappa^{-1}-1)}
		(\mathcal{E}_{z}\mathcal{G})\!\left(
		\frac{z^{\vee}}{a},\frac{q^{\vee}}{a},
		\frac{t^{\vee}}{a^{2}},(u^{\vee})^{1/\kappa}\right),\\
		\mathcal{E}_{z^{\vee}}^{\vee}\mathcal{G}^{\vee}
		&=a^{-2}(u^{\vee})^{\frac{3}{2}(\kappa^{-1}-1)}
		(\mathcal{E}_{q}\mathcal{G})\!\left(
		\frac{z^{\vee}}{a},\frac{q^{\vee}}{a},
		\frac{t^{\vee}}{a^{2}},(u^{\vee})^{1/\kappa}\right).
	\end{aligned}
\end{equation}
This proves \eqref{equation:PII Langlands BPZ dictionary}.
The scalar KZ relation
$\mathcal{K}_{\mathrm{II}}=\kappa^{2}\mathcal{E}_{z}-\mathcal{E}_{q}$
then transforms as
\begin{equation}
	\mathcal{K}_{\mathrm{II}}^{\vee}
	\quad\longleftrightarrow\quad
	a^{-2}\left(\kappa^{-2}\mathcal{E}_{q}-\mathcal{E}_{z}\right)
	=-a^{-2}\kappa^{-2}\mathcal{K}_{\mathrm{II}}.
\end{equation}

\paragraph*{Higher-order zeros and higher spins.}
The same calculation exchanges the reduced conformal weights for
higher spins and higher-order zeros.  A zero of order $\ell$ contributes charge
$-\ell\kappa$, while an unflowed spin-$\ell/2$ module contributes
charge $\ell$.  Their reduced weights are
\begin{equation}
	\Delta_{\mathrm{zero}}^{(\ell)}
	=\frac{\kappa\ell(\ell+2)}{4}-\frac{\ell}{2},
	\qquad
	\Delta_{\mathrm{spin}}^{(\ell)}
	=\frac{\ell(\ell+2)}{4\kappa}-\frac{\ell}{2}.
\end{equation}
They are exchanged by $\kappa\mapsto\kappa^{-1}$, giving the
degenerate representations for higher-order zeros and
higher spins.  The simple-zero and spin-$\frac{1}{2}$ construction
is the case $\ell=1$.

\subsection{Affine reconstruction and charge matching}
\label{appendix:PII affine reconstruction verification}

The corresponding affine map is obtained by the scalar-extraction
and reconstruction operators.  In the normalization of
Section~\ref{section:PII affine BPZ}, they act on a joint affine
solution $\mathcal{F}$ and on a reduced solution $\mathcal{G}$ as
\begin{equation}
	\begin{aligned}
		\mathscr{D}_{\kappa}\mathcal{F}
		&:=Z_{\mathrm{irr}}(q)^{-1}
		[u(z-q)]^{-\frac{1}{2}}\mathcal{F}_{1},\\
		(\mathscr{L}_{\kappa}\mathcal{G})_{1}
		&:=\sqrt{u(z-q)}\,Z_{\mathrm{irr}}(q)\mathcal{G},\\
		(\mathscr{L}_{\kappa}\mathcal{G})_{2}
		&:=\frac{Z_{\mathrm{irr}}(q)}{\sqrt{u(z-q)}}
		\left[\kappa\partial_{z}+\partial_{q}
		-(z^{2}-q^{2})+\frac{\kappa-1}{2(z-q)}\right]
		\mathcal{G}.
	\end{aligned}
\end{equation}

The first KZ row fixes the lower component displayed above.
Substitution in the second row gives
$\mathcal{K}_{\mathrm{II}}\mathcal{G}=0$, while substitution in the
upper time row gives $\mathcal{E}_{q}\mathcal{G}=0$.  Flatness fixes
the remaining time row: its possible residual has zero upper component,
and the upper KZ row multiplies its lower component by
$u(z-q)\neq0$.  Thus $\mathscr{D}_{\kappa}$ and
$\mathscr{L}_{\kappa}$ are inverse on the joint solution space with
the Cartan homogeneity of Section~\ref{section:PII affine BPZ}.

The dual maps follow by replacing every parameter with its dual value.
Consequently
\begin{equation}
	\cT_{\mathrm{aff}}
	=\mathscr{L}_{\kappa^{\vee}}\,
	\cT_{\mathrm{BPZ}}\,\mathscr{D}_{\kappa},
	\qquad
	\mathscr{D}_{\kappa^{\vee}}\cT_{\mathrm{aff}}
	=\cT_{\mathrm{BPZ}}\mathscr{D}_{\kappa},
\end{equation}
which verifies the affine intertwiner
\eqref{equation:PII Langlands affine intertwiner}.

The map relates the conformal blocks satisfying both the irregular KZ
and quantum isomonodromy equations:
\begin{equation}
	\begin{array}{ccc}
		(\mathrm{KZ}_{\kappa},\nabla_{t})
		&\xrightarrow{\ \mathscr{D}_{\kappa}\ }&
		\mathrm{BPZ}_{\kappa}\\
		{\scriptstyle\cT_{\mathrm{aff}}}\downarrow
		&&\downarrow{\scriptstyle\cT_{\mathrm{BPZ}}}\\
		(\mathrm{KZ}_{\kappa^{\vee}}^{\vee},\nabla_{t^{\vee}}^{\vee})
		&\xrightarrow{\ \mathscr{D}_{\kappa^{\vee}}\ }&
		\mathrm{BPZ}_{\kappa^{\vee}}^{\vee}.
	\end{array}
\end{equation}
Reducing either affine theory gives its BPZ equations, and the
duality transformation commutes with this reduction.  Equation
\eqref{equation:PII Langlands affine intertwiner} relates the affine
conformal blocks at the dual levels.

\subsection{Dual screening integrals and neutrality}
\label{subsection:PII dual screening representation}

\paragraph*{A dual matrix integral representation.}
The dual module has a screening realization with
$N^{\vee}=\theta^{\vee}+3/2$.  For $N^{\vee}\in\mathbb{Z}_{\geq0}$ set
$V^{\vee}(x)=2x^{3}/3+t^{\vee}x$ and define
\begin{equation}
	\label{equation:PII dual screening integral}
	\begin{aligned}
		\mathcal{I}_{N^{\vee}}^{\vee}
		={}&\frac{1}{N^{\vee}!}\int_{\Gamma^{\vee}}
		\prod_{a=1}^{N^{\vee}}\dd x_{a}\,
		\prod_{a<b}(x_{a}-x_{b})^{2/\kappa^{\vee}}\\
		&\times\prod_{a=1}^{N^{\vee}}
		(x_{a}-q^{\vee})(z^{\vee}-x_{a})^{-1/\kappa^{\vee}}
		\exp\left[-\frac{1}{\kappa^{\vee}}
		\sum_{a=1}^{N^{\vee}}V^{\vee}(x_{a})\right].
	\end{aligned}
\end{equation}
The branches and horizontally transported rapid-decay cycle obey
the conditions of Section~\ref{subsection:PII screening integral}.
The loop equations and their BPZ conjugation are derived below.

The corresponding KZ solution column is
\begin{equation}
	\label{equation:PII dual screening affine column}
	\begin{aligned}
		\cF_{1}^{\vee}
		&=(u^{\vee})^{N^{\vee}}
		\exp\left[\frac{V^{\vee}(z^{\vee})}{2\kappa^{\vee}}\right]
		\mathcal{I}_{N^{\vee}}^{\vee},\\
		\cF_{2}^{\vee}
		&=(u^{\vee})^{N^{\vee}-1}
		\exp\left[\frac{V^{\vee}(z^{\vee})}{2\kappa^{\vee}}\right]
		\left\langle\sum_{a=1}^{N^{\vee}}
		\frac{1}{(q^{\vee}-x_{a})(z^{\vee}-x_{a})}
		\right\rangle_{\vee}.
	\end{aligned}
\end{equation}
The second line equals the same exponential and $u^{\vee}$ power
times
$(\kappa^{\vee}\partial_{z^{\vee}}+\partial_{q^{\vee}})
\mathcal{I}_{N^{\vee}}^{\vee}/(z^{\vee}-q^{\vee})$.
Together with \eqref{equation:PII dual two SD equations}, this
verifies both affine equations
\eqref{equation:PII dual affine KZ} and
\eqref{equation:PII dual affine time}, by the reconstruction used in
Section~\ref{subsection:PII integral reconstruction}.  Thus the dual period realizes the affine system and
its two scalar reductions.

\paragraph*{Loop equations and BPZ covariance.}
For the dual integral \eqref{equation:PII dual screening integral},
the resolvent identity \eqref{equation:PII resolvent pair identity}
and the same integration-by-parts argument give
the two distinguished-point identities
\begin{equation}
	\label{equation:PII dual two SD equations}
	\begin{aligned}
		0=\Bigg[&
		\partial_{q^{\vee}}^{2}
		+\frac{\kappa^{\vee}}{z^{\vee}-q^{\vee}}\partial_{z^{\vee}}
		+\left(\frac{1}{z^{\vee}-q^{\vee}}
		-2(q^{\vee})^{2}-t^{\vee}\right)\partial_{q^{\vee}}\\
		&-2\kappa^{\vee}\partial_{t^{\vee}}
		+2N^{\vee}q^{\vee}\Bigg]\mathcal{I}_{N^{\vee}}^{\vee},\\
		0=\Bigg[&
		(\kappa^{\vee})^{2}\partial_{z^{\vee}}^{2}
		+\left(\kappa^{\vee}(2(z^{\vee})^{2}+t^{\vee})
		-\frac{(\kappa^{\vee})^{2}}{z^{\vee}-q^{\vee}}\right)
		\partial_{z^{\vee}}
		-\frac{\kappa^{\vee}}{z^{\vee}-q^{\vee}}\partial_{q^{\vee}}\\
		&-2\kappa^{\vee}\partial_{t^{\vee}}
		+2N^{\vee}z^{\vee}\Bigg]\mathcal{I}_{N^{\vee}}^{\vee}.
	\end{aligned}
\end{equation}
The scalar dressing is
\begin{equation}
	\mathcal{G}^{\vee}
	=(u^{\vee})^{N^{\vee}-1/2}(z^{\vee}-q^{\vee})^{-\frac{1}{2}}
	\exp\left[
	\frac{V^{\vee}(z^{\vee})}{2\kappa^{\vee}}
	-\frac{V^{\vee}(q^{\vee})}{2}\right]
	\mathcal{I}_{N^{\vee}}^{\vee}.
\end{equation}
Conjugating the two Schwinger--Dyson operators by this factor gives
$\mathcal{E}_{q^{\vee}}^{\vee}$ and
$(\kappa^{\vee})^{2}\mathcal{E}_{z^{\vee}}^{\vee}$ of
\eqref{equation:PII independently derived dual BPZ}.  The first derivatives in the coordinate of each degenerate field
cancel; the remaining simple poles, both
conformal-weight terms and the time derivative agree with the
module calculation.  The Cartan coefficient also agrees, since
\begin{equation}
	u^{\vee}\partial_{u^{\vee}}\mathcal{G}^{\vee}
	=\left(N^{\vee}-\frac{1}{2}\right)\mathcal{G}^{\vee},
	\qquad
	-2\left(N^{\vee}-\frac{1}{2}\right)+\kappa^{\vee}-2
	=-2N^{\vee}+\kappa^{\vee}-1.
\end{equation}

\paragraph*{Screening charges and the two limits.}

The dual levels also exchange the two reduced
Heisenberg screenings.  On a highest-weight state their conformal
weight is
\begin{equation}
	\Delta=\frac{a_{0}\bigl(a_{0}-2(\kappa-1)\bigr)}{4\kappa}.
\end{equation}
The two dimension-one screening charges are $a_{0}=-2$ and
$a_{0}=2\kappa$.  Their pairwise Coulomb exponents are respectively
$2/\kappa$ and $2\kappa$, and
\eqref{equation:PII Langlands reduced Heisenberg exchange}
interchanges the two charge families across the dual levels.
This gives an intrinsic origin for the Vandermonde parameters
\begin{equation}
	\beta=\frac{2}{\kappa},\qquad
	\beta^{\vee}=\frac{2}{\kappa^{\vee}}=2\kappa
	=\frac{4}{\beta}.
\end{equation}
This is the parameter inversion appearing in matrix-ensemble
dualities \cite{Desrosiers2009}.  Here the cubic potential and the
rapid-decay cycles determine the particular solutions of the two
affine systems.

The change in the screening number can be derived directly from the
Cartan power of the scalar block.  Neutrality gives
$N=\theta+3/2$, so $\mathcal{G}$ is homogeneous in $u$ of degree
$N-1/2$.  Equation~\eqref{equation:PII Langlands transformed block}
makes its dual degree
\begin{equation}
	\frac{N-\frac{1}{2}}{\kappa}
	+\frac{3}{2}(\kappa^{-1}-1)
	=\frac{N+1}{\kappa}-\frac{3}{2}.
\end{equation}
Equating this to $N^{\vee}-1/2$ gives
\eqref{equation:PII Langlands dual screening number}.
The same formula follows by substituting
$N=\theta+3/2$ and $N^{\vee}=\theta^{\vee}+3/2$ in the charge
transformation.  Thus the shift by one is fixed by the scalar-block
normalization and the Cartan transformation.

In the classical scaling
\eqref{equation:PII classical WZW theta weight},
$N=\Theta/\epsilon_{1}+1$ and
$\kappa=-\epsilon_{2}/\epsilon_{1}$.  Thus the exact charge relation is
\begin{equation}
	\label{equation:PII Langlands large small screening}
	N^{\vee}
	=-\frac{\Theta}{\epsilon_{2}}-1
	-\frac{2\epsilon_{1}}{\epsilon_{2}}.
\end{equation}
At fixed nonzero $\epsilon_{2}$ the classical limit therefore gives
\begin{equation}
	\theta^{\vee}\longrightarrow
	-\frac{\Theta}{\epsilon_{2}}-\frac{5}{2},
	\qquad
	N^{\vee}\longrightarrow-\frac{\Theta}{\epsilon_{2}}-1.
\end{equation}
This is a finite dual Cartan charge.  A fixed nonnegative integer
$N^{\vee}$ is realized by the exact charge path
\begin{equation}
	\begin{gathered}
		\theta=\kappa(N^{\vee}+1)-\frac{5}{2},
		\qquad N=\kappa(N^{\vee}+1)-1,\\
		\epsilon_{1}\theta\longrightarrow
		\Theta=-\epsilon_{2}(N^{\vee}+1).
	\end{gathered}
\end{equation}
Its finite term in $\theta$ differs from the parametrization in
Subsection~\ref{section:PII classical WZW limit}, while its leading
scaled charge and limiting position connection are the same.
The finite Cartan factor in the classical expansion is then $u^{-1}$
in place of $u$.  It changes the time equation by a scalar term, accounted for by
the normalization of the limiting connection.
When both screening numbers are integers, the path also obeys
$\kappa=(N+1)/(N^{\vee}+1)$.  The Langlands duality map acts on the affine
and BPZ solutions.

A screening cycle selects a branch of the solution, and
\eqref{equation:PII Langlands transformed block} transports it to the
dual branch.  At $\kappa=1$, $a=1$ and $N=N^{\vee}=0$, the exchanged
empty-screening block differs from the dual empty-screening block by
$\exp\left(V^{\vee}(q^{\vee})-V^{\vee}(z^{\vee})\right)$,
up to a constant branch phase.  Both solve the dual equations.
The asymptotic correspondence below uses the branches related
by the exact duality map.

\subsection{Crossed classical and NS limits}
\label{subsection:PII Langlands exchanged limits}

The deformation-parameter exchange in
\eqref{equation:PII Langlands limit dictionary} crosses the classical
and NS limits.  The finite-$N$ application in
Subsection~\ref{subsection:PII Langlands limiting opers} uses
\begin{equation}
	(\epsilon_{2}\to0,\ N\ \text{fixed},\ \epsilon_{1}\neq0)_{\mathrm{NS}}
	\quad\longleftrightarrow\quad
	(\epsilon_{1}^{\vee}\to0,\
	\epsilon_{2}^{\vee}=\epsilon_{1})_{\mathrm{classical}}.
\end{equation}
Here a finite original Bethe polynomial becomes part of the dual
Painlev\'e Lax wave function.  We also record the original-classical to dual-NS action relation.
On the critical-level side,
the position and time operators become the commuting Hamiltonians
whose joint spectrum was constructed in Subsection~\ref{section:PII NS comparison}; their oper
description is the irregular Gaudin realization
\cite{FeiginFrenkelToledanoLaredo2010}.

For the NS-to-dual-classical correspondence, use the coordinates
$q,z,t$.  The finite-coupling transformation in
\eqref{equation:PII Langlands finite coordinate map} has a singular
coordinate factor as $\epsilon_2\to0$; the finite dual Painlev\'e
coordinates are obtained after the corresponding classical-limit
renormalization.  The resulting dual apparent position and spectral coordinate are
those of \eqref{equation:PII Langlands unscaled coordinate exchange}.
Here $w$ is the isomonodromic coordinate and $\lambda$ is the
spectral argument of the dual Lax wave function.  The charge map is
\begin{equation}
	\epsilon_1^{\vee}\theta^{\vee}
	=\epsilon_2\theta^{\vee}
	\longrightarrow-\epsilon_{1}\left(\theta+\frac{5}{2}\right)
	=-\epsilon_1(N+1).
\end{equation}
The Painlev\'e II parameter is $\alpha^{\vee}=N+\tfrac{1}{2}$.

The inverse map takes the original classical connection to a dual NS
oper, with action
\begin{equation}
	\label{equation:PII Langlands limiting action exchange}
	\mathcal{W}_{\mathrm{NS}}^{\vee}(Z^{\vee},T)
	=\mathcal{W}(-Z^{\vee},T,1;\epsilon_{2})
	+\frac{(Z^{\vee})^{3}}{3}+\frac{TZ^{\vee}}{2}.
\end{equation}
The original-NS to dual-classical action is
\eqref{equation:PII Langlands limiting action exchange inverse}.

\section{Complete of  Bethe Vector}
\label{appendix:PII finite Bethe calculations}

This appendix supplies the completeness proof for
Subsection~\ref{subsection:PII NS joint spectrum completeness}
and its exceptional-point examples.

\subsection{Completeness}
\label{appendix:PII Bethe completeness proof}

We work in the sector
\eqref{equation:PII NS polynomial sector} and use the Bethe vector
\eqref{equation:PII NS Bethe vector}.

Both Hamiltonians act within this finite sector.  After removing
the powers of $u$, the two diagonal entries of $H_{t}$ are
\begin{equation}
	\begin{aligned}
		&\partial_{q}^{2}-(2q^{2}+t)\partial_{q}+2Nq+z,\\
		&\partial_{q}^{2}-(2q^{2}+t)\partial_{q}+2(N-1)q-z.
	\end{aligned}
\end{equation}
Their degree-raising terms cancel on $q^{N}$ and $q^{N-1}$,
respectively.  The off-diagonal entries are $1$ and
$2\partial_{q}$, so $H_{t}$ preserves the two degree bounds.
The spin-flip entries of $H_{1}$ are
\begin{equation}
	\left(H_{1}\right)_{12}=z-q,
	\qquad
	\left(H_{1}\right)_{21}=2(z+q)\partial_{q}-2N.
\end{equation}
The first raises the lower-component degree by at most one.
The second spin-flip term has no contribution of degree $N$ because
its two terms cancel on $q^{N}$.  Together with the diagonal terms,
this keeps both spin components within their degree bounds.
The vector \eqref{equation:PII NS Bethe vector} belongs to this space:
each term in its lower component cancels one factor of $P_{N}(q)$.
The leading coefficient of its upper component is fixed to one
after division by $u^{N}$.  This fixes the overall normalization,
and the zeros of that component recover the Bethe roots.

To count the states, set $t=-2s^{2}$ and take $s\to+\infty$ with
$z$ fixed.  Choose $r\in\{0,1\}$ with $r\leq N$ and
$n_{+},n_{-}\in\mathbb{Z}_{\geq 0}$ satisfying
\begin{equation}
	n_{+}+n_{-}=N-r.
\end{equation}
Place $n_{+}$ roots near $+s$ and $n_{-}$ roots near $-s$.
For $r=1$, place one additional root near $z$.  Its rescaled Bethe
equation gives, at leading order,
\begin{equation}
	x^{*}=z+\frac{1}{2s^{2}}+o(s^{-2}).
\end{equation}
For the distant clusters, write
$x_{a}^{*}=\pm s+s^{-\frac{1}{2}}y_{a}$.  Their leading equations are
\begin{equation}
	\pm 4y_{a}
	-2\sum_{\substack{b=1\\b\neq a}}^{n_{\pm}}
	\frac{1}{y_{a}-y_{b}}=0.
\end{equation}
The cluster near $+s$ consists of scaled Hermite roots.  Its
fluctuation matrix has a diagonal contribution $4$ and positive
pair-interaction contributions for real fluctuations, so it has
no zero mode.  Multiplying these roots by $\mathrm{i}$ gives the
cluster near $-s$; its fluctuation matrix differs by an overall
minus sign and also has no zero mode.  The root near $z$ is
nondegenerate as well.  Interactions between the separated clusters
are subleading in $s^{-\frac{1}{2}}$.  Restoring these interactions
therefore gives a smooth correction to each configuration and one
isolated Bethe solution for each $(r,n_{+},n_{-})$ at sufficiently
large $s$.

There are $N+1$ choices with $r=0$ and $N$ choices with $r=1$,
giving $2N+1$ solutions in total.  Their time eigenvalues satisfy
\begin{equation}
	E_{t}=-2s(n_{+}-n_{-})+(1-2r)z+o(1).
\end{equation}
For $r=0$, the differences $n_{+}-n_{-}$ run from $-N$ to $N$
in steps of two.  For $r=1$, they run from $-(N-1)$ to $N-1$
in steps of two.  These two sets have opposite parity, so all
$2N+1$ leading slopes are distinct.  Thus $H_{t}$ has $2N+1$
separated levels for sufficiently large $s$.  The corresponding
Bethe states are linearly independent and exhaust the $2N+1$
degrees of freedom in $\mathcal{V}_{N}$.  Since
$[H_{1},H_{t}]=0$, they diagonalize both Hamiltonians.
The case $N=0$ is included: the empty root set gives the single
vector $(1,0)^{T}$.

Starting from these separated levels, we follow the eigenstates
and Bethe roots as $z,t$ vary.  At finite $N$, the Hamiltonian
matrices have polynomial entries in $z,t$.  Level coalescence and
singular root configurations therefore occur only when polynomial
conditions on these parameters are satisfied.  They are absent
in the large-negative-$t$ regime just described.  For a generic
fixed $t$, they consequently occur at only finitely many values
of $z$.  Away from these exceptional values, the $2N+1$ states
remain independent and give the complete spectrum in this sector.
This is the genericity condition on the external spectral variable
$z$; it places no restriction on the wave-function argument $q$.

There is also no overcounting of states.  A nonzero state cannot
have a vanishing upper component, by the upper KZ equation.  This
equation also forces
the lower spin component to have degree at least one less than the
upper component.  The leading terms of the lower row then fix the
upper degree to $N$.  Once its leading coefficient is normalized
to one, the upper component is $P_{N}(q)$ and determines the root
configuration.  Since each nondegenerate level labels a single
eigenstate, another root configuration could not give an additional
state at the same level.  The $2N+1$ Bethe states therefore give
the complete simultaneous spectrum for generic $z$.

\subsection{Exceptional points}
\label{appendix:PII Bethe exceptional points}
Bethe states can merge at special spectral parameters, so the
genericity condition is necessary already for $N=1$.  The one-root Bethe equation is
\begin{equation}
	2(x^{*})^{3}-2z(x^{*})^{2}+tx^{*}+1-tz=0,
	\qquad x^{*}\neq z.
\end{equation}
Two Bethe solutions merge when the discriminant vanishes:
\begin{equation}
	\begin{aligned}
		&\operatorname{disc}_{x}
		\bigl(2x^{3}-2zx^{2}+tx+1-tz\bigr)\\
		&\qquad=4\bigl(-8tz^{4}+8z^{3}-8t^{2}z^{2}
		+36tz-2t^{3}-27\bigr).
	\end{aligned}
\end{equation}
For every fixed $t$, the expression on the right is a nonzero
polynomial in $z$.  The one-screening sector therefore has three
independent Bethe states except at the finitely many values where
this expression vanishes.
In the ordered basis
$\bigl((u,0)^{T},(uq,0)^{T},(0,1)^{T}\bigr)$ of
$\mathcal{V}_{1}$, the time Hamiltonian is
\begin{equation}
	H_{t}=\begin{pmatrix}
		z&-t&1\\
		2&z&0\\
		0&2&-z
	\end{pmatrix},
	\qquad
	\det(E_{t}\mathbf{1}-H_{t})
	=(E_{t}+z)\bigl((E_{t}-z)^{2}+2t\bigr)-4.
\end{equation}
For $t=0$ and $z=\frac{3}{2}$, the Bethe polynomial factors as
\begin{equation}
	2(x^{*})^{3}-3(x^{*})^{2}+1
	=(x^{*}-1)^{2}(2x^{*}+1).
\end{equation}
Only two distinct Bethe solutions remain,
$x^{*}=1$ and $x^{*}=-\frac{1}{2}$, with time eigenvalues
$-\frac{1}{2}$ and $\frac{5}{2}$.  More precisely,
\begin{equation}
	\det(E_{t}\mathbf{1}-H_{t})
	=\left(E_{t}+\frac{1}{2}\right)^{2}
	\left(E_{t}-\frac{5}{2}\right).
\end{equation}
The two eigenvectors at $E_{t}=-\frac{1}{2}$ coalesce into a single
state, giving a Jordan block of size two.  Together with the state
at $E_{t}=\frac{5}{2}$, only two independent eigenstates remain in
the sector of dimension three.  At $t=0$, this collision occurs precisely at
$z^{3}=\frac{27}{8}$; the displayed positive real value is one example.

A triple collision occurs whenever
\begin{equation}
	16(x^{*})^{3}=1,\qquad z=3x^{*},\qquad t=6(x^{*})^{2}.
\end{equation}
At these parameter values the Bethe polynomial becomes
$2(x-x^{*})^{3}$.  Only one Bethe saddle remains, and
\begin{equation}
	\det(E_{t}\mathbf{1}-H_{t})=(E_{t}-x^{*})^{3},
	\qquad
	(H_{t}-x^{*}\mathbf{1})^{3}=0,
	\qquad
	(H_{t}-x^{*}\mathbf{1})^{2}\neq 0.
\end{equation}
Thus all three levels and their eigenvectors coalesce: $H_{t}$
has one Jordan block of size three and only one independent
eigenstate.

\section{Finite-\texorpdfstring{$N$}{N} reconstruction and normalization}
\label{appendix:PII action normalization}

\subsection{Adjacent-size quotient}
\label{appendix:PII quotient identity}

Let $V(x;t)=\frac{2}{3}x^{3}+tx$, and for a linear combination
$\gamma=c_{+}\gamma_{+}+c_{-}\gamma_{-}$ of the two normalized
rapid-decay cycles set
\begin{equation}
\label{equation:PII normalization seed}
f(t):=\int_{\gamma}\exp\left(-V(x;t)\right)\dd{x},
\qquad f''+\frac{t}{2}f=0.
\end{equation}
The monic normalization is generated by the scaled determinant
\begin{equation}
\label{equation:PII normalization tau}
\tau_{N}(t):=\det\left[\partial_{t}^{i+j}f(t)\right]_{i,j=0}^{N-1},
\qquad
\widehat{\tau}_{N}(T):=\tau_{N}\left(\epsilon_{1}^{-2/3}T\right),
\qquad
\widetilde{\tau}_{N}(T):=\epsilon_{1}^{N^{2}/3}\widehat{\tau}_{N}(T).
\end{equation}
It determines the Painlev\'e data through
\begin{equation}
\label{equation:PII normalization hierarchy}
w_{N}=\partial_{t}\log{\frac{\tau_{N}}{\tau_{N+1}}},
\qquad
p_{N}=w_{N}'+w_{N}^{2}+\frac{t}{2},
\qquad
s_{N}=-\partial_{t}\log{\tau_{N}}.
\end{equation}
We write $z=-w_{N}$ and $R_{N}(z)=z'+z^{2}+\frac{t}{2}$, with
$z'=-w_{N}'$ taken along the trajectory, so that $R_{N}=P_{N}'/P_{N}$
and $s_{N}=\sum_{a}x_{a}^{*}$ are the quantities of
\eqref{equation:PII Langlands Bethe regularity in PII}.

Toda condensation fixes both the recurrence coefficients and the
monic normalization,
\begin{equation}
\label{equation:PII tau polynomial recursion}
\begin{gathered}
p_{r-1}=-2\frac{\tau_{r+1}\tau_{r-1}}{\tau_{r}^{2}},
\qquad P_{0}(q)=1,\qquad P_{1}(q)=q-w_{0},\\
P_{r+1}(q)=(q-w_{r})P_{r}(q)
+\frac{p_{r-1}}{2}P_{r-1}(q),\qquad r\geq1,
\end{gathered}
\end{equation}
each $P_{r}$ belonging to the member with moving insertion
$z_{r}=-w_{r}$.  At the final insertion $z=-w_{N}$ the lowering
identity for $N\geq1$ is
\begin{equation}
\label{equation:PII tau polynomial lowering}
P_{N}'(q)=p_{N-1}\left(
P_{N}(q)+(z-q)P_{N-1}(q)\right).
\end{equation}
With the root discriminant
$\Delta_{N}:=\operatorname{Disc}{(P_{N})}
=\prod_{a<b}(x_{a}^{*}-x_{b}^{*})^{2}$ and
$\Delta_{0}=\Delta_{1}=1$, the same hierarchy gives
\begin{equation}
\label{equation:PII tau discriminant identity}
\frac{P_{N}(z)}{\Delta_{N}}
=(-2)^{-N}\frac{\tau_{N}^{N+1}}{\tau_{N+1}^{N}}.
\end{equation}

The monic polynomial obeys the scalar equation
\begin{equation}
\label{equation:PII unscaled polynomial ode}
P_{N}''+\left(\frac{1}{z-q}-2q^{2}-t\right)P_{N}'
+\left(2Nq+2s_{N}-\frac{R_{N}(z)}{z-q}\right)P_{N}=0,
\end{equation}
and multiplying the Bethe equations by $x_{a}^{*}$ and summing
gives
\begin{equation}
\label{equation:PII unscaled cubic sum}
2\sum_{a=1}^{N}(x_{a}^{*})^{3}+ts_{N}=zR_{N}(z)+N(N-2).
\end{equation}
The cubic identity evaluates the finite critical action as
\begin{equation}
\label{equation:PII tau finite action discriminant}
\frac{\mathcal{W}_{\mathrm{NS}}}{\epsilon_{1}}
=\frac{zR_{N}(z)}{3}+\frac{2ts_{N}}{3}
+\frac{N(N-2)}{3}
+\log{\frac{P_{N}(z)}{\Delta_{N}}}
-\frac{z^{3}}{3}-\frac{tz}{2}.
\end{equation}
Evaluating the lowering identity at the zeros of $P_{N}$ gives
\begin{equation}
\label{equation:PII tau resultant}
\Delta_{N}=(-1)^{N(N-1)/2}p_{N-1}^{N}P_{N}(z)
\operatorname{Res}{(P_{N},P_{N-1})},
\qquad
\operatorname{Res}{(P_{N},P_{N-1})}
=\prod_{r=1}^{N-1}\left(\frac{p_{r-1}}{2}\right)^{r},
\end{equation}
and substitution of the Toda ratios in
\eqref{equation:PII tau polynomial recursion} telescopes to
\eqref{equation:PII tau discriminant identity}.  Inserting this
ratio into \eqref{equation:PII tau finite action discriminant}, and
using $R_{N}=z'+z^{2}+\frac{t}{2}$ together with
$s_{N}=-\partial_{t}\log{\tau_{N}}$, gives
\begin{equation}
\label{equation:PII unscaled W reconstructed}
\begin{aligned}
\frac{\mathcal{W}_{\mathrm{NS}}(z,t)}{\epsilon_{1}}
={}&\frac{z}{3}(z'-t)
-\frac{2t}{3}\partial_{t}\log{\tau_{N}}
+\frac{N(N-2)}{3}\\
&+\log{\left(\frac{\tau_{N}^{N+1}}{(-2)^{N}\tau_{N+1}^{N}}\right)}.
\end{aligned}
\end{equation}
This is the adjacent-size quotient: the action is determined by
$\tau_{N}$ and $\tau_{N+1}$ at the same $\epsilon_{1}$ and time, the
adjacent determinant carrying total scaled mass
$\Theta+2\epsilon_{1}$.  Rescaling
$t=\epsilon_{1}^{-2/3}T$, $z=\epsilon_{1}^{-1/3}Z$ turns
\eqref{equation:PII unscaled W reconstructed} into
\eqref{equation:PII exact scaled action}.

\paragraph*{Direct finite-size examples.}
Set $w_{0}=-f'/f$ and $p_{0}=2w_{0}^{2}+t$.  The Airy equation gives
\begin{equation}
\label{equation:PII tau first determinants}
\frac{\tau_{1}}{f}=1,\qquad
\frac{\tau_{2}}{f^{2}}=-\frac{p_{0}}{2},\qquad
\frac{\tau_{3}}{f^{3}}=-\frac{2w_{0}p_{0}+1}{4}.
\end{equation}
For $N=1$ the insertion equation is the cubic $p_{0}(z-w_{0})=1$,
and for $N=2$ the quintic
\begin{equation}
\label{equation:PII tau N2 seed quintic}
(z+w_{0})p_{0}(2w_{0}p_{0}+1)+2w_{0}p_{0}+1-p_{0}^{3}=0,
\end{equation}
with $p_{0}(2w_{0}p_{0}+1)\neq0$.  The determinant ratios and
\eqref{equation:PII tau polynomial recursion} then give
\begin{equation}
\label{equation:PII tau N2 polynomial}
p_{1}=\frac{2(2w_{0}p_{0}+1)}{p_{0}^{2}},
\qquad
P_{2}(q)=q^{2}+\frac{q}{p_{0}}+\frac{t}{2}-\frac{w_{0}}{p_{0}},
\end{equation}
and
\begin{equation}
\label{equation:PII tau N2 action}
\left.\frac{\mathcal{W}_{\mathrm{NS}}}{\epsilon_{1}}\right|_{N=2}
=\frac{z}{3}\left(p_{1}-z^{2}-\frac{3t}{2}\right)
-\frac{2t}{3p_{0}}
+\log{\left(-\frac{2}{p_{0}p_{1}^{2}}\right)},
\end{equation}
in agreement with
\eqref{equation:PII unscaled W reconstructed}.  Implicit
differentiation of the quintic, including the dependence
$w_{0}(z,t)$ of the selected seed, gives
\begin{equation}
\label{equation:PII tau N2 action derivatives}
\left.\frac{1}{\epsilon_{1}}\partial_{z}
\mathcal{W}_{\mathrm{NS}}\right|_{N=2}
=p_{1}-z^{2}-\frac{t}{2},
\qquad
\left.\frac{1}{\epsilon_{1}}\partial_{t}
\mathcal{W}_{\mathrm{NS}}\right|_{N=2}
=-\frac{1}{p_{0}}-\frac{z}{2},
\end{equation}
which are the NS action derivatives at independent $z,t$.

\subsection{Logarithmic subtraction}
\label{appendix:PII log subtraction}

The subtraction in
\eqref{equation:PII chapter7 action expansion convention} is fixed
by this normalization.  The fixed-filling expansion of the scaled
determinant, followed by the sum over nearby integer fillings,
reads
\begin{equation}
\label{equation:PII normalization expansion}
\begin{aligned}
\log{\widetilde{\tau}_{N}}
={}&-\frac{\mathcal{W}_{0}}{\epsilon_{1}^{2}}
+N_{+}\log{c_{+}}+N_{-}\log{c_{-}}
+N\log{(2\pi)}+\frac{1}{6}\log{\epsilon_{1}}\\
&+\mathcal{F}_{1}^{(\tau)}+\log{\vartheta_{N}}
+\epsilon_{1}R_{N}^{(\tau)}+O\left(\epsilon_{1}^{2}\right),
\end{aligned}
\end{equation}
where $\mathcal{F}_{1}^{(\tau)}$ and $R_{N}^{(\tau)}$ denote the
genus-one and first subleading coefficients of the tau logarithm;
only the $\frac{1}{6}\log{\epsilon_{1}}$ term is used below.
The sum over nearby fillings gives the theta factor
\cite{BorotGuionnet2013}
\begin{equation}
\label{equation:PII normalization theta}
\begin{gathered}
\eta_{N}:=\log{\frac{c_{+}}{c_{-}}}
-\frac{\mu_{+}-\mu_{-}}{\epsilon_{1}},
\qquad
\vartheta_{N}:=\vartheta(\eta_{N}),\\
\vartheta(\eta):=\sum_{j\in\mathbb{Z}}
\exp\left[-\frac{j^{2}}{2}
\left(\partial_{\nu_{+}}-\partial_{\nu_{-}}\right)^{2}\mathcal{W}_{0}
+j\eta\right],
\end{gathered}
\end{equation}
with $\mu_{\pm}:=\partial_{\nu_{\pm}}\mathcal{W}_{0}$.  The curve
quantities on the right are first evaluated at the scaled masses
$(\epsilon_{1}N_{+},\epsilon_{1}N_{-})$ and then expanded about
$(\nu_{+},\nu_{-})$.

By \eqref{equation:PII exact scaled action}, the
$\log{\epsilon_{1}}$ content of $\mathcal{W}_{\mathrm{NS}}$ is
carried by
$\epsilon_{1}\left[(N+1)\log{\widehat{\tau}_{N}}
-N\log{\widehat{\tau}_{N+1}}\right]$.  Since
$\widehat{\tau}_{N}=\epsilon_{1}^{-N^{2}/3}\widetilde{\tau}_{N}$,
this combination equals
\begin{equation}
\frac{\epsilon_{1}N(N+1)}{3}\log{\epsilon_{1}}
+\epsilon_{1}\left[
(N+1)\log{\widetilde{\tau}_{N}}
-N\log{\widetilde{\tau}_{N+1}}\right].
\end{equation}
The first term is the coordinate factor
\begin{equation}
\label{equation:PII normalization coordinate factor}
\frac{\epsilon_{1}N(N+1)}{3}\log{\epsilon_{1}}
=\mathcal{W}_{\mathrm{scale}}
+\left(\Theta+\epsilon_{1}\right)\log{\epsilon_{1}},
\end{equation}
as follows from $N=\Theta/\epsilon_{1}+1$ and the definition of
$\mathcal{W}_{\mathrm{scale}}$ in
\eqref{equation:PII chapter7 scaled action}.  In the second term
the $\frac{1}{6}\log{\epsilon_{1}}$ of
\eqref{equation:PII normalization expansion} contributes a further
$\epsilon_{1}\log{\epsilon_{1}}/6$, while the other coefficients
contribute at higher order.  The total coefficient of
$\log{\epsilon_{1}}$ is therefore
\begin{equation}
\label{equation:PII normalization subtraction}
\left(\Theta+\epsilon_{1}\right)+\frac{\epsilon_{1}}{6}
=\Theta+\frac{7\epsilon_{1}}{6},
\end{equation}
which is the subtraction displayed in
\eqref{equation:PII chapter7 action expansion convention}.

\IfFileExists{JHEP.bst}{\bibliographystyle{JHEP}}{\bibliographystyle{unsrt}}
\bibliography{biblio}

\providecommand{\href}[2]{#2}\begingroup\raggedright\begin{thebibliography}{10}

\bibitem{Alday:2009aq}
L.F.~Alday, D.~Gaiotto and Y.~Tachikawa, \emph{{Liouville Correlation Functions
  from Four-dimensional Gauge Theories}},
  \href{https://doi.org/10.1007/s11005-010-0369-5}{\emph{Lett. Math. Phys.}
  {\bfseries 91} (2010) 167} [\href{https://arxiv.org/abs/0906.3219}{{\ttfamily
  0906.3219}}].

\bibitem{Argyres:1995jj}
P.C.~Argyres and M.R.~Douglas, \emph{{New phenomena in SU(3) supersymmetric
  gauge theory}},
  \href{https://doi.org/10.1016/0550-3213(95)00281-V}{\emph{Nucl. Phys. B}
  {\bfseries 448} (1995) 93}
  [\href{https://arxiv.org/abs/hep-th/9505062}{{\ttfamily hep-th/9505062}}].

\bibitem{Gaiotto:2009ma}
D.~Gaiotto, \emph{{Asymptotically free $\mathcal{N} = 2$ theories and irregular
  conformal blocks}},
  \href{https://doi.org/10.1088/1742-6596/462/1/012014}{\emph{J. Phys. Conf.
  Ser.} {\bfseries 462} (2013) 012014}
  [\href{https://arxiv.org/abs/0908.0307}{{\ttfamily 0908.0307}}].

\bibitem{Bonelli:2011aa}
G.~Bonelli, K.~Maruyoshi and A.~Tanzini, \emph{{Wild Quiver Gauge Theories}},
  \href{https://doi.org/10.1007/JHEP02(2012)031}{\emph{JHEP} {\bfseries 02}
  (2012) 031} [\href{https://arxiv.org/abs/1112.1691}{{\ttfamily 1112.1691}}].

\bibitem{Gaiotto:2012sf}
D.~Gaiotto and J.~Teschner, \emph{{Irregular singularities in Liouville theory
  and Argyres-Douglas type gauge theories, I}},
  \href{https://doi.org/10.1007/JHEP12(2012)050}{\emph{JHEP} {\bfseries 12}
  (2012) 050} [\href{https://arxiv.org/abs/1203.1052}{{\ttfamily 1203.1052}}].

\bibitem{Bonelli:2016qwg}
G.~Bonelli, O.~Lisovyy, K.~Maruyoshi, A.~Sciarappa and A.~Tanzini, \emph{{On
  Painlev\'e/gauge theory correspondence}},
  \href{https://doi.org/10.1007/s11005-017-0983-6}{\emph{Lett. Matth. Phys.}
  {\bfseries 107} (2017) 2359}
  [\href{https://arxiv.org/abs/1612.06235}{{\ttfamily 1612.06235}}].

\bibitem{GrassiGu2019}
A.~Grassi and J.~Gu, \emph{{Argyres--Douglas} theories, {Painlev{\'e} II} and
  quantum mechanics},
  \href{https://doi.org/10.1007/JHEP02(2019)060}{\emph{JHEP} {\bfseries 02}
  (2019) 060} [\href{https://arxiv.org/abs/1803.02320}{{\ttfamily
  1803.02320}}].

\bibitem{Dedushenko:2018bpp}
M.~Dedushenko, S.~Gukov, H.~Nakajima, D.~Pei and K.~Ye, \emph{{3d TQFTs from
  Argyres\textendash{}Douglas theories}},
  \href{https://doi.org/10.1088/1751-8121/abb481}{\emph{J. Phys. A} {\bfseries
  53} (2020) 43LT01} [\href{https://arxiv.org/abs/1809.04638}{{\ttfamily
  1809.04638}}].

\bibitem{Gaiotto:2011nm}
D.~Gaiotto and E.~Witten, \emph{{Knot Invariants from Four-Dimensional Gauge
  Theory}}, \href{https://doi.org/10.4310/ATMP.2012.v16.n3.a5}{\emph{Adv.
  Theor. Math. Phys.} {\bfseries 16} (2012) 935}
  [\href{https://arxiv.org/abs/1106.4789}{{\ttfamily 1106.4789}}].

\bibitem{Hu2016}
S.~Hu and P.~Liu, \emph{Kauffman polynomial from a generalized yang--yang
  function}, \href{https://doi.org/10.1007/s00023-015-0426-9}{\emph{Annales
  Henri Poincar{\'e}} {\bfseries 17} (2016) 1145}
  [\href{https://arxiv.org/abs/1406.1851}{{\ttfamily 1406.1851}}].

\bibitem{Gu:2023bjg}
X.~Gu and B.~Haghighat, \emph{{Liouville conformal blocks and Stokes
  phenomena}}, \href{https://doi.org/10.4310/BPAM.2024.v1.n1.a6}{\emph{Beijing
  Journal of Pure and Applied Mathematics} {\bfseries 1} (2024) 373}
  [\href{https://arxiv.org/abs/2301.07957}{{\ttfamily 2301.07957}}].

\bibitem{HaghighatLiuReshetikhin2023}
B.~Haghighat, Y.~Liu and N.~Reshetikhin, \emph{Flat connections from irregular
  conformal blocks},
  \href{https://doi.org/10.1007/s00220-025-05317-2}{\emph{Commun. Math. Phys.}
  {\bfseries 406} (2025) 138}
  [\href{https://arxiv.org/abs/2311.07960}{{\ttfamily 2311.07960}}].

\bibitem{Gu:2026mlf}
X.~Gu, B.~Haghighat and P.~Putrov, \emph{{On the monodromy of KZ-connections
  with irregular singularities}},
  \href{https://arxiv.org/abs/2603.24098}{{\ttfamily 2603.24098}}.

\bibitem{RibaultTeschner2005}
S.~Ribault and J.~Teschner, \emph{{$H_{3}^{+}$ correlators from Liouville
  theory}}, \href{https://doi.org/10.1088/1126-6708/2005/06/014}{\emph{JHEP}
  {\bfseries 06} (2005) 014}
  [\href{https://arxiv.org/abs/hep-th/0502048}{{\ttfamily hep-th/0502048}}].

\bibitem{HikidaSchomerus2007}
Y.~Hikida and V.~Schomerus, \emph{{$H_3^+$} {WZNW} model from {Liouville} field
  theory}, \href{https://doi.org/10.1088/1126-6708/2007/10/064}{\emph{JHEP}
  {\bfseries 10} (2007) 064} [\href{https://arxiv.org/abs/0706.1030}{{\ttfamily
  0706.1030}}].

\bibitem{GaiottoLamyPoirier2013}
D.~Gaiotto and J.~Lamy-Poirier, \emph{Irregular singularities in the
  {$H_{3}^{+}$} {WZW} model},
  \href{https://arxiv.org/abs/1301.5342}{{\ttfamily 1301.5342}}.

\bibitem{FrenkelGukovTeschner2015}
E.~Frenkel, S.~Gukov and J.~Teschner, \emph{Surface operators and separation of
  variables}, \href{https://doi.org/10.1007/JHEP01(2016)179}{\emph{JHEP}
  {\bfseries 01} (2016) 179}
  [\href{https://arxiv.org/abs/1506.07508}{{\ttfamily 1506.07508}}].

\bibitem{DrinfeldSokolov1985}
V.G.~Drinfeld and V.V.~Sokolov, \emph{Lie algebras and equations of
  korteweg--de vries type},
  \href{https://doi.org/10.1007/BF02105860}{\emph{Journal of Soviet
  Mathematics} {\bfseries 30} (1985) 1975}.

\bibitem{FeiginFrenkel2007}
B.~Feigin and E.~Frenkel, \emph{Quantization of the {Drinfeld--Sokolov}
  reduction}, \href{https://doi.org/10.1016/0370-2693(90)91310-8}{\emph{Physics
  Letters B} {\bfseries 246} (1990) 75}.

\bibitem{BanerjeeHaghighatLatifi2026}
S.~Banerjee, B.~Haghighat and A.~Latifi, \emph{Quantum flat connections, {KZ}
  equations, and integrability},
  \href{https://arxiv.org/abs/2604.26159}{{\ttfamily 2604.26159}}.

\bibitem{ArakawaFrenkel2019}
T.~Arakawa and E.~Frenkel, \emph{Quantum {Langlands} duality of representations
  of {$W$}-algebras},
  \href{https://doi.org/10.1112/S0010437X19007553}{\emph{Compos. Math.}
  {\bfseries 155} (2019) 2235}
  [\href{https://arxiv.org/abs/1807.01536}{{\ttfamily 1807.01536}}].

\bibitem{Resh-KZ}
N.~Reshetikhin, \emph{The knizhnik-zamolodchikov system as a deformation of the
  isomonodromy problem},
  \href{https://doi.org/10.1007/BF00420750}{\emph{Letters in Mathematical
  Physics} {\bfseries 26} (1992) 167}.

\bibitem{JimboNagoyaSun2008}
M.~Jimbo, H.~Nagoya and J.~Sun, \emph{Remarks on the confluent {KZ} equation
  for {$\mathfrak{sl}_{2}$} and quantum {Painlev{\'e}} equations},
  \href{https://doi.org/10.1088/1751-8113/41/17/175205}{\emph{J. Phys. A}
  {\bfseries 41} (2008) 175205}.

\bibitem{Nagoya_2010}
H.~Nagoya and J.~Sun, \emph{Confluent primary fields in the conformal field
  theory}, \href{https://doi.org/10.1088/1751-8113/43/46/465203}{\emph{Journal
  of Physics A: Mathematical and Theoretical} {\bfseries 43} (2010) 465203}
  [\href{https://arxiv.org/abs/1002.2598}{{\ttfamily 1002.2598}}].

\bibitem{NagoyaSunConfluent2011}
H.~Nagoya and J.~Sun, \emph{Confluent {KZ} equations for {$\mathfrak{sl}_N$}
  with {Poincar{\'e}} rank 2 at infinity},
  \href{https://doi.org/10.1088/1751-8113/44/28/285205}{\emph{J. Phys. A}
  {\bfseries 44} (2011) 285205}
  [\href{https://arxiv.org/abs/1002.2273}{{\ttfamily 1002.2273}}].

\bibitem{Nagoya:2015cja}
H.~Nagoya, \emph{{Irregular conformal blocks, with an application to the fifth
  and fourth Painlev\'e equations}},
  \href{https://doi.org/10.1063/1.4937760}{\emph{J. Math. Phys.} {\bfseries 56}
  (2015) 123505} [\href{https://arxiv.org/abs/1505.02398}{{\ttfamily
  1505.02398}}].

\bibitem{Felder:2020ugk}
G.~Felder and G.~Rembado, \emph{{Singular modules for affine Lie algebras, and
  applications to irregular WZNW conformal blocks}},
  \href{https://doi.org/10.1007/s00029-022-00821-y}{\emph{Selecta Math.}
  {\bfseries 29} (2023) 15} [\href{https://arxiv.org/abs/2012.14793}{{\ttfamily
  2012.14793}}].

\bibitem{FeiginFrenkelToledanoLaredo2010}
B.~Feigin, E.~Frenkel and V.~Toledano~Laredo, \emph{Gaudin models with
  irregular singularities},
  \href{https://doi.org/10.1016/j.aim.2009.09.007}{\emph{Adv. Math.} {\bfseries
  223} (2010) 873} [\href{https://arxiv.org/abs/math/0612798}{{\ttfamily
  math/0612798}}].

\bibitem{Nagoya2011}
H.~Nagoya, \emph{Hypergeometric solutions to {Schr{\"o}dinger} equations for
  the quantum {Painlev{\'e}} equations},
  \href{https://doi.org/10.1063/1.3620412}{\emph{J. Math. Phys.} {\bfseries 52}
  (2011) 083509} [\href{https://arxiv.org/abs/1109.1645}{{\ttfamily
  1109.1645}}].

\bibitem{Dijkgraaf:2009pc}
R.~Dijkgraaf and C.~Vafa, \emph{{Toda Theories, Matrix Models, Topological
  Strings, and N=2 Gauge Systems}},
  \href{https://arxiv.org/abs/0909.2453}{{\ttfamily 0909.2453}}.

\bibitem{Nishinaka:2012kn}
T.~Nishinaka and C.~Rim, \emph{{Matrix models for irregular conformal blocks
  and Argyres-Douglas theories}},
  \href{https://doi.org/10.1007/JHEP10(2012)138}{\emph{JHEP} {\bfseries 10}
  (2012) 138} [\href{https://arxiv.org/abs/1207.4480}{{\ttfamily 1207.4480}}].

\bibitem{Rim:2012tf}
C.~Rim, \emph{{Irregular conformal block and its matrix model}},
  \href{https://arxiv.org/abs/1210.7925}{{\ttfamily 1210.7925}}.

\bibitem{Choi:2014qha}
S.K.~Choi, C.~Rim and H.~Zhang, \emph{{Virasoro irregular conformal block and
  beta deformed random matrix model}},
  \href{https://doi.org/10.1016/j.physletb.2015.01.003}{\emph{Phys. Lett. B}
  {\bfseries 742} (2015) 50} [\href{https://arxiv.org/abs/1411.4453}{{\ttfamily
  1411.4453}}].

\bibitem{Rumanov2014}
I.~Rumanov, \emph{{$\beta$} ensembles, quantum {Painlev{\'e}} equations and
  isomonodromy systems},  in \emph{Algebraic and Analytic Aspects of Integrable
  Systems and {Painlev{\'e}} Equations}, vol.~651 of \emph{Contemporary
  Mathematics}, pp.~125--155, American Mathematical Society (2015),
  \href{https://doi.org/10.1090/conm/651/13036}{DOI}
  [\href{https://arxiv.org/abs/1408.3847}{{\ttfamily 1408.3847}}].

\bibitem{GukovHaghighatLiuReshetikhin2024}
S.~Gukov, B.~Haghighat, Y.~Liu and N.~Reshetikhin, \emph{Irregular {KZ}
  equations and {Kac--Moody} representations},
  \href{https://doi.org/10.1088/1751-8121/ae35ec}{\emph{J. Phys. A} {\bfseries
  59} (2026) 055202} [\href{https://arxiv.org/abs/2412.16929}{{\ttfamily
  2412.16929}}].

\bibitem{Haghighat2025}
B.~Haghighat, \emph{{Liouville} {CFT}, matrix models and constrained {WZW}},
  \href{https://arxiv.org/abs/2508.06901}{{\ttfamily 2508.06901}}.

\bibitem{GaiottoTeschner2024}
D.~Gaiotto and J.~Teschner, \emph{Quantum analytic {Langlands} correspondence},
  \href{https://doi.org/10.21468/SciPostPhys.18.4.144}{\emph{SciPost Physics}
  {\bfseries 18} (2025) 144}
  [\href{https://arxiv.org/abs/2402.00494}{{\ttfamily 2402.00494}}].

\bibitem{EtingofFrenkelKirillov1998}
P.I.~Etingof, I.B.~Frenkel and A.A.~Kirillov, Jr., \emph{Lectures on
  Representation Theory and {Knizhnik--Zamolodchikov} Equations}, vol.~58 of
  \emph{Mathematical Surveys and Monographs}, American Mathematical Society,
  Providence, RI (1998),
  \href{https://doi.org/10.1090/surv/058}{10.1090/surv/058}.

\bibitem{SchechtmanVarchenko1990}
V.V.~Schechtman and A.N.~Varchenko, \emph{Hypergeometric solutions of
  {Knizhnik--Zamolodchikov} equations},
  \href{https://doi.org/10.1007/BF00626523}{\emph{Lett. Math. Phys.} {\bfseries
  20} (1990) 279}.

\bibitem{Kohno2014}
T.~Kohno, \emph{Local systems on configuration spaces, {KZ} connections and
  conformal blocks},
  \href{https://doi.org/10.1007/s40306-014-0088-6}{\emph{Acta Math. Vietnam.}
  {\bfseries 39} (2014) 575}.

\bibitem{NekrasovShatashvili2009}
N.A.~Nekrasov and S.L.~Shatashvili, \emph{Quantization of integrable systems
  and four dimensional gauge theories},  in \emph{Proceedings of the XVIth
  International Congress on Mathematical Physics}, pp.~265--289, World
  Scientific, 2010, \href{https://doi.org/10.1142/9789814304634_0015}{DOI}
  [\href{https://arxiv.org/abs/0908.4052}{{\ttfamily 0908.4052}}].

\bibitem{Li1997}
H.~Li, \emph{The physics superselection principle in vertex operator algebra
  theory}, \href{https://doi.org/10.1006/jabr.1997.7126}{\emph{J. Algebra}
  {\bfseries 196} (1997) 436}.

\bibitem{Jimbo_2008}
M.~Jimbo, H.~Nagoya and J.~Sun, \emph{Remarks on the confluent {KZ} equation
  for {$\mathfrak{sl}_{2}$} and quantum {Painlev{\'e}} equations},
  \href{https://doi.org/10.1088/1751-8113/41/17/175205}{\emph{Journal of
  Physics A: Mathematical and Theoretical} {\bfseries 41} (2008) 175205}.

\bibitem{KacRoanWakimoto2003}
V.G.~Kac, S.-s.~Roan and M.~Wakimoto, \emph{Quantum reduction for affine
  superalgebras},
  \href{https://doi.org/10.1007/s00220-003-0926-1}{\emph{Communications in
  Mathematical Physics} {\bfseries 241} (2003) 307}
  [\href{https://arxiv.org/abs/math-ph/0302015}{{\ttfamily math-ph/0302015}}].

\bibitem{GiribetNakayamaNicolas2008}
G.~Giribet, Y.~Nakayama and L.~Nicolas, \emph{Langlands duality in
  {Liouville}--{$H_{3}^{+}$} {WZNW} correspondence},
  \href{https://doi.org/10.1142/S0217751X09044607}{\emph{Int. J. Mod. Phys. A}
  {\bfseries 24} (2009) 3137}
  [\href{https://arxiv.org/abs/0805.1254}{{\ttfamily 0805.1254}}].

\bibitem{Frenkel2005LanglandsLectures}
E.~Frenkel, \emph{Lectures on the {Langlands} program and conformal field
  theory},  in \emph{Frontiers in Number Theory, Physics, and Geometry {II}},
  pp.~387--533, Springer (2007),
  \href{https://doi.org/10.1007/978-3-540-30308-4_11}{DOI}
  [\href{https://arxiv.org/abs/hep-th/0512172}{{\ttfamily hep-th/0512172}}].

\bibitem{FeiginFrenkel1992}
B.~Feigin and E.~Frenkel, \emph{Affine {Kac--Moody} algebras at the critical
  level and {Gelfand--Dikii} algebras},
  \href{https://doi.org/10.1142/S0217751X92003781}{\emph{International Journal
  of Modern Physics A} {\bfseries 7} (1992) 197}.

\bibitem{Frenkel2005Wakimoto}
E.~Frenkel, \emph{Wakimoto modules, opers and the center at the critical
  level}, \href{https://doi.org/10.1016/j.aim.2004.08.002}{\emph{Advances in
  Mathematics} {\bfseries 195} (2005) 297}
  [\href{https://arxiv.org/abs/math/0210029}{{\ttfamily math/0210029}}].

\bibitem{Desrosiers2009}
P.~Desrosiers, \emph{Duality in random matrix ensembles for all {$\beta$}},
  \href{https://doi.org/10.1016/j.nuclphysb.2009.02.019}{\emph{Nucl. Phys. B}
  {\bfseries 817} (2009) 224}
  [\href{https://arxiv.org/abs/0801.3438}{{\ttfamily 0801.3438}}].

\bibitem{BorotGuionnet2013}
G.~Borot and A.~Guionnet, \emph{Asymptotic expansion of {$\beta$} matrix models
  in the multi-cut regime},
  \href{https://doi.org/10.1017/fms.2023.129}{\emph{Forum of Mathematics,
  Sigma} {\bfseries 12} (2024) e13}
  [\href{https://arxiv.org/abs/1303.1045}{{\ttfamily 1303.1045}}].

\end{thebibliography}\endgroup
\end{document}